\documentclass[enabledeprecatedfontcommands,bibliography=totoc,listof=totoc,index=totoc,twoside=true,BCOR=12mm,DIV=12,openany]{scrbook}
\usepackage{cite}
\usepackage[utf8]{inputenc}                       
\usepackage[bachelor]{mdsg}
\lmutitle{Linear and non-linear machine learning attacks on physical unclonable functions}
\lmustudentone{Michael Lachner}
\lmuprofone{Prof. Dr. Claudia Linnhoff-Popien}
\lmuadvisorone{Steffen Illium}
\lmuadvisortwo{Markus Friedrich}
\lmuadvisorthree{Prof. Dr. Dr. Ulrich Rührmair}
\lmudeadline{12.07.2021}                      
\usepackage{cmap}                                 
\usepackage[T1]{fontenc}                          
\usepackage{lmodern}
\usepackage[english]{babel}                       
\usepackage{tabularx}                             
\usepackage{booktabs}                             
\usepackage{rotating}                             
\usepackage{multirow}                             
\usepackage{amssymb,amsmath}                      
\usepackage{hyperref}                             
\lmuhypersetup                                    
\usepackage{flafter}                              
\usepackage{subfig}                               
\usepackage{pdflscape}                            
\usepackage{hyphenat}                             
\usepackage[all]{hypcap}                          
\usepackage{url}                                  
\usepackage{enumitem}                             
\usepackage{graphicx} 
\usepackage{subfig} 
\usepackage{caption}

\graphicspath{{./pictures/}}                      
\begin{document}
    \setlist{noitemsep}                           
    \lmufront                                     
    \newpage
    \cleardoublepage
    \newpage
    \cleardoubleemptypage
    \thispagestyle{empty}
    \vspace*{2cm}

\begin{center}
    \textbf{Abstract}
\end{center}

\vspace*{1cm}

\noindent In this thesis, several linear and non-linear machine learning attacks on optical physical unclonable functions (PUFs) are introduced. For this purpose, a simulation of such a PUF is implemented to generate a variety of datasets that differ in several factors, to find the superior simulation setup, as well as to investigate the behaviors of the machine learning attacks under different circumstances. The main focus lies on the number of bits of the applied challenges and multiple challenge restrictions, which are intended to increase the input-output-complexity of the simulated PUF. All the datasets are evaluated with respect to the individual samples and their correlations among each other. In the following, both linear and deep learning approaches are used to attack these PUF simulations to comprehensively study the influence of the varying factors on the datasets with respect to their security level against adversaries. An additional focus will lie on the different behaviors of both attacking methods, i.e. the major differences in their performances and which approach should be preferred under which circumstances. Several independent metrics are used to highlight these differences from varying perspectives. Furthermore, multiple enhancements to these first machine learning models and new attacks that fall into the two categories will successively be introduced and investigated, while aiming for gradually better modeling performances. This leads to the development of an attack that is able to almost perfectly predict the outputs of the simulated PUF. Also, data from a real optical PUF will be investigated and compared to those from the simulation. This data is further used to see how the introduced machine learnings models would perform in the real world. Here, quite impressive results could be found, such that all the models fulfilled the defined criterion for a successful machine learning attack. For all the datasets, there are quite large differences between both the linear and non-linear attacking approaches, which this thesis will try to thoroughly elaborate on and give further insights into the benefits of differing architectures.
    \thispagestyle{empty}
    \frontmatter                                  
    \tableofcontents                              
    \mainmatter                                   
%
%
    \chapter{Introduction}\label{sec:Introduction}
In our technologically steadily advancing society, electronic devices have become ubiquitous in almost all parts of our everyday life. In the course of this, cryptography has taken an exceedingly important role in the security of the transportation of information. In general, such a cryptographic algorithm should be very easy and fast to compute, but virtually infeasible to invert. Thus, the overhead of the encryption process should be minimized, while at the same time the effort to decrypt the message for outside parties should be maximized. Many modern approaches use a binary key-pair, which is composed of a \textit{private key} and a \textit{public key}, which is generated using the private key. The latter is publicly accessible and can be used by anyone to encrypt messages, such that decrypting the message requires the private key that was used to create the public key in the first place. Given an adversary has no other knowledge about the process, they can only apply brute-force attacks to guess the private key, which, needless to say, will statistically not produce any results in a reasonable time frame. Therefore, most attacks instead focus on acquiring the private key used in the public key generation. Once the private key is known, all encryptions carried out via this key-pair can effortlessly be read out by the adversary. There already exist dozens of approaches, such as physical key extraction or viruses, which have been successfully applied to attack such systems \cite{eisenbarth2008power, kasper2010all, boneh1998attack}.

One approach to avert these types of attacks is not to use a digital key for the encryption process, but instead a so-called \textit{physical unclonable function} (PUF). PUFs take advantage of the properties of a physical structure, which are tightly bound to the instance of this system and vary strongly between different instances. This is supposed to make the PUF resilient to cloning and reverse engineering, hence the name physical \textit{unclonable} function. In general, a PUF is presented with some kind of specific external stimulus, often called Challenge $C_i$, which the PUF reacts upon with an output called Response $R_i$. The PUF can therefore be thought of as a function $f$ that maps a domain of challenge $C$ onto a domain of responses $R$ in constant time. Since no digital keys are used in this process, the classical aforementioned attacks cannot be applied on PUFs.

A concrete example of such a PUF would be a so-called \textit{optical PUF}, which relies on speckle patterns that are the result of the interference of several coherent wavefronts \cite{dainty2013laser}. Typically, a laser beam that propagates through an inhomogenous material, most of the time several randomly placed microstructures, is used. These structures scatter the incoming light, which is an uncontrolled process that is based on highly complex interactions between all wavefronts that emerge during this process and is strongly dependent on the initial incoming electromagnetic field as well as the position, size and shape of the used scatterers. \cite{pappu} Creating a model of this type of PUF would require to divide the whole system into voxels of the size of the laser beam's wavelength and to exactly solve Maxwell's equation, which is a numerically overly laborious task \cite{ruhrmair2013optical}. However, there have been reports of vulnerabilities of these types of systems to machine learning attacks \cite{ruhrmair2013optical, atakhodjaev2018machine}, which can be attributed to the linearity in the process of these systems. In such a scenario, the adversary gains access to the speckle pattern outputs of the PUF as well as the applied challenges and trains a model to learn an approximation to the underlying function. This knowledge is then used to further predict the patterns that will emerge for new challenges.

In this thesis, different machine learning attacks on optical PUFs will be tested, including several linear as well as non-linear attacks. The necessary data will be created using a software specialized on optical interference problems by simulating the behavior of a real optical PUF as close as possible. While the generation of large datasets on real optical PUFs can be a laborious and expensive operation, simulating the data allows for a comparably easy and fast computation and facilitates the construction of multiple varying datasets. Therefore, the machine learning attacks will be used on several datasets with different properties to draw conclusions on the strengths and weaknesses of the attacks themselves as well as the properties of the datasets and the underlying simulation.

    \chapter{Related Work}\label{sec:related-work}
\subsubsection*{Physical Unclonable Functions (PUFs)}
One of the most well known works in the optical PUF research landscape is called \textit{"Physical One-Way Functions (POWFs)"} from  Pappu et al. (2002) \cite{pappu}, where they presented one of the first PUFs and the first version of an optical PUF. They suggested a one-way function usable in modern cryptography by exploiting highly complex physical interactions. On this account, they proposed to take advantage of the properties that are shown by light when it propagates through disordered media. They used a light source that emits coherent radiation, which propagates through a manufactured token that contains many microscopic particles, which scatter the incoming light into many different directions. This creates a set of coherent wavefronts, which produce highly complex interference patterns throughout the whole medium. At the opposing end of the device, an unique speckle pattern emerges, which is highly dependent on the exact events that happened within the PUF and is very sensitive to tiny changes within the system. The positioning of the scattering particles is carried out uncontrolled by distributing them completely randomly and therefore makes it virtually close to impossible to copy the physical system.

Another work on this topic is from Tuyls and Skoric \cite{petkovic2007security} from 2007, who discussed how optical PUFs can be used for security aspects on data management systems. On this account, they elaborate on the unclonability of the system itself, the cost-effective storage of the cryptographic key information as well as the PUF as a system for strong authentication. Furthermore, they suggested an integrated version of an authentication token that uses an optical PUF, combining the challenge application, scattering process and the response detection into a single small, integrated package. However, their suggestion was only theoretical, and no real experimental prototypes were constructed. Tuyls and Skoric also further explored secure key storage with PUFs in \cite{tuyls2007secure} and \cite{tuyls2007security}.

In 2013, Rührmair et al. suggested three new types of integrated optical PUFs and actually manufactured one of these in \cite{ruhrmair2013optical}. Their integrated version directly allows to miniaturize optical PUFs, and, in contrast to previous miniaturized suggestions as in \cite{skoric2007experimental}, works without any moving components. This removes the need for a complex readout mechanism as it was necessary for the optical PUF of Pappu et. al. \cite{pappu}. Furthermore, they proposed an attacking algorithm on their integrated optical PUF and showed that, under the assumption that the adversary has access to the raw speckle patterns, the PUF can successfully be modelled with extremely high accuracy using machine learning.

Additionally, further types of enhancements and other variants of optical PUFs have been published. These include reconfigurable optical PUFs \cite{kursawe2009reconfigurable, horstmeyer2015physically, jacinto2021utilizing}, optical PUFs using noise-immune nanostructures to increase the robustness of responses \cite{lu2018cmos} and optical PUFs using a single optical waveguide as physical token to even further decrease physical clonabiliy \cite{mesaritakis2018physical}. Furthermore, PUFs have been used to provide optical channel communication for vehicle to vehicle authentication within vehicular environments \cite{dolev2016optical} and for anti-counterfeiting of manufactured metallic goods \cite{dachowicz2018optical}.

In addition to \textit{optical} PUFs, numerous different types of PUFs based on other physical systems have been suggested. A large focus has been set on PUFs that take advantage of uncontrollable variations during operations in electrical circuits, such as ring-oscillators \cite{maiti2009improving, yin2009temperature}, arbiter-based PUFs \cite{arbiterPUF, gassend2002silicon}, SRAM PUFs \cite{xu2015reliable, holcomb2008power} and many more \cite{chen2011bistable,chen2009analog,ruehrmair2012method,ruhrmair2010towards,jaeger2010random,katzenbeisser2011recyclable,delvaux2019machine,nedospasov2013invasive,koeberl2013memristor,chatterjee2018rf,mazady2015memristor,zhang2014survey,shi2019approximation,maes2008intrinsic,yu2016lockdown,delvaux2013side,grubel2017silicon,horstmeyer2013physical,buchanan2005fingerprinting,dejean2007rf,lakafosis2010rf,zalivaka2018reliable,liu2017acro,john2021halide,rosenfeld2010sensor,rose2013hardware,tang2016securing,majzoobi2012slender,rostami2014robust,maes2009soft,lugli2013physical,ruhrmair2010applications,ruhrmair2012simpl,sauer2017sensitized,kappelhoff2022strong,langhuth2011strong,csaba2010application,ruhrmair2015virtual,ruhrmair2011simpl,ruhrmair2009simpl,jin2020erasable,jin2015playpuf,orosa2022spyhammer,eliezer2022exploiting,jin2022programmable,chen2011circuit,gao2018efficient,csaba2009chip,pavanello2021recent}.  Interested readers are referred to some of the existing PUF-surveys and tutorials for further studies \cite{ruhrmair2014pufs,ruhrmair2019towards,ruhrmair2020sok,ruhrmair2022secret,ruhrmair2012security,ruhrmair2009foundations}.

\subsubsection*{Machine Learning and PUF Attacks}
Due to their applicability in countless varying domains, the field of machine learning is significantly broader and can be divided into numerous subcategories. Linear models such as linear regression, which form one of these categories, are among the most widely used statistical techniques and have been intensively optimized to fit to different problems. Detailed information on these types of models and their modifications can be found in \cite{searle2016linear}, \cite{mccullagh2019generalized} and \cite{rencher2008linear}. The range of applications of linear models is huge and include genetic studies \cite{roehe2000estimation, pirinen2013efficient}, economics \cite{chursin2017linear, henshaw1966application} and even human behavior \cite{brehmer1994psychology, dawes1974linear}. A further important category of machine learning is deep neural networks, which have significantly risen in popularity over the last two decades. Nowadays, they are quite common in various fields, such as speech recognition, image recognition, natural language processing and so on. Deep neural networks themselves can again be split into several further categories, one of which are generative neural networks. These types of networks have been used in dialogue generation \cite{serban2015hierarchical, serban2016building}, image reconstruction \cite{mosser2017reconstruction, rivenson2018phase} and even unconventional fields such as quality prediction for work-in-progress products \cite{wang2019generative} or wind speed forecasting \cite{khodayar2018interval}. Another branch of these networks that has become increasingly popular is image generation \cite{gregor2015draw, yan2016attribute2image, taigman2016unsupervised, oord2016conditional}.

So far, there have been several reports of successful attacks on electrical PUFs, including modeling attacks \cite{nguyen2015efficient, modellingRuehrmair, modellingRuehrmair2, ruhrmair2014puf,sehnke2010policy} as well as side-channel attacks \cite{sideChannelRuehrmair, sideChannelTajik,ruhrmair2013power,ruhrmair2014special} or even physical cloning in the case of SRAM PUFs \cite{helfmeier2013cloning}. However, the amount of research on the topic of attacks on optical PUFs is rather sparse. While integrated versions of optical PUFs have been successfully attacked \cite{ruhrmair2013optical}, surprisingly, no successful machine learning attacks on a PUF following the principle of Pappu et. al. \cite{pappu} have been reported up to this point \cite{ruhrmair2013optical,atakhodjaev2018machine}.  

To complete this section, another central Strong PUF attack vector should not be missed, namely PUF protocol attacks.  In the latter, not the security features of the PUF itself are attacked, but rather their specific use in a given protocol or application.  This attack type has emerged over the last ten years, and actually constitutes one of the most efficient and inexpensive attacks  forms.  Since it is not our main concern in this thesis, interested readers are again referred to the literature for further investigations on PUF protocols and PUF protocol attacks \cite{ruhrmair2013pufs,ruhrmair2010strong,ruhrmair2010oblivious,ruhrmair2013practical,ruhrmair2012practical,ruhrmair2011physical,van2014protocol,ruhrmair2016security,dachman2014feasibility,ostrovsky2013universally}.  

\subsubsection*{Contributions of This Thesis}
This work will make the following contributions to this field of research: Firstly, an integrated optical PUF similar to the one from Rührmair et al. \cite{ruhrmair2013optical} will be simulated to artificially generate several datasets of challenge-response-pairs (CRPs). The results of this simulation will be thoroughly investigated, including different restrictions on the challenge generation to evaluate their influence on the output complexity of the system. Furthermore, the CRPs will be used for different machine learning algorithms, which consist of several linear and non-linear modeling attacks. In the beginning, the focus will lie on simple linear regression and one generative neural network, while adding further modifications of the linear model and further deep learning models later on as well. It will be investigated which algorithms work better on which datasets and which factors may be responsible for the differences that may be found. The algorithms will be tried to be optimized in the course of this thesis to achieve the best possible results with the models at hand. In addition to the evaluation of the simulation and the attacks, a dataset from a real integrated optical PUF will be investigated as well, to be able to draw comparisons across the results from a simulated and a real optical PUF where it is applicable.

    \chapter{Background}\label{sec:background}
To grasp how precisely optical PUFs work and to later understand what the benefits and disadvantages of using them are, the first part of this chapter will dive deeper into the processes that take place within these systems. Afterwards, the focus will switch to the machine learning algorithms that will be used later on to attack the optical PUF. Lastly, further insights into important metrics that will be used in this thesis will be provided.

\section{Optical PUFs}
The type of optical PUF that will be the subject of this thesis follows the previously explained principle established by Pappu et al. \cite{pappu}. Here, light propagates through a disordered medium to create an unique speckle pattern that is highly dependent on several factors, such as the structure, positioning and number of the spheres or the positioning of the laser and the readout system for the speckle pattern. To ensure the necessary complexity, the light has to undergo multiple scattering events before exiting the token. The scatterers are therefore randomly distributed in all three dimensions. Furthermore, it is important that there is enough space between two scatterers, so that optical interference effects can take place. Specifically, this distance needs to be larger than the wavelength of the incoming light. For the optical PUFs that will be used in this thesis, this inequality will easily be fulfilled. \cite{pappu}

Pappu et al. \cite{pappu} realized this proposition in the following way: Several glass spheres ranged in size from 500 microns to 650 microns were stirred into a transparent epoxy, taking care to not create any patterns. This mixture was poured into an 10 mm wide and 2.54 mm thick pre-cut square aperture in the center of a sheet of plexiglass and waited for to set. A laser beam with a diameter of 1 mm was used to illuminate the microstructure, and the emerging speckle pattern was captured with a camera. Additionally, a complex system called \textit{reader} was built to allow moving the laser beam precisely to specific positions. This ensures a high readout stability, i.e. preserves the relative positions of all parts, such that carrying out the same experiment will result in the same speckle pattern. A challenge to the system is realized by the position and angle of the laser beam. \cite{pappu}

\begin{figure}
  \centering
  \includegraphics[width=1\textwidth]{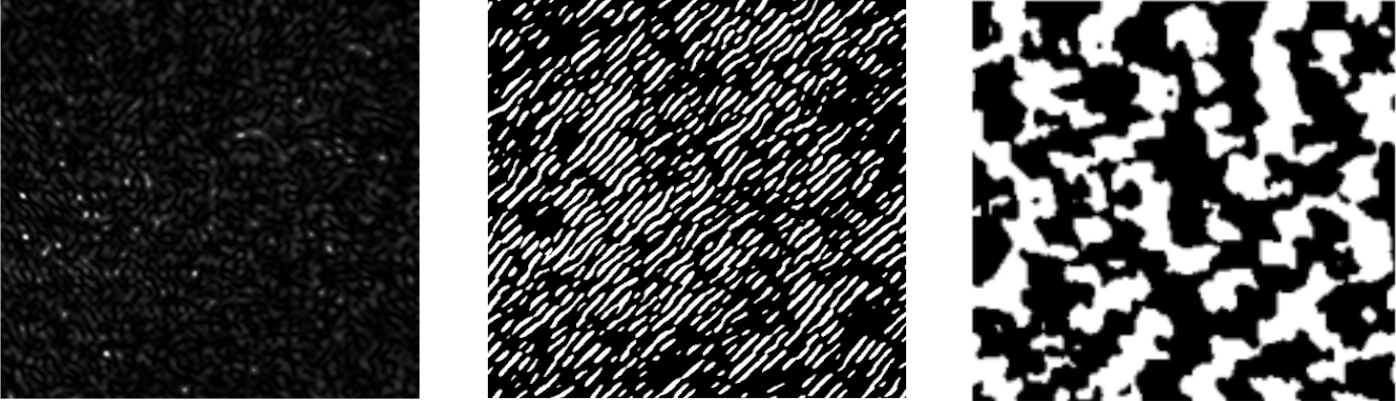}\\
  \caption[Examples of gabor-transformed images.]{An example speckle pattern and two images of a gabor-transformed speckle pattern for different parameters.}\label{fig:gabor_examples}
\end{figure}

The raw speckle patterns are usually too large to directly use them for security purposes. Additionally, they can be influenced by outside factors such as illumination, and inaccuracies during measuring sessions can lead to small spatial transformations of the speckles. Therefore, Pappu et al. \cite{pappu} suggest not to use them as the direct response and instead apply a post-processing algorithm, the \textit{2D Gabor transformation} proposed by Daugman \cite{gabor_daugman}, to the raw output. This transformation is mostly insensitive to the aforementioned effects and can furthermore be used to scale down the image. An example of how these transformed images may look like can be found in Figure \ref{fig:gabor_examples}. Using the value 0 as a threshold, the gabor-transformed image is further transformed into a 2D binary array, which is flattened to a 1D bitstring and finally represents the response of the PUF. Now working with bitstrings, the \textit{Fractional Hamming Distance} (FHD) was chosen as indicator for the decorrelation between responses. The FHD is defined as the Hamming distance of two bitstrings divided by their length. A FHD of $0.5$ indicates virtually decorrelated Gabor images and is the ideal value when comparing responses of the PUF, as this indicates that the two responses do not share any significant patterns and therefore cannot be distinguished from a randomly generated bitstring of the same length. \cite{pappu} Altogether, they report that the PUF can be presented with roughly $2.37 \times 10^{10}$ challenges for which the gabor-transformed outputs are uncorrelated.

\begin{figure}
  \centering
  \includegraphics[width=1\textwidth]{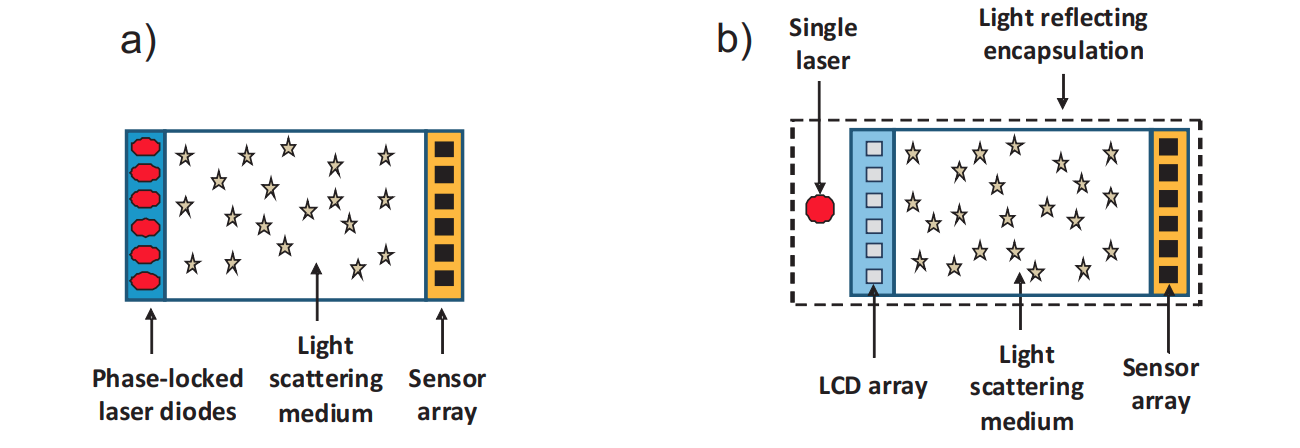}\\
  \caption[Integrated PUFs suggested by Rührmair et al.]{Two versions of integrated PUFs as suggested by Rührmair et al. \cite{ruhrmair2013optical}.} \label{fig:integrated_pufs}
\end{figure}

However, the whole structure cannot be easily transported, and miniaturization at some point becomes rather difficult due to the necessary precision during measuring. On this account, Rührmair et al. \cite{ruhrmair2013optical} proposed alternative versions called \textit{integrated optical PUFs}, with the intent to overcome these downsides. Similar to Pappu et al. \cite{pappu}, they used a medium containing several randomly distributed scattering particles. They gave two suggestions for the light source: Either a single fixed laser is used to illuminate a LCD array, or an array of phase-locked lasers is used directly. An illustration can be found in Figure \ref{fig:integrated_pufs}. In both cases, there are $k$ units which can be switched on and off by either opening or closing the corresponding pixel of the LCD array, or by turning the corresponding laser on or off. The system is attached in front of the medium containing the light scattering particles. On the opposite site, i.e. behind the medium, they use a sensor that registers the incoming scattered light. A challenge to this system corresponds to one configuration of the light source array, hence the system provides a challenge space of size $2^k$. The response is the bitstring that can be retrieved from the transformed version of the speckle pattern registered by the sensor. \cite{ruhrmair2013optical}

The concept of Rührmair et al. \cite{ruhrmair2013optical} works without any moving parts and can be manufactured quite easy. It also allows for miniaturization. However, they report that while finding no vulnerabilities for a setup similar to what Pappu et al. \cite{pappu} used, they were able to successfully attack the integrated version and predict the raw speckle patterns with very high accuracy. \cite{ruhrmair2013optical}

\section{Machine learning}\label{sec:backgroundML}
\textit{Machine learning} (ML) refers to a group of algorithms that attempt to detect meaningful patterns in data in an automated matter. In most cases, these patterns are too complex for a human to explicitly tell the program how to extract them, or sometimes even what the exact patterns to look for are. Nowadays, ML has become ubiquitous in many completely different fields, such as search engine optimization, e-mail filtering, security software, face and voice recognition, automated driving and much more. \cite{shalev2014understanding}

Two of the largest subcategories are \textit{supervised} and \textit{unsupervised learning}. The main difference here is that in the former, models are trained with datasets where the samples already have correctly assigned labels or are correctly categorized, while in the latter, models are presented with uncategorized/unlabeled data \cite{sathya2013comparison}. Since the common ML attacks on PUFs belong to the supervised learning category, this thesis is going to focus on these.

Specifically speaking, the goal in supervised learning is to infer a function $f: \mathcal{X} \rightarrow \mathcal{Y}$ from a training set $T$ of ordered pairs $(x_i, y_i) \in (\mathcal{X} \times \mathcal{Y})$, where $\mathcal{X}$ refers to the input and $\mathcal{Y}$ to the output space. It is preconditioned that the samples of the training data are independent and identically distributed pairs. To allow the model to measure how accurate its predictions with respect to the real outputs are, a \textit{loss function} $L: \mathcal{Y} \times \mathcal{Y} \rightarrow \mathbb{R}^+$ is used. This function describes the error between $f(x_i)$ and $y_i$ for a pair $(x_i,y_i)$, which the model tries to minimize. The choice for $L$ is highly dependent on the training data and the exact problem the model is working on. \cite{cunningham2008supervised} In regression problems in particular, \textit{quadratic loss functions} of the form $(f(x_i) - y_i)^2$ are frequently used, since they are simple to use and symmetric, i. e. an error above and below the target causes the same loss. In practice, a loss function is often applied on a whole batch of $n$ pairs. In such a case, for the two vectors $\bar{y}$ of the real and $\bar{x}$ of the corresponding predicted values, the \textit{Mean Squared Error} is a commonly used quadratic loss function and is defined as
$$MSE = {\frac {1}{n}}\sum _{i=1}^{n}(\bar{y} - \bar{x})^{2}.$$
For regression problems and linear systems, \textit{linear regression} (LR) is a common approach to infer the function $f$. LR assumes that there is a linear relationship between $\mathcal{X}$ and $\mathcal{Y}$, and each $y_i \in \mathcal{Y}$ can therefore be calculated as
$$y_i = \beta_1 x_{i1} + \beta_2 x_{i2} \dots + \beta_p x_{ip} + \epsilon_i,$$
where $\epsilon_i$ is called \textit{error variable} and represents the influence on $y$ that cannot be covered by the linear equation. This is relevant if the relationship is not completely linear and the regression is looking for the best linear approximation of $f$. Often, all $n$ equations of the $n$ pairs are stacked together and instead written in matrix notation as
$$y = X \beta + \epsilon,$$
where
$$y = \begin{pmatrix}y_1 \\ \vdots \\ y_n\end{pmatrix}, \quad
  X =\begin{pmatrix}x_1^{\mathsf{T}} \\ \vdots \\ x_n^{\mathsf{T}} \end{pmatrix} = \begin{pmatrix}x_{11} & \dots & x_{1p} \\ \vdots & & \vdots \\ x_{n1} & \dots & x_{np} \end{pmatrix},\quad
  \beta = \begin{pmatrix} \beta_1  \\ \vdots \\ \beta_p \end{pmatrix},\quad
  \epsilon = \begin{pmatrix} \epsilon_1 \\ \vdots \\ \epsilon_n \end{pmatrix}.$$
Here, $y$ is denoted as the \textit{dependent} variable and $x_1,...,x_p$ as the \textit{independent} variables. The elements of $\beta$ are the \textit{coefficients} for the independent variables. Regularly, there exists one independent variable which takes on the constant value $1$, such that the corresponding coefficient is called the \textit{intercept}. The goal of the regression is to model the relationship between the independent and dependent variables by estimating the coefficients. Standard LR models are usually fitted using a version of the aforementioned quadratic loss function, which is called \textit{ordinary least squares} (OLS) and aims to minimize the function
$$\sum_{i=1}^{n}(y_{i}-\sum_{j=1}^{p}X_{ij}\beta _{j})^{2} = \Vert y - X \beta \Vert^2.$$
From this, the estimator
$$\hat{\beta} = (X ^{\mathsf{T}} X)^{-1}X ^{\mathsf{T}}y$$
for the coefficients can be derived. \cite{gross2012linear}

Note that this model is restricted to only one dependent variable. When predicting multiple dependent variables, the corresponding model is called \textit{multivariate linear regression} and takes the form
$$y_{ij} = \beta_{1j} x_{i1} + \beta_{2j} x_{i2} + \dots + \beta_{pj} x_{ip} + \epsilon_{ij}$$
\textit{for each} independent variable $j$ and observation $i$. Since the coefficients are specific for each independent variable, the multivariate linear regression is equivalent to computing a standard linear regression for each $j$. \cite{yuan2007dimension}

\begin{figure}
  \centering
  \includegraphics[width=1\textwidth]{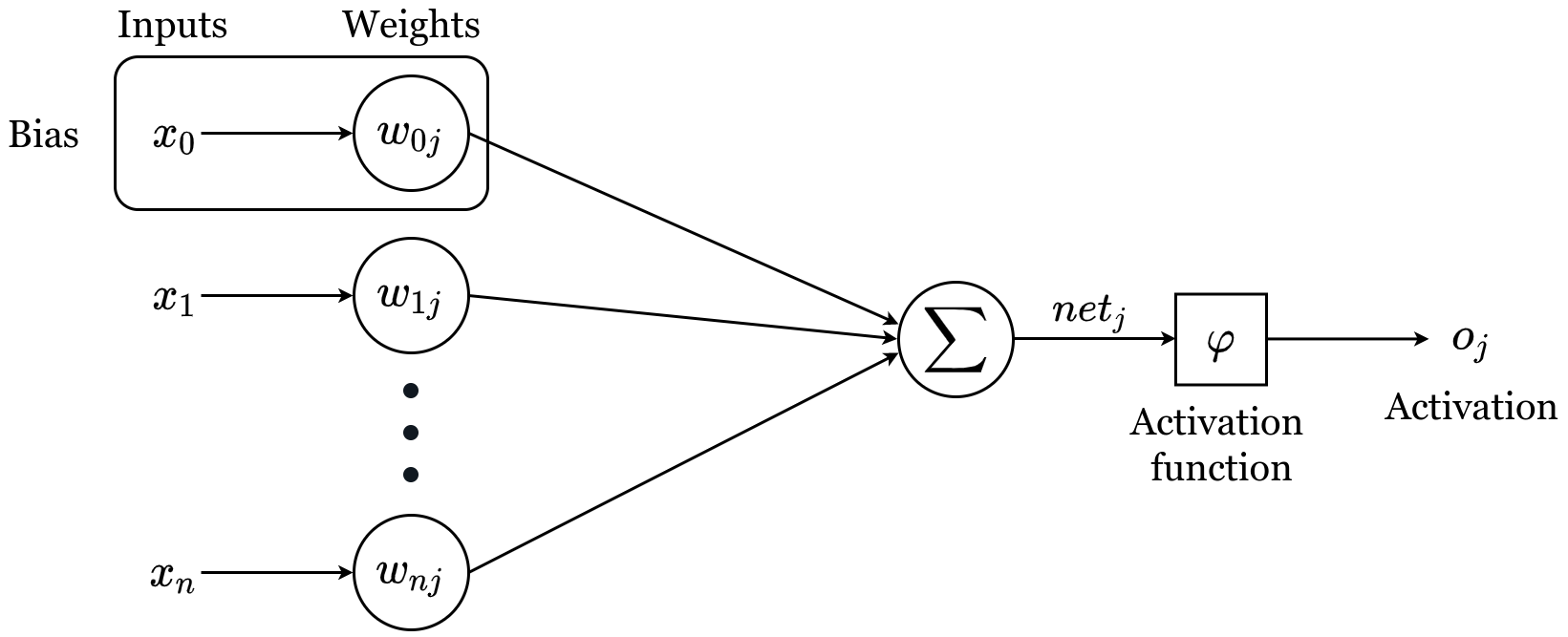}\\
  \caption[Concept of an artificial neuron.]{An illustration of the parts of an artificial neuron.}\label{fig:artificial_neuron}
\end{figure}

Another, more general approach to ML is \textit{deep learning} (DL), which takes the biological nervous system as role model and uses so-called \textit{artificial neural networks} (ANNs) to model complex interactions of large numbers of small components. An ANN consists of several concatenated \textit{layers}, which in turn consist of several \textit{artificial neurons}. Generally, a neuron $j$ receives $n$ numerical inputs $x_1,...,x_n$, which are each weighted with an input-specific value $w_1,...,w_n$. By applying the dot product to the vectors of the inputs and weights, the input to the neuron is transformed into a single numerical value and can be depicted as
$$net_j = x \cdot w = \sum_{i=1}^{n} x_{i}w_{ij}.$$
This value is then handed over to an activation function $\varphi$, which calculates the output $o_j$ of the neuron, also called \textit{activation}. There are numerous implementations for this function, but often the activation has to exceed a specific threshold. An example for this would be ReLU, which is defined as $f(x) = max(0,x)$. To avoid having to redefine the activation functions for different thresholds, the threshold is often realized by an additional input $x_0$ called \textit{bias} with the constant value $x_0 = 1$, while the threshold itself is realized in its weight $w_0$. \cite{priddy2005artificial} An illustration of this concept can be found in figure \ref{fig:artificial_neuron}.

The type of ANNs used in DL consist of several successive layers, hence the name \textit{deep} learning. To summarize, the ANN is supposed to approximate a certain function and can be given an input, which is propagated through the various layers of the network to produce a specific output.
This thesis is going to focus on \textit{convolutional neural networks} (CNNs), or \textit{transposed CNNs} in particular, which are \textit{feedforward neural networks}, i.e. the information only passes the network in one direction, so there are no cycles. Note that there are numerous other kinds of ANNs, e.g. \textit{recurrent neural networks}, in which data can theoretically flow in any direction. \cite{idoko2020deep}

CNNs receive their name from the mathematical \textit{convolution} operation and are most commonly used for analyzing image data. The core of these networks lies in the convolutional layers, which consist of several learnable \textit{kernels} that form a \textit{filter}. These kernels are convolved across the whole image, producing a feature map of that kernel for each area of the image it is applied to. As the network learns, filters become able to detect specific features in the input. There are many options to customize a convolutional layer. Firstly, the number and size of the filters can be adjusted. It is also possible to change the \textit{stride}, which defines how many pixels a kernel moves between the convolutions and therefore impacts the amount of overlap between the extracted features. A \textit{padding} adds artificial pixels around the image and can help in detecting features at the borders of the image. \cite{albawi2017understanding}

Whereas regular convolutional layers usually create smaller representations of its input with each layer, there are also \textit{transposed convolutional layers}, which carry out convolutions in the opposite direction and increase the shape by each step. This allows to create shapes and even whole images from an abstract representation. \cite{dumoulin2016guide}

To allow feedforward networks to learn, an algorithm called \textit{backpropagation} is applied. As the inputs to a neuron can be described by a vector, the whole layer can be described as a matrix. This allows to define the whole network as a combination of function compositions and matrix multiplications, using a single function $g$. Just like in LR, a loss function $L$ evaluates the quality of the output $g(x_i)$ with respect to the corresponding label $y_i$ for a sample $(x_i, y_i)$ of the training set. Backpropagation then computes the gradient of the loss function for a fixed sample with respect to the weights. Since $g$ is a composition of differentiable functions, the \textit{chain rule} can be applied to compute the individual components of the gradient for each layer recursively, propagating through the network backwards from the output to the input layer. Lastly, the weights are updated accordingly by moving in the direction of the negative gradient, aiming to decrease the value of the loss function for the next iteration and therefore increasing the performance of the network. \cite{Goodfellow-et-al-2016}

\section{Important metrics}\label{sec:metrics}
In this thesis, there will be several occasions were certain metrics are used to evaluate the results. This section quickly summarizes what those metrics are and how they are defined.

First, the so-called \textit{Shannon entropy}, often simply called entropy, is a numerical value that represents the amount of information within a variable. For a random variable $X$ and all possible outcomes $x_1,...,x_n$, which occur with the probability $P(x_1),...,P(x_n)$, the entropy is defined as
$$-\sum_{i=1}^{n}P(x_i)logP(x_i),$$
where the base of the logarithm corresponds to the choice of the unit for measuring information and is commonly chosen as either $2$, $e$ or $10$. \cite{shannon2001mathematical}

Another metric is the \textit{Pearson correlation coefficient} (PCC), which, given a sample of pairs $(x_1,y_1),...,(x_n,y_n)$, is defined as
$$PCC = {\frac {\sum _{i=1}^{n}(x_{i}-{\bar {x}})(y_{i}-{\bar {y}})}{{\sqrt {\sum _{i=1}^{n}(x_{i}-{\bar {x}})^{2}}}{\sqrt {\sum _{i=1}^{n}(y_{i}-{\bar {y}})^{2}}}}},$$
where $n$ is the size of the sample, $x_i$ and $y_i$ are the values of the $i$-th entry in the sample, and $\bar{x}= \frac{1}{n} \sum_{i=1}^{n}x_i$ as well as $\bar{y}= \frac{1}{n} \sum_{i=1}^{n}y_i$ are the means of the samples. It is a common measurement for the linear correlation between two datasets. The PCC can vary in the interval $[-1,1]$, where $PCC > 0$ indicates a positive and $PCC < 0$ a negative correlation between both datasets. \cite{lee1988thirteen} When referring to the PCC, the ``\textit{coefficient}'' is often omitted from the term, so the abbreviation PC will be commonly used in this thesis as well.

Lastly, the \textit{Structural Similarity Index} (SSIM) is a perceptual metric to quantify the similarity between two given images. It compares three components, namely the luminance $l$, contrast $c$ and structure $s$. For two images $x$ and $y$, these are defined as
$$ l(x,y)={\frac {2\mu _{x}\mu _{y}+C_{1}}{\mu _{x}^{2}+\mu _{y}^{2}+C_{1}}},\quad
   c(x,y)={\frac {2\sigma _{x}\sigma _{y}+C_{2}}{\sigma _{x}^{2}+\sigma _{y}^{2}+C_{2}}},\quad
   s(x,y)={\frac {\sigma _{xy}+C_{3}}{\sigma _{x}\sigma _{y}+C_{3}}}.$$
Here, $\mu _{x}$ and $\mu _{y}$ are the means and $\sigma _{x}$ and $\sigma _{y}$ the variances of $x$ and $y$. Further, $\sigma _{xy}$ is the covariance of $x$ and $y$ and $C_1$, $C_2$ and $C_3$ are constants, which are used to avoid instabilities when the corresponding parts in the denominator are close to $0$. Specifically speaking, these constants are defined as
$$C_1=(K_1L)^2, \quad
  C_2=(K_2L)^2, \quad
  C_3=\frac{C_2}{2},$$
where $L$ is the dynamic range of the pixel values, e.g. 255 for 8-bit grayscale images, and $K_1 \ll 1$ and $K_2 \ll 1$ are again constants, usually $K_1 = 0.01$ and $K_2=0.03$. These three components are combined into a single value, defining the SSIM as
$$ SSIM(x,y)=\left[l(x,y)^{\alpha }\cdot c(x,y)^{\beta }\cdot s(x,y)^{\gamma }\right],$$
where $\alpha$, $\beta$ and $\gamma$ are different weights for each of the components. When setting these to $\alpha = \beta = \gamma = 1$, the whole formula can be simplified to
$$ SSIM(x,y)=\frac {(2\mu_x \mu_y + C_1)(2\sigma_{xy} + C_2)}{(\mu_x^2+\mu_y^2 + C_1)(\sigma_x^2 + \sigma_y^2 + C_2)}.$$
While it is possible to calculate the SSIM globally on the whole image at once, it usually operates using several windows of the same size, which move pixel-by-pixel across the image and compute the SSIM locally for each window. The resulting SSIM is the mean of all of the individual values for all windows.  In this case, the SSIM is also referred to as \textit{Mean Structural Similarity Index} (MSSIM), but is often still simply called SSIM. The resulting values move in the interval $[-1, 1]$, where a higher value indicates a higher similarity between the images. \cite{wang2004image}

    \chapter{Concept}\label{sec:Concept}
The first idea of this thesis is about simulating an optical PUF, so that there is no need to manufacture any physical parts or to carry out the laborious measurements. The kind of PUF that will be simulated is closest to the second integrated optical PUF suggested by Rührmair et al. \cite{ruhrmair2013optical}, where a single laser is used as light source that illuminates a LCD array. Different configurations of the LCD array, where certain blocks are turned on or off, correspond to a challenge to the system. The light will then propagate through a cluster of spheres, where it scatters into multiple directions and interference effects can take place. Behind this cluster, the intensity of the outgoing electric field will be measured, and the emerging speckle pattern will be transformed using the Gabor transformation to receive a bitstring as the response. Due to restrictions imposed by the simulation software, there will be no light reflecting encapsulation around the simulated PUF. Hereinafter, when there is no further information given, the simulated version of this optical PUF will also simply be referred to as PUF.

\begin{figure}[t]
  \centering
  \includegraphics[width=0.4\textwidth]{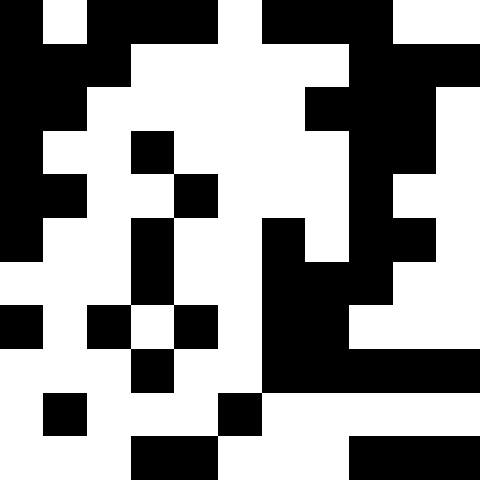} \hspace{1cm}
  \includegraphics[width=0.4\textwidth]{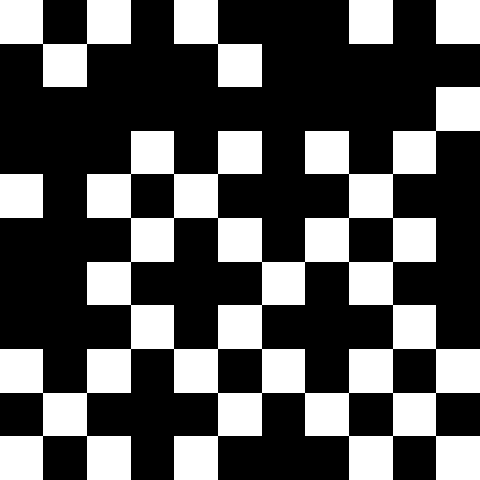}
  \caption[Examples challenges for the simulated PUF.]{An example of a possible challenge of $11 \times 11$ blocks of type A (left) and type B (right).}
  \label{fig:sample_challenges}
\end{figure}

The simulation will be carried out multiple times with different setups, focusing on the compositions of the challenges. It will be investigated whether, and if so how the results differ for varying challenge sizes and which kinds of downsides come along with increasing the challenge space. Furthermore, for each of these challenge sizes, there will be four approaches regarding the compositions of the challenges. The first one will be simply not to have any restrictions, i.e. all blocks can in theory be activated. This type of dataset will be referred to as \textit{type A}. The following approaches will each restrict the composition of the challenges in some way. The first restriction will be to use only every second bit in a row of the challenge mask and to disable the rest, alternating whether to use even or odd positions by each row. This prevents directly vertically or horizontally neighboring blocks from being activated simultaneously and is hoped to increase the overall decorrelation between responses, as the simulation shows that the influence on the speckle pattern appears to be smaller when light travels through neighboring pixels, compared to locally unrelated pixels. The corresponding datasets will be called \textit{type B}. An example of how such a challenge could look like in comparison to one without restrictions can be found in figure \ref{fig:sample_challenges}. The other restriction will be to define an upper boundary for how many blocks can be activated simultaneously, while there is no local restriction on which blocks these can be. It was observed that when a large number of blocks is activated, the amount of light that travels through the PUF already extensively illuminates the whole structure, so further challenges with more activated blocks lead to strongly correlating speckle patterns, since the light that travels through further opened blocks does not greatly influence the internal events. The upper boundary is supposed to work against this and decrease the overall correlation between CRPs. There will be two of these restrictions: one, where only half of the blocks and one, where only two-thirds of the blocks can be activated, referred to as \textit{type C} and respective \textit{type D}.

For all of these datasets, two machine learning attacks of two different categories will be applied: a linear modeling attack and a deep learning attack. The linear modeling attack is realized by a linear regression and the deep learning attack by a model here referred to as \textit{generator}, which closely relates to the principle of the generator of a GAN \cite{goodfellow2014generative} or the decoder of an autoencoder \cite{Goodfellow-et-al-2016}. Both approaches will be trained on all available datasets, to see if there are any fundamental differences in the complexity of the systems, as well as to compare both concepts and see if there are any differences in their modeling performances.

For the evaluation of the simulated PUF, the main metric that will be used is the \textit{fractional Hamming distance} (FHD). The FHD of a dataset will be computed by taking a random subset of all available CRPs, calculating the FHD between each of the responses and taking the mean. The optimal value, in this case, would be $0.5$, which indicates complete decorrelation between the responses. While the FHD is used as the metric to compare two responses, the \textit{Shannon entropy} will be used to measure the quality of a single response. In this thesis, the outcomes $x_1,...,x_n$ of the random variable used in the entropy computation correspond to all different values of the pixels of the response, and $P(x_i)$ is the probability of a pixel having the value $x_i$ based on the frequency of its appearance. The entropy of a dataset is computed by calculating the entropy of all responses and again taking the mean. While it may be rather difficult to draw meaningful conclusions from the absolute entropy value, a comparison across the different setups may portray useful implications regarding the amount of information stored in a response.

\begin{figure}
\centering
  \includegraphics[width=1\linewidth]{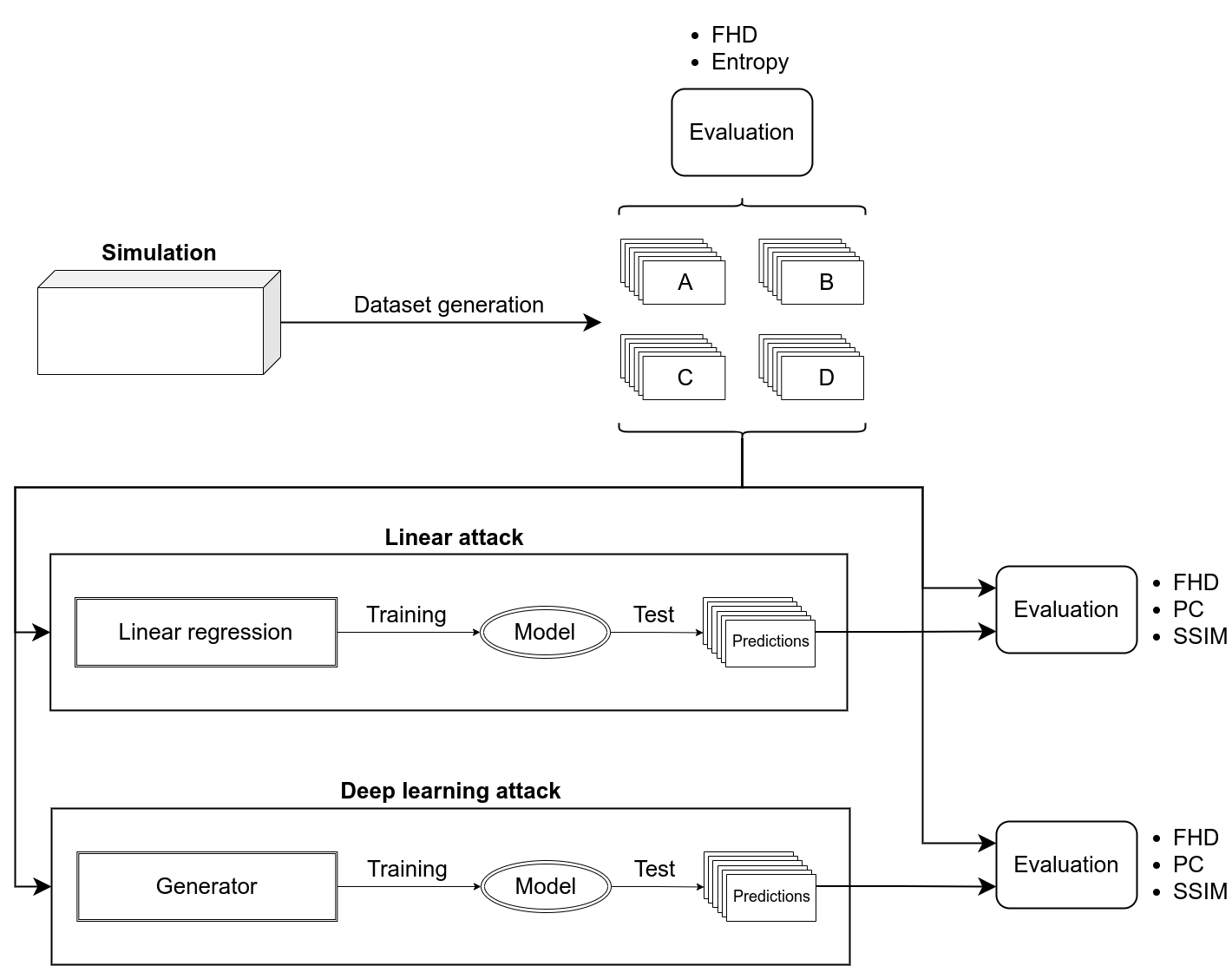}
\caption[Illustration of the concept.]{An illustration that shows all components and the procedure of the concept of this thesis.}
\label{fig:concept}
\end{figure}

The FHD will also be used for evaluating the ML attacks, by calculating the FHD between the real response of the simulation and the generated response of the ML model. In this case, a FHD of $0$ would imply that the model perfectly recreates the response. Furthermore, there will be two supplementary metrics used to evaluate the quality of the predictions. The first is the \textit{Pearson correlation coefficient} (PC), where here a sample corresponds to a pair of a real and a predicted response. For the previously defined formula, here, $n$ is the number of pixels, $x_i$ and $y_i$ are the values of the $i$-th pixel, and $\bar{x}$ as well as $\bar{y}$ are the means of all pixel values. The ideal value for the PC would be 1, which implies a perfect positive correlation and therefore a perfect prediction. The last metric that is used is the \textit{Structural Similarity Index} (SSIM), specifically the \textit{Mean Structural Similarity Index}, which will be simply referred to as SSIM, as is common practice. Just as for the PC, a value of 1 would be the optimal case. Each of these three metrics will be computed for all pairs of real and predicted responses, taking the mean as the final value. An illustration of the whole concept can be found in figure \ref{fig:concept}.

    \chapter{Implementation}\label{sec:implementation}
\begin{figure}
\centering
  \includegraphics[width=1\linewidth]{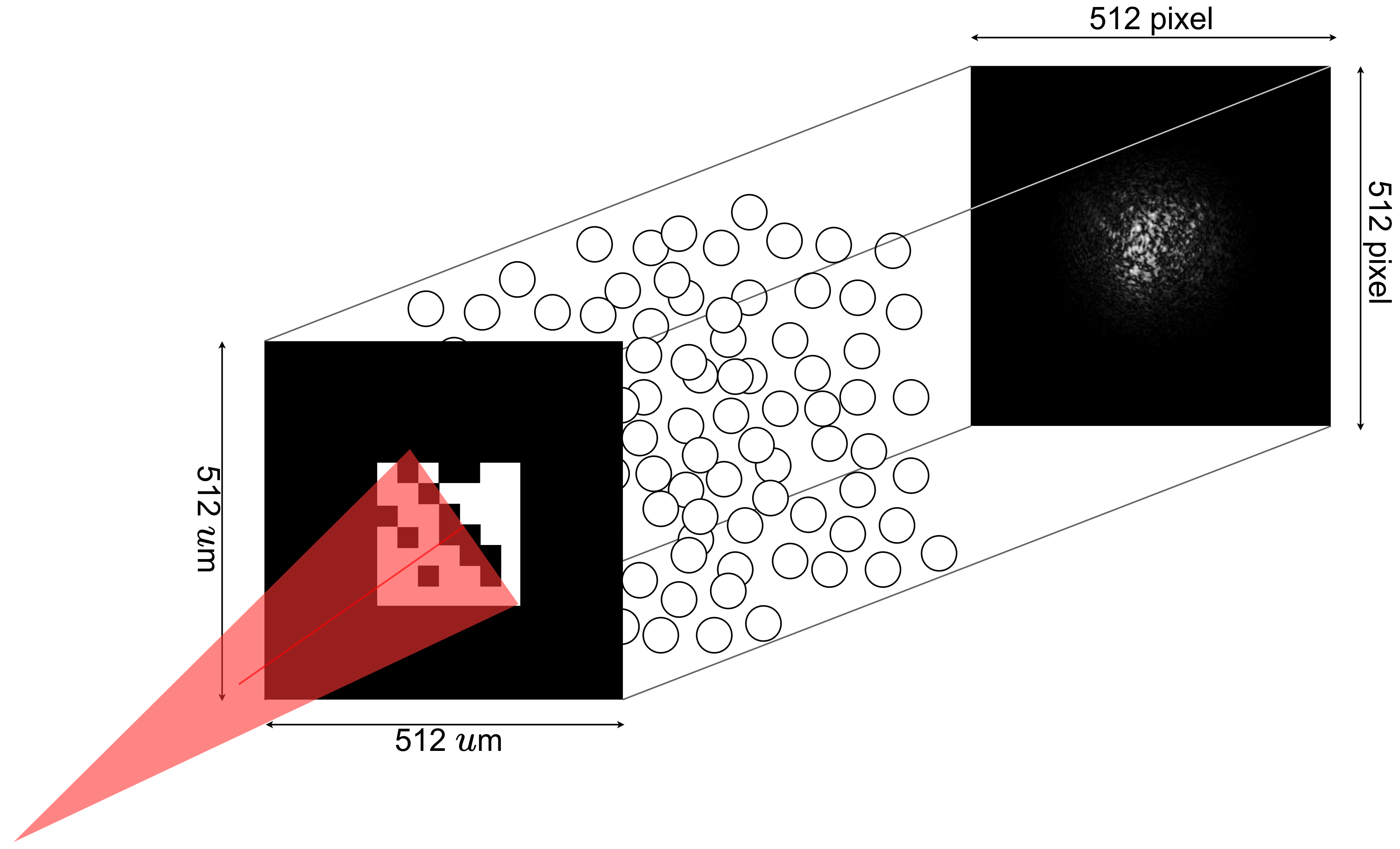}
\caption[Setup of the simulated PUF.]{An illustration of the setup that will be used for the simulation of an optical PUF.}
\label{fig:puf_setup}
\end{figure}

The aforementioned methods were implemented in several programming languages fitting the associated needs. For the simulation, the Python library \textit{Diffractio} \cite{diffractio}, a package for diffraction and interference optics, was used. The PUF consists of 1179 spherical scatterers of sizes randomly varying between 10$\mu$m and 15$\mu$m with a refractive index of 1.52. Each sphere has a padding of 10$\mu$m, resulting in a distance of at least 20$\mu$m between all scatterers. These spheres are randomly placed within the boundaries of a square with the length of 512$\mu$m for each side. For the challenge, a two-dimensional quadratic mask of length 512$\mu$m is placed right in front of the PUF. A gaussian beam is placed perpendicular to the mask and illuminates a circular area in the center, such that a square of length 128$\mu$m is constantly illuminated. Challenges are formed by splitting this inner square into a given number of blocks and either allowing or prohibiting light from traveling through the individual blocks. The propagation of the incoming light is computed via the \textit{Beam Propagation Method} \cite{poon2017engineering}. On the opposite site, the responses are generated by evaluating the intensity in the XY-plane 300$\mu$m behind the boundary of the PUF. The responses are of the size 512$\mu$m as well and are evaluated with a resolution of 1 pixel per $\mu$m, leading to images of $512 \times 512$ pixels. An illustration of what the setup looks like can be found in figure \ref{fig:puf_setup}.

The different challenge sizes are created by dividing the challenge mask into $l \times l$ blocks, such that the size of the LCD array stays the same, but the size per block decreases proportional to $l$. The challenges will be of the sizes $5$, $7$, $9$, $11$, $13$ and $15$ blocks per row, such that for $l$ blocks per row, the challenges are of the length $l^2$, and for each setup, there will be $2^l$ possible challenges. Due to computational restrictions and for reasons of simplicity, only the cases where $l$ is an uneven number will be examined in this thesis. Tests were made to ensure that using an even or uneven number of blocks per row has no influence on the complexity of the PUF. The challenges are created by generating a random bit vector, where each bit is equally likely to be activated with a chance of $50\%$. The bit vector is then transformed into a matrix of size $l \times l$, where each bit determines whether the corresponding block in the challenge is activated. For datasets of type A, no restrictions will be made and therefore the vector will be of length $l^2$ and simply be transformed into the challenge. For type B, the generated bit vector will be only of size $\frac{l + 1}{2}$, inserting a 0 at every odd position to again achieve a vector of size $l^2$. Type C and D are realized by a bit vector of size $l^2$, where bits are randomly set to 0 if more bits than allowed are activated, until the respective threshold is reached. For each of the resulting 24 datasets, 1000 randomly chosen CRPs will be computed.

Since the FHD of the generated CRPs is highly dependent on the type of transformation that is used to retrieve a bitstring from the response, the Gabor transformation is applied two times with two different kernels of sizes $35 \times 35$ and respective $9 \times 51$. This is to make sure that the results are not only depending on the kind of transformation that is used. Additionally, since the speckle pattern of the simulation emerges only in the center of the response, a center square of $128 \times 128$ pixels is cut out and used for the FHD calculation. This ensures that the whole image is filled with the speckle pattern and that there is no surrounding black area, which would be the same for all responses and therefore lead to a generally lower FHD. While this does cut off some of the outer speckle pattern, the square was picked in a way such that this cut off area is as small as possible. After transforming the response, the binary conversion is executed with 0 as threshold, such that all values greater or equal 0 will turn into 1 and all values lower into 0. For the calculation of the FHD of a dataset, 300 CRPs are randomly selected, and the FHDs between each response and all others are calculated.

For the ML attacks, each dataset is randomly split into a training and a test set of 900 and respective 100 CRPs. For the fitting of the linear models, no special amendments were made and the implementation follows the aforementioned general principle of a multivariate linear regression. The generator is built to take a bitstring as an input, reshape it into a 2-dimensional representation and create a response using several consecutive transposed convolutions (T-Conv2D). After each convolution, a batch normalization (BN) is run and for the activation function, LeakyReLU with a slope of $0.2$ was used. The detailed structure of the layers can be found in table \ref{tab:generatorLayers}. The model is trained with the ADAM optimizer, using the values $\alpha = 0.002$, $\beta_1 = 0.5$ and $\beta_2 = 0.999$, and uses a loss function that tries to minimize the MSE and maximize the SSIM. For the parameters of the SSIM, the common values of $\alpha = \beta = \gamma = 1$, $K_1 = 0.01$ and $K_2=0.03$ were chosen. Using a batch size of $32$, each model is trained for $100$ epochs, after which the performance of the predictions with respect to the FHD did not improve any more for a precision of three decimal places. All hyperparameters were empirically analyzed, such that the presented set of values leads to the best performances.

After training the ML models, the metrics will be measured on the predictions of the models for the test set. The retrievals of the bitstrings are executed in the same way as for the dataset evaluation. The FHD, PC and SSIM are all calculated between the real and generated response pairs. The parameters of the SSIM are the same as for the loss function.

\begin{center}
\begin{table}[htbp]
{\small
\begin{center}
\begin{tabular}[center]{cccccr}
\toprule
Nr. &  Layer & kernel & stride & padding & parameters \\
\midrule
1 & T-Conv2D, BN, LeakyReLU & 4 & 2 & 1 & 411,648 \\
2 & T-Conv2D, BN, LeakyReLU & 4 & 2 & 1 & 8,389,632 \\
3 & T-Conv2D, BN, LeakyReLU & 4 & 2 & 1 & 2,097,664 \\
4 & T-Conv2D, BN, LeakyReLU & 4 & 2 & 1 & 524,544 \\
5 & T-Conv2D, BN, LeakyReLU & 4 & 2 & 1 & 131,200 \\
6 & T-Conv2D, BN, LeakyReLU & 8 & 4 & 2 & 131,136 \\
7 & T-Conv2D, Tanh & 8 & 4 & 2 & 2,048 \\
\bottomrule
& & & & & 11,687,872 \\
\end{tabular}
\end{center}
}
\caption[Structure of the generator.]{Structure of the generator for the DL attack.}
\label{tab:generatorLayers}
\end{table}
\end{center}

    \chapter{Results}\label{sec:results}
After following the procedure described in the previous chapter, all datasets have been generated and evaluated with the defined metrics. One ML model for each approach has been trained on each dataset in the previously mentioned manner. In the following, the results of these works will be reported.

For the most important results, a boxplot showing the data will be directly provided. This plot follows the standard conventions and shows the first and third quartile at the lower and upper border of the box as well as the median as a line within the box. The whiskers reach up to the largest (or respective lowest) observed value within 1.5 times the interquartile range and outliers are depicted with a small circle. All the plots that are not directly provided can be found in the appendix under \ref{sec:appendix_datasets} and \ref{sec:appendix_attacks}.

\section{Analysis of the generated datasets}\label{sec:x1_results}
\subsubsection*{Fractional Hamming distance}
\begin{figure}[t]
\centering
  \includegraphics[width=1\linewidth]{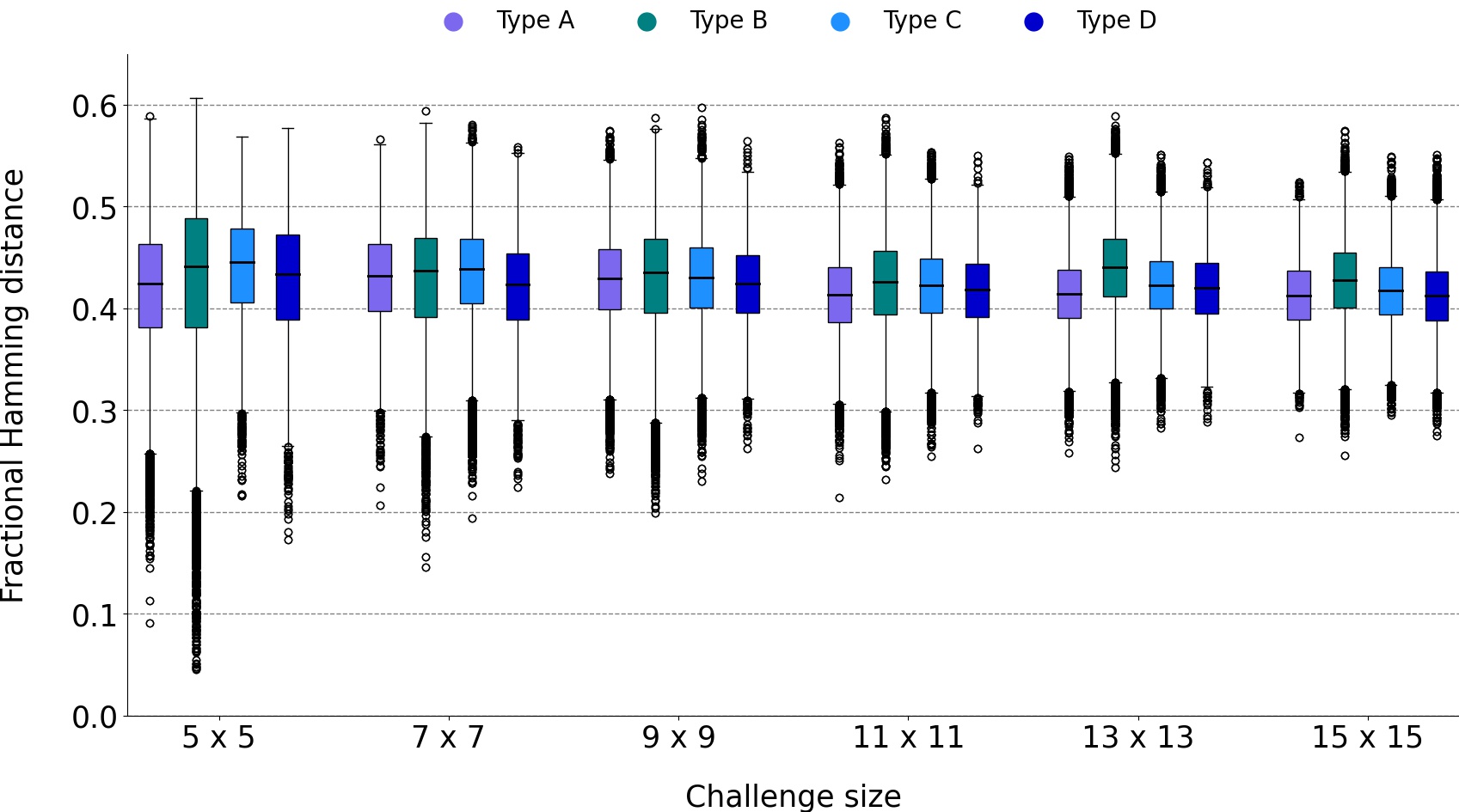}
\caption[FHDs of the generated datasets.]{The FHDs of all generated datasets using the first Gabor transformation.}
\label{fig:x1_FHD_box}
\end{figure}

Using the \textbf{first Gabor transformation}, the mean \textbf{FHD} across all datasets lies at $0.424$. The individual means move in the interval $[0.412, 0.439]$ and therefore stay about the same across all datasets. While there are some outliers, there appears to be a trend of a decreasing mean for an increase in the number of blocks $n$. For the individual FHDs of the pairs within a dataset, around half of the computed values lies in an interval between around $0.4$ and $0.45$. However, there is a wide gap between the smallest and the largest values, which applies in particular to the datasets with a smaller $n$. Here, FHDs below a value of even $0.1$ could be observed, with the smallest recorded value at $0.045$ for the type B dataset of size $5 \times 5$. The values of the lowest FHDs tend to gradually increase with an increasing $n$, but still reach values as low as $0.25$, even for the dataset with the largest $n$. In general, the distributions start to gradually shrink, such that the extreme FHDs approach the mean value with fewer outliers near the lower boundaries and more outliers near the upper boundaries.
The datasets of type B seem to have the widest range of FHDs, always including most of the largest and most of the smallest FHDs within one dataset group of challenge size $n$. Additionally, datasets of type C always lead to the best or second best mean FHD within a dataset group of the same size and always score for a higher mean than the type D alternative. However, note that since the means move within a very small interval, the differences are only minor. A plot containing all these values can be found in figure \ref{fig:x1_FHD_box}.

With the \textbf{second Gabor transformation}, the average mean of $0.352$ across the datasets is significantly lower than for the previous transformation. The means for all datasets again do not stray apart from each other very far and lie in the interval $[0.340, 0.376]$. The majority of all individual FHDs lie between the values $0.3$ and $0.4$. While there are a few exceptions, in general, all relative relationships that have been observed for the datasets using the first Gabor transformation are valid in this case as well, i.e. no new consistent observations could be made. The smallest recorded FHD lies at $0.027$.

\subsubsection*{Shannon entropy}
\begin{figure}[t]
\centering
  \includegraphics[width=1\linewidth]{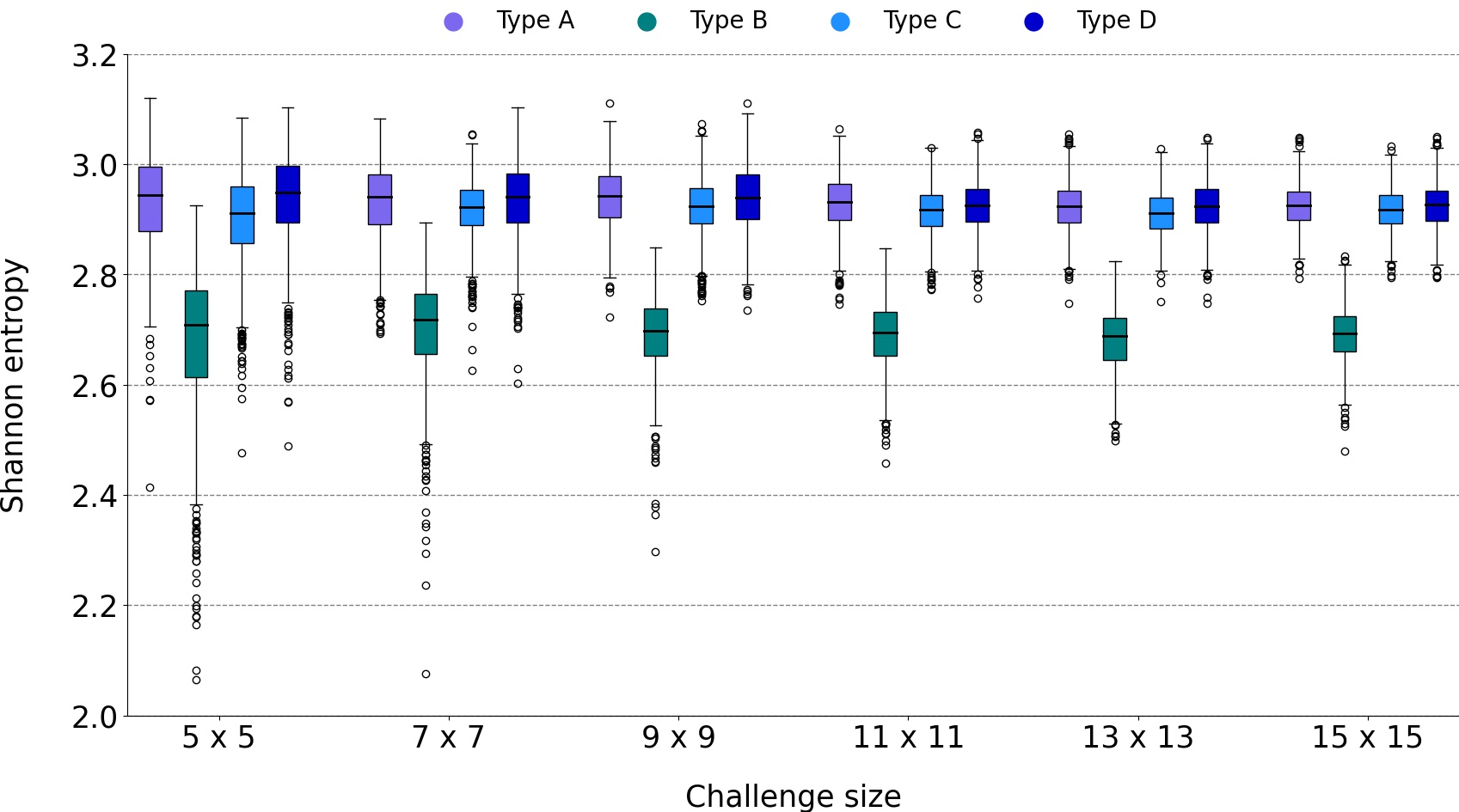}
\caption[Shannon entropies of the generated datasets.]{The Shannon entropies of all generated datasets.}
\label{fig:x1_Entropy_box}
\end{figure}

Regarding the \textbf{entropy} of the datasets, the means lie in the interval $[2.681, 2.939]$ with an average of $2.866$. The smallest recorded entropy is $1.491$, which is a rather large step in comparison to the second smallest value that lies at $2.065$. This extreme outlier was found to be the only challenge where only a single bit is activated. This time, the entropy means do not significantly change with an increase in $n$, but similar to the FHD, the distributions tend to shrink with fewer outliers at the lower boundary. One especially prominent aspect is that the mean entropy as well as the individual entropies for datasets of type B are significantly lower than these of all other types, which in turn rather seem to closely stick together. When leaving out the datasets of type B and instead only accounting for those of type A, C and D, the mean entropy rises to $2.925$. When in turn only looking at the type B datasets, it drops to $2.690$. Also, both previously mentioned lowest entropies are of a type B dataset, which also accounts for the lowest value within all groups of datasets with the same $n$. The first smallest outlier that does not belong to a dataset of type B has a value of $2.414$. Furthermore, the datasets of type A and D always take the highest and second highest values of the mean entropy, but are only very slightly above these of type C. A plot containing all these values can be found in figure \ref{fig:x1_Entropy_box}.

\section{Performance of the machine learning attacks}
Again following the previously described principles, all datasets were attacked using a LR and the generator. In the following, the results of these attacks will be reported. Note that in contrast to the evaluation of the datasets, from the point of a ML attack, the FHD will be tried to minimize.

\subsection{Linear regression}
\begin{figure}[t]
\centering
  \includegraphics[width=1\linewidth]{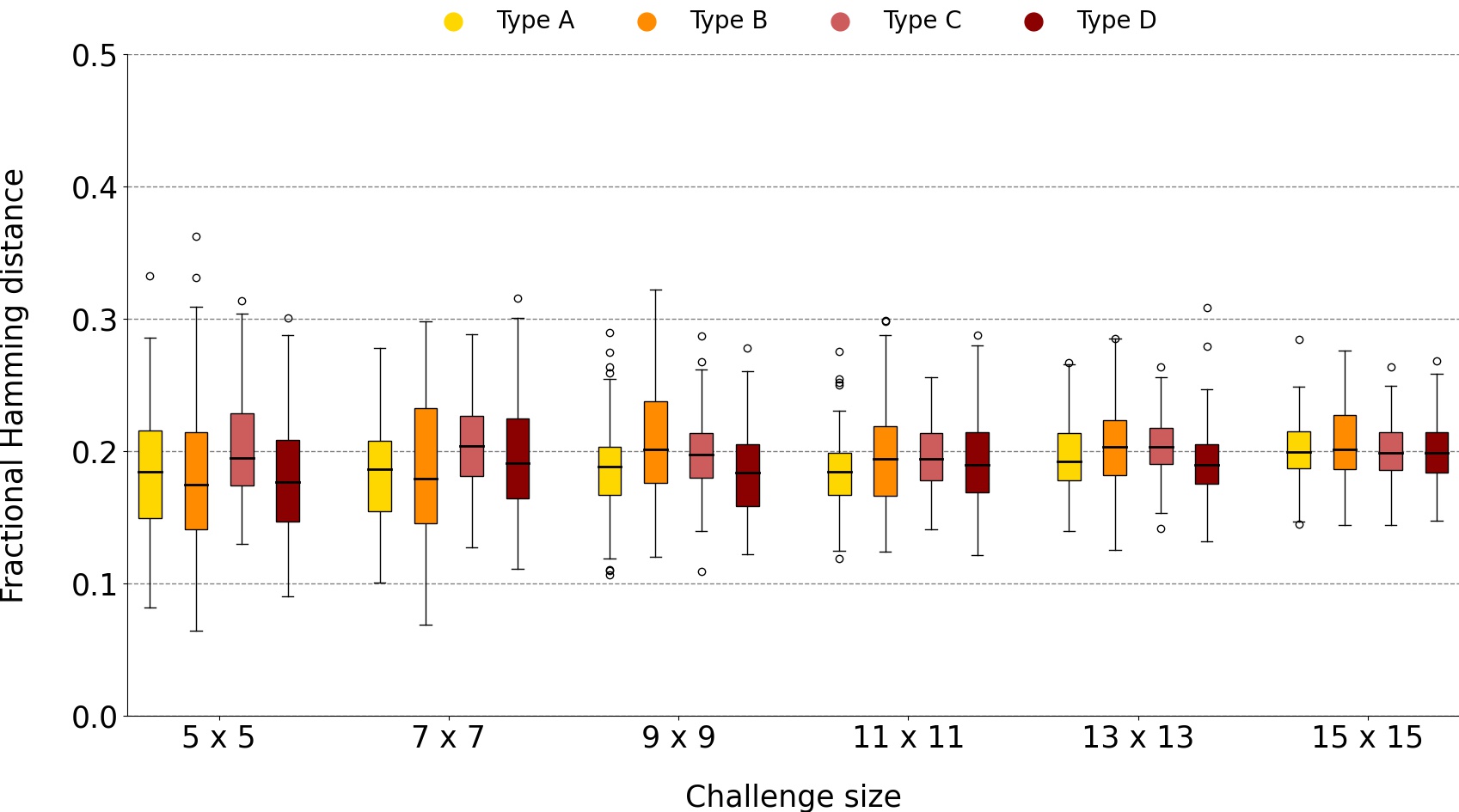}
\caption[FHDs of the LR attack.]{The FHDs between the real and predicted responses of the linear regression using the first Gabor transformation.}
\label{fig:x1_FHD_LR_box}
\end{figure}

Evaluating with the \textbf{first Gabor transformation}, the mean \textbf{FHDs} of all LR attacks lie in the range $[0.179,0.206]$, which makes for an average of $0.194$. The lowest recorded FHD across all datasets lies at $0.064$, while the largest is at $0.362$. Even though this maximal value is relatively high, the overwhelming majority of all computed FHDs lie between $0.15$ and $0.25$. There also seems to be a trend to have a slightly increasing FHD proportional to $n$.
The distributions again shrink with an increasing $n$, such that the number of very accurately predicted, but at the same time also the number of very poorly predicted responses decrease slightly. Even though these differences are rather small, the predictions of the model become more stable for larger $n$.
A comparison of the different dataset types shows that most of the time, datasets of type A as well as D show the lowest mean FHDs, while type B and C tend to produce the higher FHDs. Additionally, the distributions of the FHDs of type B are the widest, while for the rest, no consistent patterns emerge. There are outliers across all datasets that appear rather random, showing no meaningful pattern. A plot containing all these values can be found in figure \ref{fig:x1_FHD_LR_box}.

With the \textbf{second transformation}, only a generally lower FHD could be found, similar to the observations for the evaluation of the datasets. There are no noteworthy consistent differences compared to the results of the first Gabor transformation.

The mean \textbf{PCs} lie at quite high values between $[0.943,0.965]$ with a mean of $0.959$ and do not change much between the datasets. The lowest PC was recorded for the $5 \times 5$ dataset of type B and lies at $0.788$, which is a rather large exception, since even the second lowest computed value of $0.849$ is a lot higher. Starting from the $11 \times 11$ datasets, there appears to be a small trend of a decreasing mean PC. Just as for the FHD, the distributions start so shrink with an increase in $n$, though the differences are not as significant as for the FHD. Also, the distributions of the type B datasets are again most of the time a lot wider than those of the other ones, especially at the lower boundary. The mean for these datasets is, starting with the $9 \times 9$ datasets, always remarkably lower than for the other types. The datasets of type C show a generally lower PC than the two others as well, but the differences are not as large as for those of type B.

Lastly, the mean \textbf{SSIMs} stays at a similarly high value as the PC, moving in the interval $[0.973,0.982]$ with a mean of $0.977$. Therefore, the results vary even less across the datasets than for the other two metrics before. Even the lowest SSIM of $0.950$ is relatively close to all other computed values, i.e. there are no significant outliers as could be found for the PC. A general decrease in the mean as well as shrinking distributions can be found with an increase in $n$ as well, even though the differences are much smaller than before. This time, the highest mean SSIMs can be found for datasets of type B, while those of type C consistently show the lowest means. Also, in contrast to the other metrics, the distributions of the type B datasets are not consistently wider than the others and are instead actually more narrow for the datasets with smaller $n$. When zooming in, there is a pattern of type B datasets showing the highest and the type C datasets the lowest mean, with the other two alternating in between. However, these differences only amount to values even less than $0.01$. Altogether, with respect to the results of the LR, the predictions appear to be most stable for the SSIM.

\subsection{Deep learning}
\begin{figure}[t]
\centering
  \includegraphics[width=1\linewidth]{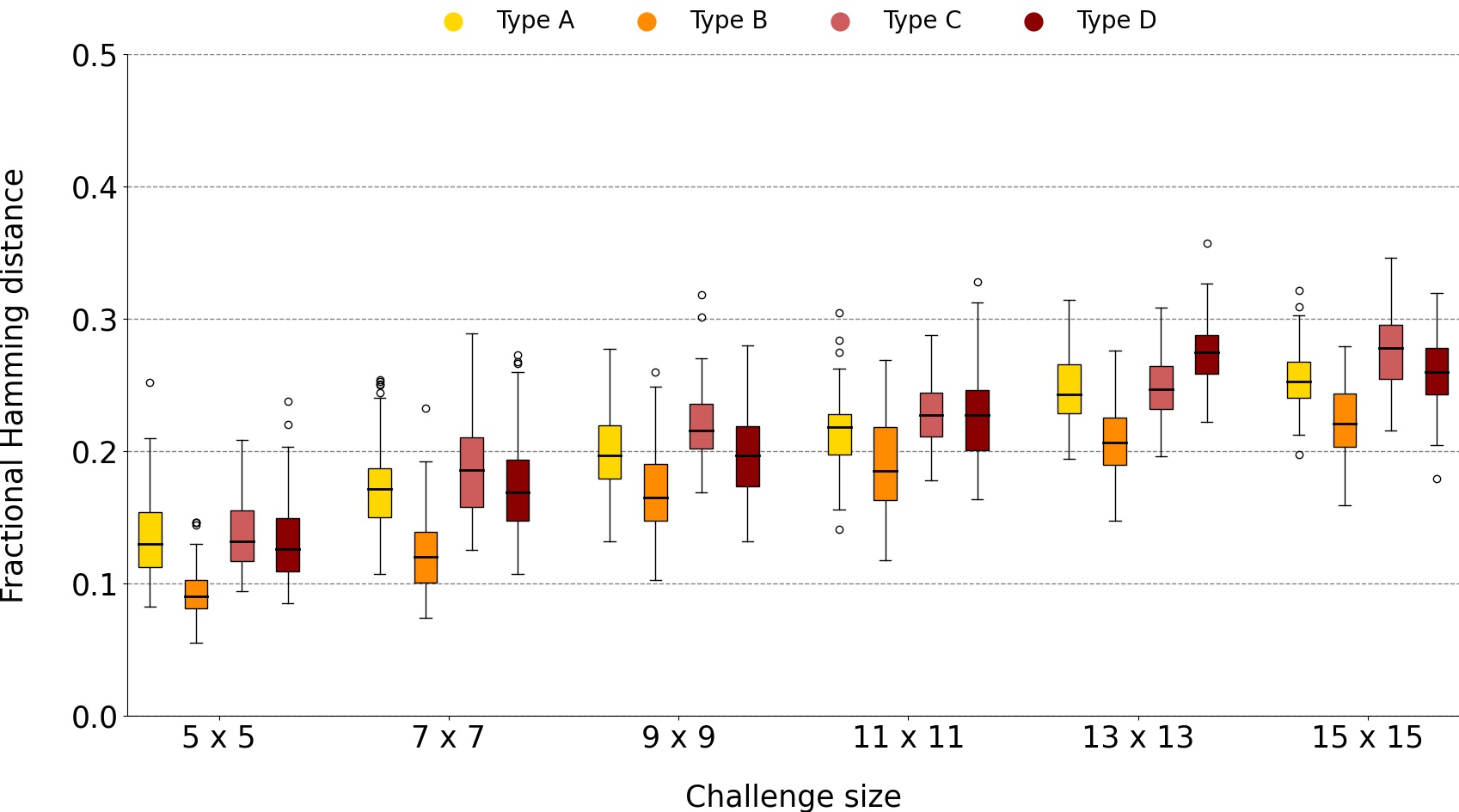}
\caption[FHDs of the generator attack.]{The FHDs between the real and predicted responses of the generator using the first Gabor transformation.}
\label{fig:x1_FHD_DL_box}
\end{figure}

Using the \textbf{first Gabor transformation}, the \textbf{FHDs} show a very clear gradual increase proportional to $n$ that can be found for both the means and the distributions. The means move in the range $[0.093,0.278]$, so compared to the linear model, they are better for the datasets with smaller $n$, but are, in turn, worse for those with larger $n$. On average, the mean lies at $0.199$. With a lowest value of $0.055$ and largest of $0.357$ for any individual FHD, both extremes show only slightly better values than what was found for the linear model. Also, the distributions here do not widely differ in size for different $n$ and are comparable to the distributions of the $15 \times 15$ datasets of the linear attack. The relationships between the mean FHDs for the different types are quite similar for varying $n$, showing a tendency for type A, C and D to stay very close together and type B to have a lower mean FHD than all the others. Except for the $13 \times 13$ datasets, the mean FHD is constantly highest for type C. The ranges of all distributions vary by only a small amount, showing no clear patterns. Outliers are rather rare and appear almost always above the upper boundary. A plot containing all these values can be found in figure \ref{fig:x1_FHD_DL_box}.

Again, the \textbf{second Gabor transformation} shows a constantly lower FHD, but no other differences.

Similar to the LR, the generator scores a steadily high mean \textbf{PC}, though the decrease coming along with a larger $n$ is a lot more noticeable here, which matches the observations for the FHD. Again, the results are initially better than they were for the linear attack, but become worse when $n$ increases, and the distributions do not gradually shrink in range. There are fewer extreme outliers and the lowest recorded PC lies at $0.870$, while the linear model had a few samples where the PCs were below this threshold. Most of the time, the PC for the type B datasets is slightly higher, but except for this, no patterns emerge.

In contrast to the other two metrics, compared to the LR, the \textbf{SSIM} is a lot worse for this DL model in almost all cases and goes as low as a value of $0.904$, which is $0.046$ below the lowest value of the linear model. While this may not sound much, this in fact almost doubles the difference between the largest and smallest SSIMs. Again, there is a very clear gradual decrease across datasets with increasing $n$ and no patterns regarding the distribution ranges. Also, noticeably higher values for the type B datasets were measured. However, it is important to note that the SSIM cannot be considered as an independent measurement for the generator, since it is part of the loss function and therefore biased by the training process.

    \chapter{Discussion}\label{sec:discussion}
After having thoroughly investigated the results of the carried out experiments, these results will now be interpreted to better understand the reasons why these findings were made. On this account, first, some general observations on the experimental setup will be made, before diving deeper into the analysis on the results themselves.

\section{Applicability of the Gabor transformations}
All findings show that the results are not solely dependent on the applied Gabor transformation, since only the absolute values of the resulting FHDs differ between the first and second transformation, keeping the relative relationships constant. In general, it seems that the first transformation is better at actually extracting information of the responses that are generated by the simulation. This is directly visible from the overall difference in the FHD of the datasets, where the first transformation consistently creates a higher average value. Since no other parameters except for the type of the transformation vary, it is plausible that the second transformation creates less independent bits from a response. As a matter of course, this also causes seemingly better results for the ML attacks, since the FHDs between the real and generated responses will on average be smaller if the FHD between arbitrarily chosen responses is smaller in the first place as well. This does not mean that the ML models actually perform better, only the metric for the evaluation of the results is less suitable for the type of responses from the simulation. Due to this fact, in the following, the focus will lie on the first transformation, since the type of transformation will be chosen from the point of view of the PUF holder, who will try to increase the number of independent bits when transforming the responses. However, in general, all relationships between the FHDs of the datasets and of the ML results are valid for both types of transformation, which shows that the conclusions do not only depend on the type of transformation that is used, but are instead influenced by the underlying responses themselves.

\section{Distributions of the number of activated bits within a challenge}\label{sec:cDist}
As previously mentioned, each bit is equally likely to be activated and therefore each possible challenge is equally likely to be generated. When neglecting any benefits to the complexity of the underlying physical system for certain challenge compositions, from the point of view of security, this would be the most rational decision, since any influence on the challenge generation may be used by adversaries to facilitate an attacking process. However, in this case, this has proven not to be the optimal choice for an analysis on the PUFs complexity, as the uniform distribution of the challenge generation leads to a significantly higher probability to pick a challenge, for which the number of simultaneously activated bits $b_{act}$ equals half of all bits. This is simply due to the fact that there are more challenges in general that fall into this category. Essentially, $b_{act}$ is equivalent to the famous example of the number of heads after tossing a coin $n$ times. $b_{act}$ therefore follows a binomial distribution with $n$ experiments for $n$-bit challenges and a probability of ``success'' of $0.5$.

\begin{figure}
\centering
  \includegraphics[width=1\linewidth]{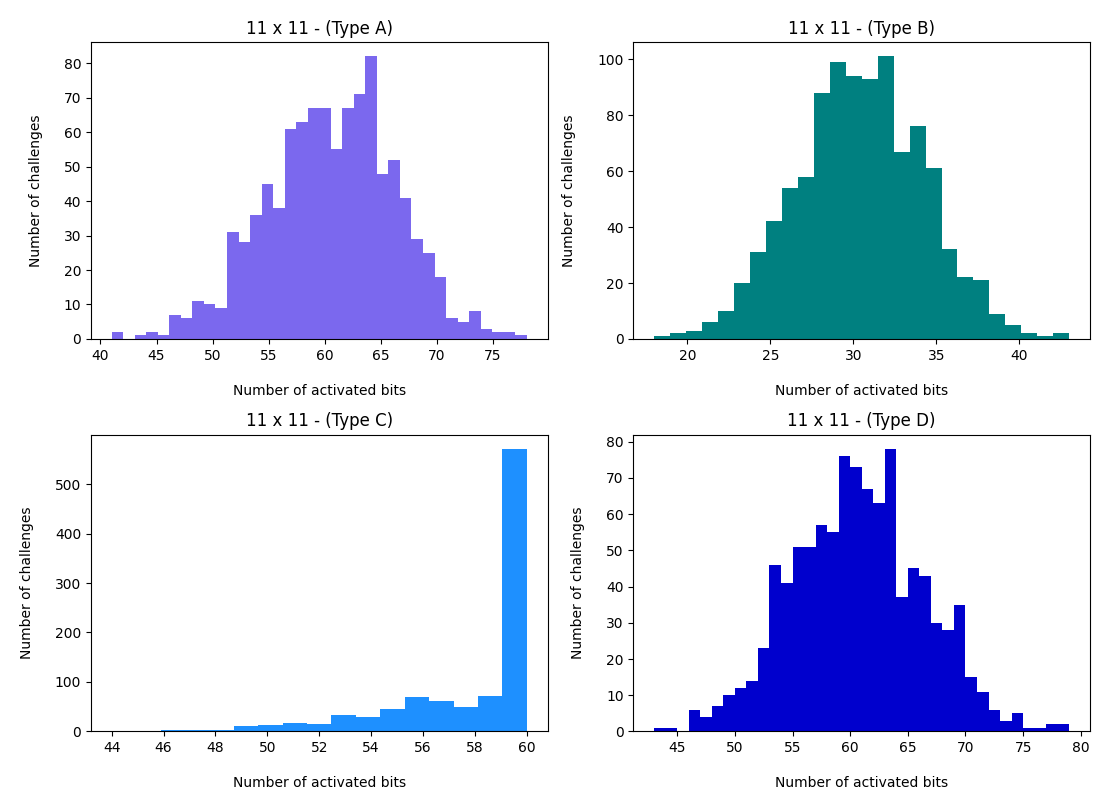}
\caption[Example of the distributions of the number of activated bits in a challenge.]{The distributions of the number of activated bits in the computed challenges for all $11 \times 11$ datasets.}
\label{fig:challenge_dist}
\end{figure}

Since the number of challenges gradually increases when $b_{act} \rightarrow \frac{n}{2}$, the overwhelming majority of computed challenges were found to revolve around this value. Naturally, following the law of large numbers, this effect becomes even stronger for larger $n$. An example of the distributions of $b_{act}$ for the datasets with $n = 121$ blocks clearly shows this tendency and can be seen in figure \ref{fig:challenge_dist}. Note that for type B datasets, the expected value of the distribution changes to $\frac{n}{4}$, as the maximum value for $b_{act}$ is cut in half before the bit vector is generated. On the other hand, the generation of the bit vector for type C datasets is the same as for type A, except that for all cases where $b_{act} > \frac{n}{2}$, bits are deactivated until $b_{act} = \frac{n}{2}$, which explains the large spike at this value. This occurrence can in theory be found for type D for challenges with $b_{act} > \frac{2n}{3}$ as well, however, since no challenges fulfilling this criterion were generated, this does not become apparent in this example. This is simply due to the fact that the chance of generating such a challenge in the first place is rather low, especially for larger $n$. An investigation on all generated datasets revealed that starting with $n = 81$, the datasets of type A and D do not show any differences regarding $b_{act}$ any more, which, as this was the only difference between these datasets, makes the approaches practically the same. They only differ due to the randomness of the challenge sampling, not showing any significant differences imposed by the restriction of type D.

Altogether, this finding means that the majority of the generated challenges do not fall into the categories where most of the bits are either activated or deactivated. These are the edge cases where the simulation seemed to produce unwanted behaviors such as strongly correlated speckle patterns. All observations that were made and all interpretations that follow will therefore only refer to the case of a ``regular'' challenge sampling, as it would normally be done during the usage of the PUF. Therefore, no extensive statements regarding the behavior of the simulation on certain edge cases can be derived from the results that were shown.

\section{Evaluation of the generated datasets}\label{sec:dataset_eval}
First, the results of the datasets from the simulation will be elaborated on, before further investigating the results of the ML attacks, as they form the basis for the attacks themselves in the first place. For this reason, this part will focus on the previously reported results and give further insights into why the corresponding observations may have been made this way.

Note that a large number of CRPs for a lot of different setups were needed for the ML attacks, so using a simulation greatly reduces the necessary effort and makes it possible to build a simple interface that allows for cheap modifications on the whole setup and unrivaled readout stability and precision. Modifying the quality, quantity and positions of the scatterers of the simulated PUF, as well as varying the challenge application and the readout setup are a lot easier and cheaper to carry out with a simulation, and at the same time there is no need to be wary of any unwanted external influences. Furthermore, the exact position of each part and the same relative positioning between different readouts can be guaranteed with 100\% precision, even over an arbitrary number of sessions. However, as a matter of course, the simulation is not supposed to be a substitution for a real optical PUF. Obviously, this would defy the whole purpose of moving away from a binary key to a physical system in the first place and there are better cryptographic alternatives when working with computer algorithms. However, the simulation is still worth to take a closer look at and provides significant benefits regarding the reduction of effort within the scope of this thesis.

\subsection*{Fractional Hamming distance}
It appears that even though there is a small trend of a decreasing mean FHD proportional to $n$, the overall influence is rather small and even negligible, especially when taking into account that in turn, the number of challenges that can be applied to the PUF increases drastically. Also, the distributions shrink proportionally to $n$, so the number of response pairs with a low FHD gradually decreases. This may lead to the assumption that a larger $n$ should always be preferred, however, one should keep in mind that when picking random challenge out of a larger challenge space, the chance of getting closely related challenges (i.e. challenges, where most of the bits are the same) naturally becomes smaller. Additionally, the chance of retrieving challenges where the number of activated bits is close to 0 or close to $n$, which have been previously identified as bad samples with respect to the FHD, also decreases with an increasing $n$. If one were to exhaustively generate all CRPs of one PUF, it would be expected that the boundaries of the distributions do not greatly differ with varying $n$.
In this thesis, the number of CRPs for one dataset amounts to only $1000$ and those used in the FHD calculation to only $300$, which is a significantly low number with respect to the whole possible output of a PUF. Since $n$ is inversely proportional to the amount of light that travels through each block (and therefore the influence of switching a single bit in the response), for sufficiently large numbers of CRPs, there will be a lot of almost visually indistinguishable responses for the datasets with larger $n$. Even for the datasets with smaller $n$, where the influence of switching single bits is the largest, there are still response pairs that show a FHD of below $0.1$, which clearly shows that by far not all CRPs can virtually be classified as independent. So considering all possible CRPs, this will not become better for larger $n$ with even less influence of single bits. Only picking a very small subset might mislead to the assumption that a larger $n$ is always better, while there actually is a tradeoff between the input-output complexity of the PUF and $n$.

As further investigations, additional datasets of $1000$ CRPs of type A using $25 \times 25$ as well as $35 \times 35$ blocks were generated. The computed mean FHDs are $0.400$ and respective $0.363$, showing that the trend of a decreasing mean FHD continues beyond the investigated number of blocks. Of course, this trend will not go on forever and likely settles at some value. One last additional dataset of $40 \times 40$ blocks was investigated and presented a mean FHD of $0.367$, which indicates that this settling value is most likely somewhere close to that of the $35 \times 35$ blocks dataset, since no decrease could be found any more.

A comparison of the FHDs of the different types shows patterns that are emerging from the challenge compositions themselves, though the differences are only minor. The wider distributions for type B datasets may as well only emerge from the resulting lower challenge space, since only allowing every second block to be activated virtually cuts $n$ in half (or half plus one for uneven $n$). As stated previously, the smaller $n$, the more likely the chance to pick related challenges or challenges where $b_{act} \rightarrow 0$. One may think that the restriction of type C to only use half of all blocks reduces the challenge space in the same manner, however, this is actually not the case. While for type B datasets, only around half of all blocks are usable, type C datasets allows for all blocks to be activated and only excludes those configurations, where more than half of the blocks are simultaneously activated. This still allows for more configurations than actually deactivating half of the blocks. Another explanation might be that certain parts of the PUF will never be as extensively illuminated as by the other variants, since no light will enter through specific blocks of the mask, resulting in a lower overall complexity. However, this hypothesis would require further analyses on the simulation including a significantly large subset of all possible CRPs, which would require a tremendous number of computational resources and was therefore not possible in the scope of this thesis. It is to be expected that across the whole challenge space, the mean FHD will be slightly better for the restricted datasets, since the restrictions act as a countermeasure to the previously mentioned problems of the simulation, but for a subset significantly smaller than the possible challenge space, the differences are negligible.

To give further insights into the quality of the simulated datasets, two additional datasets from two different \textbf{real optical PUFs} were investigated. The first PUF uses liquid crystals as scatterers and was provided by an Italian research group\footnote{Francesco Riboli (CNR-INO), Sara Nocentini (INRiM, LENS), Giuseppe Emanuele Lio (CNR-INO, LENS)}. The challenges are realized in the same way as in this thesis, using a LCD array of $16 \times20$ blocks. The speckle patterns are images of $200 \times200$ pixels, from which a center square of $128 \times 128$ pixels was cut out for the computation of the FHD as well. This ensures that the number of bits retrieved from the bit transformation is the same across both setups. The computed FHDs for the real PUF lie in the interval $[0.32, 0.511]$ with a mean of $0.426$ for the first Gabor transformation, which is quite similar to the values computed for the simulated data with a comparably large number of blocks. Here, the second Gabor transformation creates very similar values with an interval of $[0.315, 0.52]$ and the mean $0.419$, showing that the transformation is a better fit for these speckle patterns than for the simulated ones.

The second PUF uses spray-painted ZnO nanoparticles as scatterers and was published as part of the \textit{BOOST} (Break Our Optical Security Technology) challenge\footnote{\url{https://security1.win.tue.nl/Id_IoT/}, Organizers: Boris Škorić (TU Eindhoven, The Netherlands), Teddy Furon (IRISA, France), Slava Voloshynovskiy (Univ. Geneva, Switzerland)}. This setup uses a Spacial Light Modulator (SLM) of $80 \times 60$ blocks, using $963$ of these blocks simultaneously for a challenge. The SLM does not simply either allow or prohibit light from travelling, but also modulates the phase of the incoming light differently for each block. The responses are of the size $801 \times 821$, where, similar to the simulated datasets, a circular speckle pattern emerges in the center of the image. For the computation, again only the center square was used for better comparability. This resulted in FHDs in the range $[0.39, 0.608]$ and a mean of exactly $0.500$ for the first, as well as $[0.377, 0.625]$ and respective $0.497$ for the second transformation. While the mean is clearly higher for this setup, the ranges are again not that much different from the simulated version. However, one should be careful in interpreting these direct comparisons due to the very different challenge setups.

All in all, the results show that at least for randomly chosen CRPs in the size regime of this setup, the FHDs for the simulated data are in a similar range to these of real optical PUFs. However, no clear statements regarding the difference of the FHDs between the simulated and real optical PUFs for only edge cases (e.g. almost all/no bits are activated) or for a significantly larger number of CRPs can be made with the available data.

\subsection*{Shannon entropy}\label{sec:discussion_entropy}
\begin{figure}
\centering
  \includegraphics[width=.49\linewidth]{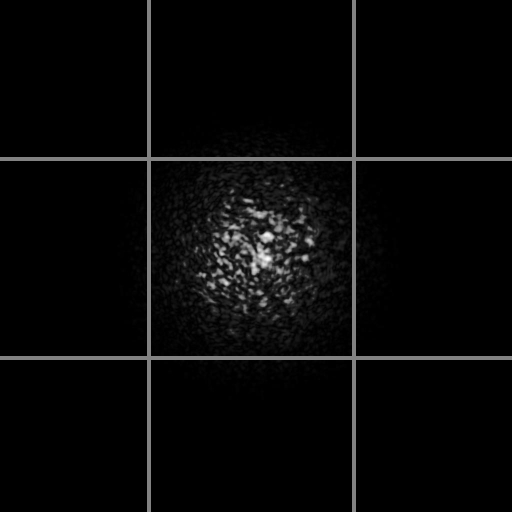}
  \includegraphics[width=.49\linewidth]{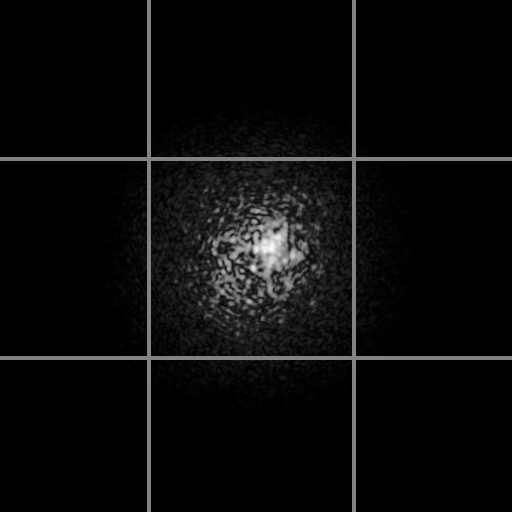}
\caption[Differences in speckle patterns with varying number of activated bits.]{Two speckle patterns for a $9 \times 9$ blocks dataset, showing the differences in size and brightness coming along with a varying number of activated bits. $10$ bits were activated in the left and $70$ bits in the right image.}
\label{fig:speckles_var_bits}
\end{figure}

There is an important aspect of the simulation that should be kept in mind when interpreting the entropy measurements: First, the speckle patterns that emerge when a lot of light travels through the system are by far superior in size and brightness compared to the patterns when only a small amount of light is exposed to the system. Also, with an increasing size of the speckle pattern, the dark area around it becomes smaller. Both these phenomena can be seen in figure \ref{fig:speckles_var_bits}. Since the entropy is calculated on the whole image, this dark area has a large impact on the computed value and results in an overall smaller entropy. Naturally, if the speckle pattern increases and the dark area decreases in size, the entropy will increase as well, since there is more information stored in the picture. Now, since the restriction of type B only allows half of the bits to be activated, there will on average less light be travelling through the system per CRP, resulting in on average smaller speckle patterns and therefore a smaller entropy, which explains the discrepancy found between this type and the others. Also, within the smaller challenge space, there is a higher chance to produce challenges where only a very small number of bits is activated, which may be why close to all outliers with an exceptionally low entropy are from type B datasets. In fact, all of the extreme outliers for the type B dataset of size $5 \times 5$ were found to have a significantly low number of activated bits and therefore only a very subtle speckle pattern. While it is possible that other effects influence these observations as well, it is highly likely that these two consequences imposed by the restriction already have a rather large impact. Again, the different methods with which the challenges are generated for each type are the reason why this same behavior was not found for type C datasets. To verify the above assumptions, the entropy was calculated again, this time following the preprocessing of the calculation for the FHD, i.e. to only use the center square of $128 \times 128$ pixels for the evaluation. This type of response will from now on be referred to as \textit{cropped response} and is supposed to protect against the undesired influence of the dark area. The obtained results show exactly what would be expected: All entropy values greatly increased and the means started to align, moving within the minuscule interval of $[7.618,7.672]$. In general the entropies of the individual responses do not vary much and revolve around $\pm 0.2$ of their mean. In contrast to the previous evaluation, the type B datasets now actually show a distribution with slightly higher values than the others, though the differences are very small. This indicates that the previously mentioned smaller size of speckle patterns that were found for type B datasets were the reason for the initially lower entropy.

Furthermore, the previously presented datasets from \textbf{real optical PUFs} were investigated regarding their entropies as well. The responses of the PUF from the italian group scored a greatly higher mean of $11.709$ and all values move in a comparably high range of $[11.350, 12.031]$. Since the speckle patterns already fill the whole image, the results for a cropped response are almost the same. For the second PUF, only the entropies for the cropped responses were investigated due to the dark area again decreasing the entropy. Here, a mean of $6.230$ and the range $[6.042, 6.519]$ was found, which is significantly lower than the first PUF's and even lower than the simulated PUF's results.

Hence, it appears that the \textit{Shannon} entropy of the responses of this simulation is somewhere between those of real optical PUFs as well. However, the rather large difference between both optical PUFs already indicates that the entropy measurement does not give a clear answer regarding the true complexity of the response, since the responses of the second real optical PUF are not less complicated than those of the first one. Furthermore, only the pixel values and no spatial information is taken into account, which would actually be a rather important factor when evaluating speckle patterns. Other entropy measurements may lead to completely different conclusions. While in general the entropy evaluation is a non-trivial matter for which no universal solution has been found so far, some other popular approaches include the NIST Statistical Test Suite \cite{rukhin2001statistical}, Diehard Tests \cite{marsaglia2008marsaglia} or TESTU01 \cite{10.1145/1268776.1268777}. All three methods were designed to measure the quality of random number generators and perform more complex evaluations than the Shannon entropy.

\subsection*{Choosing the challenge size and applicability of restrictions}
All in all, the differences of the metrics between all datasets are quite small or even negligible for the number of CRPs that were investigated. As stated a few times, many of the differences were found to be at least to some point caused by the challenge generation and do not only emerge from changes in the complexity of the PUF. While it may be possible that more significant differences start to appear for considerably larger numbers of CRPs, from the currently available data, no such ideas can be inferred. The only tendency that already became apparent even for the small number of CRPs was a decrease in the mean FHD for larger $n$. It would therefore be suggested to choose a reasonably large challenge space, somewhere in the order of $15 \times 15$ blocks. This still allows for $2^{225} \approx 5.39 \cdot 10^{67}$ different challenges without suffering too much from the decreasing mean FHD. Furthermore, even though it did not become clear from the results of this thesis, the kind of restrictions of type C and D would still be recommended, since the responses at some point do become very similar for a large number of activated blocks. Further analyses that could not be covered so far would be necessary to decide on a fitting threshold for the allowed number of simultaneously activated bits.

\section{Evaluation of the machine learning attacks}\label{sec:ml_eval}
Now that the observations for the simulation have been elucidated, the focus will switch towards the results of the ML attacks. There is one important aspect to keep in mind during the following interpretations: The simulation used in this thesis is completely deterministic without any type of noise, i.e. for the same challenge, the according response will in all cases be exactly the same, even across an arbitrary number of measurements. Therefore, it is rather difficult to define a point at which a ML attack can be viewed as successful. Pappu et. al. \cite{pappu} defined a possible candidate for this value as follows: First, they calculated the distributions of the FHDs between a set of collected CRPs, where the mean ideally lies at $0.5$, i.e. the responses are on average completely decorrelated. Next, they collected the same set of CRPs again and this time calculated the distributions of the FHDs between all of the CRPs that used the same challenge, to measure how much the responses differ between the two measuring sessions. In an ideal scenario, all of these FHDs would be $0$, i.e. the responses are identical for the same challenge. However, due to external influences such as noise in the system or measuring inaccuracies, this is not the case, and there are in fact differences between the responses for the same challenge. This raises the need to define a decision criterion for when two responses are viewed to originate from the same challenge, which was defined to be at the cross-over value of the two distributions. In their case, this value was found to be at approximately $0.41$. This value leads itself to be used as the criterion for the ML attack as well, so if the predictions of the attack are below this threshold, the attack can be classified as successful. However, since the responses of the simulation will always be identical for the same challenge, this threshold will always be found to lie at $0$. This makes interpreting the computed values a bit more complex, since all pairs of responses that exceed a FHD of $0$ can be regarded as different, but this would require the ML attacks to be able to perfectly model the simulation, which is rather difficult to achieve. At the same time, it is no solution to simply use a threshold found for real optical PUFs for the simulation as well, as ML attacks in the real case would have to work with noisy data, which can be expected to decrease the performance compared to data without any noise. Therefore, while the results of the ML attacks will be further discussed, it will be refrained from classifying the ML attacks as successful or unsuccessful.

\subsection{Linear regression}\label{sec:eval_lr}
\begin{figure}
\includegraphics[width=1\linewidth]{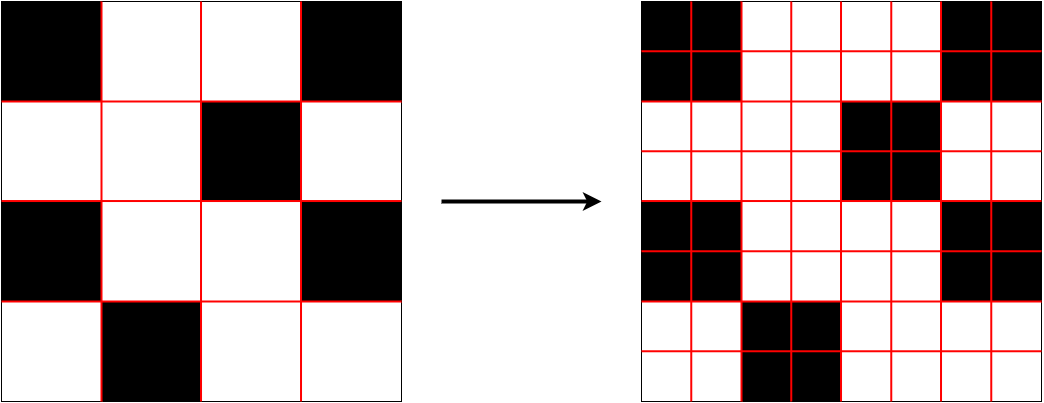}
\caption[Concept of the challenge transformation into a larger challenge space.]{The concept of the challenge transformation into a larger challenge space. Here, a $4 \times 4$ challenge is transformed into an equivalent $8 \times 8$ challenge.}
\label{fig:challenge_transformation}
\end{figure}

Before diving further into the results of the LR, consider the following analysis: Since the linear attack works with the bitstrings of the challenges, the number of independent variables rises accordingly to the increase in $n$. However, there is no actual change in the underlying scattering system. Only the challenge mask is divided into smaller blocks, which allows to more precisely adjust the area the incoming light can pass through, but all properties of the beam, the scatterers and the relative positioning stay constant.
Therefore, it is possible to define a transformation for all challenges of a given $n$ into challenges with larger $n$, by simply dividing all blocks into several smaller quadratic blocks of equal size. If each block is split into 4 smaller blocks, as is depicted in figure \ref{fig:challenge_transformation}, the resulting challenges will be of the size $4n$. This does not raise the need to carry out any changes on the simulation or to provide any additional information, but still maintains the same relationship between each CRP. This also quadruples the number of independent variables used in the regression. The same process can in theory be applied for the final regression model in a similar way: As previously described, the multivariate LR model takes the form $Y = XB + E$. Here, $X$ is a $p \times n$ matrix, such that $p = 900$, i.e. the number of training CRPs, and $n$ corresponds to the number of blocks. Now, for an arbitrary observation $i$ as well as an arbitrary value of the response $Y_{ij}$, any variable $X_{ik}$ can be split into four variables $X_{ik_{1}},X_{ik_{2}},X_{ik_{3}}$ and $X_{ik_{4}}$, which always take the same value as $X_{ik}$. Accordingly, the corresponding coefficient $\beta_{kj}$ can be split into the four coefficients $\beta_{kj_{1}},\beta_{kj_{2}},\beta_{kj_{3}}$ and $\beta_{kj_{4}}$. To ensure that the results of the models do not change, the equation $X_{ik}\beta_{kj} = \sum_{m=1}^{4} X_{ik_{m}}\beta_{kj_{m}}$ has to hold, which, since all $X_{ik}$ are binary, can be reduced to $\beta_{kj} = \sum_{m=1}^{4} \beta_{kj_{m}}$. This can easily be fulfilled, for example by setting $\beta_{kj_{m}} = \frac{1}{4}\beta_{kj}$ for all $m=1,..,4$. By applying this procedure to all variables, it is possible to create a model that produces equivalent results, while taking four times as many independent variables. Obviously, the regression will not follow such a pattern as no similar assumptions are made, but this analysis shows that such a model does exist. Therefore, if the LR were independent of $n$, it should be able to find a similar model that does not show a difference in the quality of the predicted responses when $n$ is increased in this manner.

\subsubsection*{Fractional Hamming distance}
To better understand the differences in the mean FHDs of the LR, the aforementioned procedure was applied to all datasets, training a LR model each time. Every attack showed an increase in the mean FHD after applying the challenge-transformation. While this was rather low for most of the $5 \times 5$ datasets (difference of around $0.01$), the $7 \times 7$ datasets already showed a difference of $0.04$ and almost all the others showed a difference of at least $0.05$, with many even above $0.1$. Since no other variables were influenced between these attacks, it can be assumed that for these training sizes, at least part of an increase in the FHD for larger $n$ is simply due to the resulting increase in the number of independent variables used in the regression. Further analyses on this topic as well as a way to avoid this behavior by modifying the loss function of the LR will be shown in chapter \ref{sec:Ridge}.

As further investigations, the in chapter \ref{sec:dataset_eval} mentioned additional datasets of sizes $25 \times 25$ and $35 \times 35$ were attacked as well, resulting in mean FHDs of $0.264$ and $0.257$. While there is another increase in the mean FHD for the first attack, the latter shows that this increase will not go on forever. This is most likely due to the decrease in the mean FHD of the datasets themselves, which appears to cancel out the influence of the previously described effect to some degree. Furthermore, the mean FHDs of the transformed datasets were compared with the non-transformed datasets where $n$ is as similar as possible, e.g. the transformed $5 \times 5$ ($n = 100$) and the non-transformed $9 \times 9$ ($n = 82$) dataset. If the increase in the mean FHD were only due to the number of independent variables, it would be expected that the transformed dataset would show a higher mean FHD. However, the mean FHDs of the non-transformed datasets are consistently significantly higher, indicating that there are also other effects taking place. In summary, the observed trend of an increasing FHD proportional to $n$ seems to be based on a combination of the increase in the number of independent variables, as well as some additional factors of the datasets.

When comparing the distributions of the FHDs of the datasets with the FHDs of the LR, one may notice that in both cases the distributions start to shrink with an increasing $n$ and that there are some similarities regarding the types as well. This suggests some correlation which may be explained with the following: The wider distributions of the datasets mostly affect the lower boundary, i.e. there are more CRPs that are strongly correlated. This means that there are on average more visually similar responses in the training sets, which may bias the ML model and increase the tendency towards overfitting. The over-represented visually similar responses are predicted more accurately, but in turn, the results on the other responses deterioriate. As $n$ grows, the number of closely correlated responses shrinks and therefore the model's predictions stabilize.

Lastly, one may think that the generally worse mean FHDs of the predictions for the type C datasets indicates that the LR has more trouble modeling these. Except for the $5 \times 5$ and $7 \times 7$ datasets, this observation holds for the type B datasets as well. However, there are also noticeable differences between the results for the type A and D datasets, which, as already shown in \ref{sec:cDist}, should show no differences when $n$ is larger than for the $9 \times 9$ datasets. For the $11 \times 11$ datasets, the difference between the best and worst mean FHD across all types is $0.012$, while the difference between these two virtually equivalent datasets already amounts to $0.007$. This value can only be inferred from the randomness of the challenge generation. This shows that chance also plays an important role, which, as the mean FHDs do not vary much more than these values, makes it difficult to draw precise conclusions. In general though, it can be said that even if there were fundamental differences between the types that affect the results of the LR, the influence is quite small and, depending on the usage, may even be neglected.

\subsubsection*{PC and SSIM}
The \textbf{PC} between the real and predicted responses is at a constantly high level, showing that there is a strong linear correlation between the images. The repeated decrease of the mean with an increasing $n$ substantiates the assumption that the model's predictions gradually become worse, as has been observed for the FHD as well. The differences found for the predictions after the aforementioned challenge-transformation procedure produced a similar relative decrease for the PC as was found as an increase for the FHD. This suggests a relationship between the number of independent variables and $n$ with respect to the variations in the PC, as was found for the FHD before as well.

Again, a trend towards shrinking distributions for larger $n$ can be observed, which at large matches the results of the FHD and may therefore be explained in the same manner. Even though the distributions of the type B datasets are noticeably wider here as well, there are still some differences when comparing the results for individual datasets to those of the FHD. This again shows that there are some significant differences between how the metrics measure similarity. Especially prominent is the lower correlation for the type B datasets starting with the $9 \times 9$ variant, which indicates that the model's predictions are slightly worse with respect to the PC. Due to the relatively large distance to all other types as well as its consistent occurrence, this is unlikely to only emerge from chance. While one may interpret some patterns in the results for the other types as well, the differences are so small that they may as well only originate from chance. Altogether, since the overall range the values are moving in is very small in the first place, it is difficult to draw clear conclusions from the PC in this case.

The \textbf{SSIM} is at an even higher value than the PC and shows close to no deviations in the absolute values. The pattern of a higher SSIM for type B and lower SSIM for type C datasets is rather constant and therefore quite unlikely to be only due to chance, but the differences are so small that further investigations are rather difficult to carry out. In any case, even assuming that there is a structural reason for this pattern, due to the minuscule scale they are moving in, they were deemed to be negligible.

\subsection{Deep learning}
The most prominent observation for the DL attack is probably the clear gradual deterioration of each metric with an increase in $n$. A closer look at the setup of the generator architecture shows that $n$ corresponds to the number of input channels of the first transposed convolutional layer. However, the number of output channels of this layer as well as the number of inputs and outputs of all following layers are fixed, i.e. only the input into the network varies corresponding to $n$.

In general, a two-dimensional (transposed) convolutional layer can be thought of as a tensor of shape $c' \times c \times k_w \times k_h$, such that $k_w$ and $k_h$ correspond to the kernel sizes, $c$ to the number of inputs and $c'$ to the number of output channels. This means that for each output channel, there is a tensor of shape $c \times k_w \times k_h$, which is also called filter, i.e. a collection of $c$ kernels. When carrying out the convolutions, the values of each output channel are calculated from the corresponding kernels, which operate on all input channels. So, for each output channel, a convolution is performed \textit{on each channel} on the two-dimensional input using the two-dimensional kernel, summing up all results over the input channels to again retrieve a two-dimensional tensor for each output channel. This results in a transformation of a $c \times w \times h$ tensor into a $c' \times w' \times h'$ tensor, such that $w, h$ and $w', h'$ correspond to the sizes of the inputs and outputs. These sizes are also dependent on the stride and padding of the convolution.

Now, in the case of the generator, the first layer corresponds to a tensor of shape $1028 \times n \times 4 \times 4$, that receives inputs of shape $n \times 1 \times 1$ and produces outputs of shape $1028 \times 2 \times 2$. It is already clear that while the complexity of the inputs increases by adding more channels for a larger $n$, the output of the layer is of fixed size and does not provide an accordingly larger complexity to react to the increasing number of input channels. In other words, the convolution has to gradually operate on more input channels in the above explained manner, but has to map these a fixed number of output channels. As was already seen for the LR, it is possible that other circumstances influence the decrease of the prediction quality as well.

\subsubsection*{Fractional Hamming distance}
In addition to the increase in the means for larger $n$, the model constantly produces the best results with respect to the FHD for the type B datasets. Note that for these datasets, every other bit and therefore every other input channel will in all cases have a value of $0$. This means that all corresponding kernels of the filter for these channels will not have any influence on the output of the first layer, since, due to the nature of a convolution, all results will be $0$ as well. This means that no matter how these kernels are adjusted during backpropagation, they will never influence the output of the layer. This will, of course, also impact the training process and may even result in the corresponding kernels being ignored, effectively only considering half of all input channels. This leads to a smaller complexity on the initial input of the model, which, as previously analysed, makes the model produce better results. To test this assumption, the model that was trained on the $11 \times 11$ dataset of type A was taken as an example. The model first received regular challenges from the test dataset and afterwards a modification of these challenges where \textit{every} usually deactivated bit was instead activated, to compare the predictions of the corresponding responses. And indeed, the predicted responses for the regular and modified challenges were almost identical, presenting a FHD of only around $0.006$ (PC: $0.999$, SSIM: $0.996$). If instead only half of the deactivated bits were set to $1$, the FHD shrinks to around $0.0017$ (PC: $0.999$, SSIM: $0.999$) and going even lower quickly results in completely identical predictions. This clearly shows that, as suspected, the model actually does learn to almost completely ignore the corresponding input channels.
Lastly, it seems that the model actually does perform a bit worse for the type C datasets, as this pattern is quite consistent across varying $n$ and the three metrics. No external or unwanted influence was found that could be the cause for this behavior, however, these differences are most of the time quite small to begin with.

\subsubsection*{PC and SSIM}
All previously analysed relationships for the FHD can mostly be applied to both the \textbf{PC} and \textbf{SSIM} as well. This indicates that the model's predictions actually do become worse and the observation is not only specific to the FHD. However, no further observations can be made from these metrics that would give further insights beyond what has already been found for the FHD.

\section{Comparison of linear and deep learning attacks}\label{sec:comparison_attacks}
Now, after exploring the results of both ML attacks, there are a few interesting contrasts that are worth to further elaborate on. Probably the most prominent difference is the development of the prediction quality with an increase in $n$. Even though it was shown that both approaches do perform worse for larger $n$, there is a significant discrepancy between the degrees of deterioration. When grouping the four types for each challenge size and taking the mean, the LR starts with a FHD of $0.186$ for all $5 \times 5$ datasets that increases to $0.201$ for the $15 \times 15$ datasets. On the other hand, the corresponding means for the generator go from $0.124$ up to $0.253$. In other words, while the mean FHD of the LR increases by only around $8\%$, the mean of the generator increases by $104\%$. The LR model is by far more stable with respect to the mean FHD when $n$ increases, while the generator is a lot more sensitive. This makes the LR a better pick for datasets with a larger $n$, while the generator outperforms for those with smaller $n$. This is also supported by the distributions of the FHDs for the LR, which present significantly more worse predictions for smaller $n$ and only gradually approach the rather stable distributions of the generator. Also, due to the virtually lower number of usable challenge bits, the generator should be preferred for datasets of type B with an at most medium-sized $n$ (with respect to the in this thesis investigated size range), as starting with the $13 \times 13$ datasets, the LR starts to outclass the generator even for these datasets. For all other datasets, the LR model should be preferred starting with the $9\times 9$ dataset, where it again starts outperforming the DL model. However, obviously, all of this only holds for this specific kind of linear and DL model and may be different for other architectures, as will be investigated in the next chapter.

In practice, when choosing the best ML architecture, this discrepancy between both approaches greatly outweighs all other observations on the performance that were made. Nevertheless, an additional aspect should be highlighted as well. Firstly, it was seen that the LR on average produces slightly worse results for the type B and C datsets. While for the generator, the type B datasets are biased due to the network ignoring parts of the challenge, for the type C datasets, this observation can be made as well. Since this is consistent across most of the dataset sizes and especially across both models, it seems that the restriction of type C actually makes it more difficult to accurately predict the responses using these ML attacks. However, these differences are not too large, and most of the time do not even exceed a deviation in the mean FHD of around $0.01$ to $0.02$ to the next values for the type A or D dataset. Also, as mentioned in \ref{sec:eval_lr}, the mean FHD of the predictions of the LR for two equivalent datasets (i.e. type A and D) already differed by up to $0.007$, which goes up to even $0.028$ on the $13x13$ datasets for the generator. Therefore, even the randomness of the challenge generation can already have a similarly large influence on the quality of the predictions. And finally, this still assumes a naive approach on the side of the attacker, where the information that only half of the bits can be activated is not detected or completely ignored. However, the consistency of this observation indicates that there are some fundamental differences that emerge from the restriction of type C, which may be interesting to further investigate in different experiments.

Except for these aspects, no other differences that would be worth to further investigate have been found. Both the PC and the SSIM mostly follow the observations of the FHD and therefore support the conclusions that could be drawn from this metric. While there is a considerably large drop in the absolute value of the SSIM from the LR to the generator, since it is not an independent measurement for the latter model, one should be careful in interpreting these differences.

All in all, the previous observation that the restrictions do not greatly increase the complexity of the datasets (within this size regime) also holds after investigating the ML attacks. On the contrary, the type B restriction even greatly reduced the complexity on the DL model side, as the restriction was automatically learned by the algorithm without even giving any further information. Only the type C restriction might be considered to make the PUF a bit more difficult to model, but the differences are very small and would first have to be investigated on other setups and for a larger number of CRPs.

    \chapter{Further experiments}
In the following, further experiments on the simulation as well as the ML attacks will be presented. This includes a new setup for the simulation, modifications of the old ML attacks, completely new ML models, attacks on real optical PUFs and attacks with a larger number of CRPs, both for the linear and non-linear approaches.

\section{New simulation setup} \label{sec:datasetx2}
\begin{figure}
\centering
  \includegraphics[width=1\linewidth]{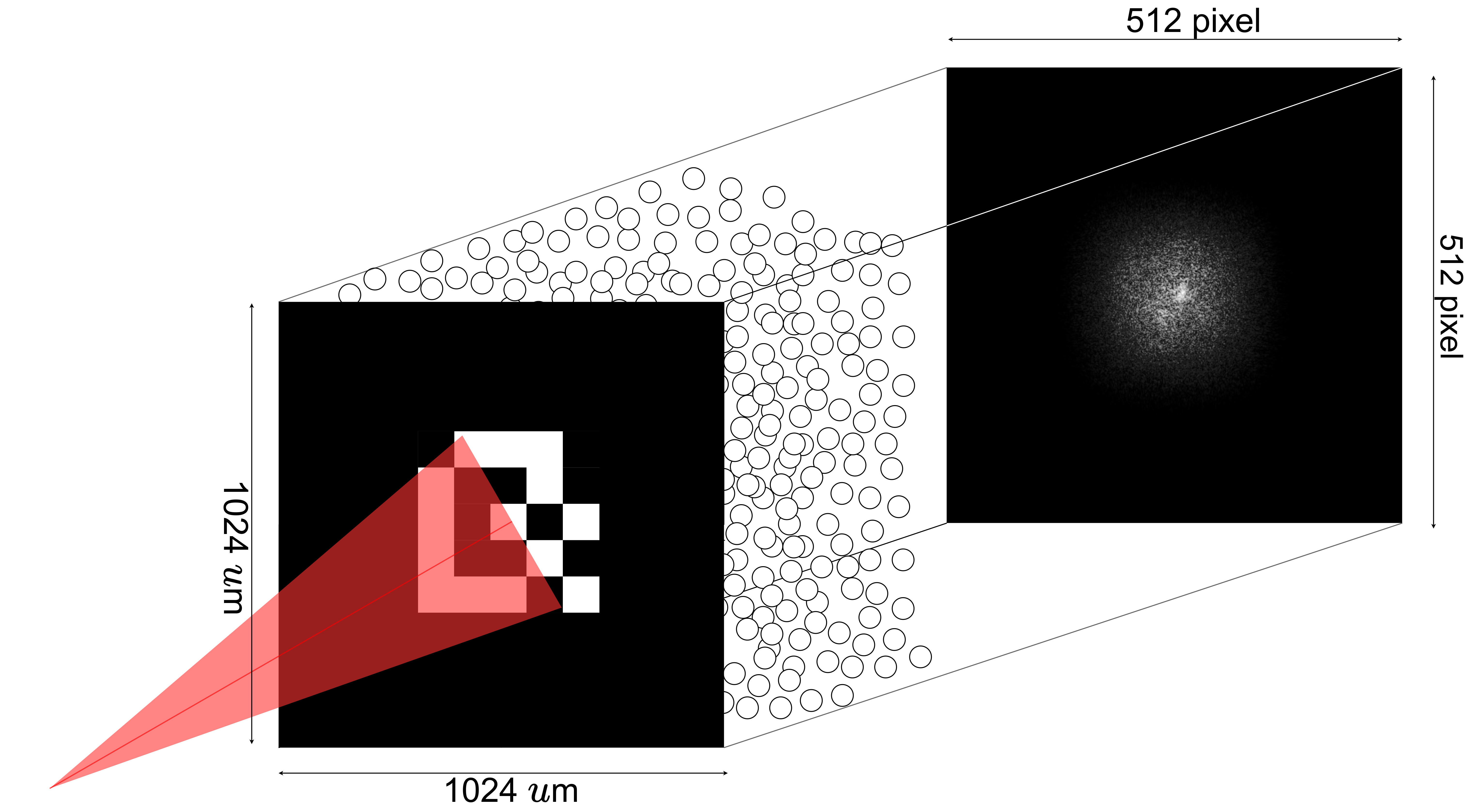}
\caption[Setup of the larger simulation.]{The new setup for the simulated PUF where the length and width are doubled.}
\label{fig:setupx2}
\end{figure}

\begin{figure}
\centering
  \includegraphics[width=.4\linewidth]{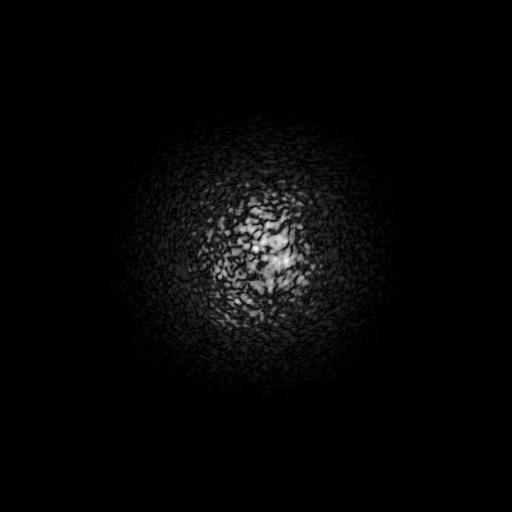}
  \hspace{1cm}
  \includegraphics[width=.4\linewidth]{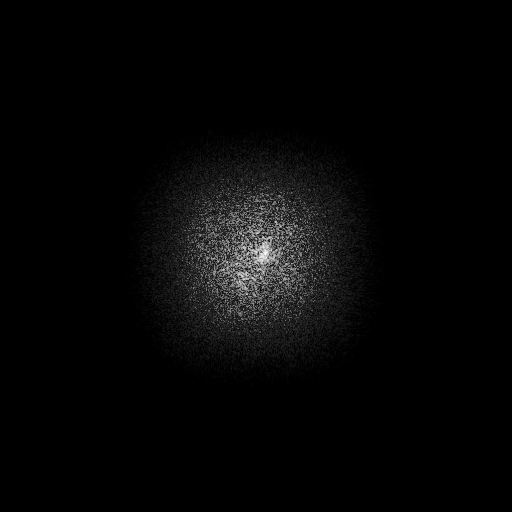}
\caption[Comparison of the responses for the initial and the larger simulated PUF.]{Two speckle patterns for a $11 \times 11$ dataset for the initial simulation setup (left) and the new setup with doubled lengths and widths (right).}
\label{fig:responsesx2}
\end{figure}

This part introduces an additional simulation setup with changes in a few of the parameters that are used. The idea is to double both the length and width of the PUF while keeping the depth constant. This virtually quadruples the area of the challenge mask and the responses, and changes the whole PUF's dimensions from $512 \mu m \times 512 \mu m \times 512 \mu m$ to $1024 \mu m \times 1024 \mu m \times 512 \mu m$. Likewise, the number of scatterers is roughly quadrupled as well to fit the resulting larger volume. All other settings are kept constant and the responses are still recorded with a size of $512 \times 512$ pixels. An illustration of how the setup looks like can be found in \ref{fig:setupx2}. It was observed that this new setup creates more fine-grained responses, as can be seen in figure \ref{fig:responsesx2}. It was hoped that these changes increase the complexity by responding to one of the flaws entailed by a larger number of blocks $n$: the absolute size of a block and therefore the amount of light that can travel through gradually decreases proportional to $n$. While this relationship of course does not change with the new setup, it may still help to increase the influence of flipping single bits, since the initial area of each block is larger compared to the initial setup. As an example: When using $4 \times 4$ blocks with the first setup, each block will be of size $(128\mu m)^2 = 16.384 mm^2$, while the new setup yields blocks of four times the size, i.e. $(256\mu m)^2 = 65.536mm^2$. Additionally, since the overall area of the speckle pattern and the total number of scatterers increases, the whole PUF is hoped to become more complex, since a significant amount of new information is introduced to the simulation. For the evaluation, the same 24 datasets as previously investigated with 1000 CRPs each are generated and evaluated in the same manner. The new setup will also be simply referred to as \textit{larger simulation}.

Even though the apparently more detailed responses may look very promising in terms of a larger complexity, the evaluation actually showed a decrease in both metrics. For the \textbf{first Gabor transformation}, the average mean \textbf{FHD} across all datasets falls from $0.424$ to $0.414$, which is largely influenced by the significantly lower mean FHD for the datasets with smaller $n$, where the lowest mean FHD lies at $0.380$ (previous minimum: $0.412$). However, the larger $n$ becomes, the more these values start to approach the previous means, i.e. there is a strong tendency to an increasing FHD proportional to $n$. This quickly falls off with the $11 \times 11$ dataset, where the values for both setups start to align. However, for the $13 \times 13$ and $15 \times 15$ datasets, there is actually still a very small increase in the mean FHD for all types except B. The distributions of the FHDs are quite similar and again shrink with an increasing $n$. This phenomenon is therefore not specific to the previously generated datasets.
Interestingly, the type B datasets seem to produce significantly worse results with the new setup and most of the time present the lowest mean FHD, which is especially prominent for the $5 \times 5$ and $7 \times 7$ datasets. For the initial setup, no similar distinction could be found and the type B datasets actually tended to have one of the higher FHDs. However, for both setups, the type C datasets produce the highest mean FHDs.

Using the \textbf{second Gabor transformation}, the results become even worse. While the mean FHD previously fell to $0.352$ and therefore already significantly decreased, this time, the mean FHD on average lies at only $0.284$, with the largest value only at $0.295$. However, the same relationships for the types can be seen again. It is highly likely that the more fine-grained responses are the cause of this deterioration and that the second Gabor transformation is an even worse fit for the larger simulation than it already was for the initial one.

Similarly, there is a rather large drop in the mean \textbf{Shannon entropy} from $2.866$ to $2.368$. In contrast to the FHD, this is not caused by a part of the datasets, but is instead consistent across all of them, i.e. each mean is about $0.4$ to $0.5$ lower than for the initial datasets. All relative relationships still hold, except that the distance of the type B datasets to the others is a bit lower. However, as already stated in \ref{sec:discussion_entropy}, the Shannon entropy is not the best fit for estimating the complexity of a response and one therefore should be careful with interpreting these values.

In general, the new setup did not show many improvements compared to the initial setup and instead even produced on average worse results. However, a small increase in the mean FHDs for larger $n$ could be observed. This may indicate that the increase in the size of the blocks actually does help at some point, where the influence of a single block starts to become too small for the initial simulation. Altogether, from this evaluation, the new setup should not be considered as superior, but instead as an alternative that has its own benefits and downsides compared to the initial one. Again, further investigations regarding special challenge compositions, a significantly larger number of CRPs and maybe even other types of Gabor transformations or entropy estimations should be considered to fully understand the potential of this setup.

\section{Further attacks on simulated data}
So far, only the standard datasets have been the subject of the modeling attacks. While there already were a lot of variations introduced by using different challenge sizes and restrictions on the challenge compositions, the underlying PUF setup as well as the type of response did not change at all. In the following, the results of attacks on the previously introduced larger simulation and on only the cropped responses will be investigated.

\subsection{Attacks on the new simulation setup}
As previously seen in chapter \ref{sec:datasetx2}, the larger simulation should not be regarded as a simple improvement to the initial one, but instead as a separate type of PUF. Both the generator and LR were trained on this new type of data, to see how the approaches perform compared to initial setup. Note that the objective of this section is not to find the best type of attack on the new data, but instead to see how the already implemented models perform when they are used on a similar but still somewhat different type of data. Therefore, no changes to the models or any hyperparameters were made and both attacks were carried out in the exact same manner as already explained in chapter \ref{sec:implementation}.

This time, the \textbf{LR} shows a clear gradual decrease in its prediction quality with respect to the \textbf{FHD}. The best mean FHD was registered at $0.146$ (previously: $0.179$) for a $5 \times 5$ and the worst at $0.216$ (previously: $0.206$) for a $15 \times 15$ dataset, i.e. compared to the initial setup, the results are better for smaller, but worse for larger $n$. The distributions follow this pattern of an increasing FHD and again slightly decrease in range quite similar to how it was investigated before. Instead of a decrease in the number of observations on both ends of the distribution, only the number of very accurately predicted responses declines, while the upper boundary of the distribution stays about the same. However, compared to the attack on the regular dataset, the distributions are generally a bit more narrow.
Most of the time, the model produces the best results for type B and the worst for type C datasets, with the other two in between. This matches what could be inferred from the evaluation of this dataset for the four different types.

The \textbf{PC} shows fewer differences to the regular datasets' results. All mean values move in a very similar size range of $[0.943,0.965]$, where the lower values appear starting with the $11 \times 11$ dataset, where a small general decrease can be found. This decrease has already been observed before, but became slightly larger for the new setup. However, the absolute values of these differences are again quite small in the first place, so the actual quality with respect to the PC does not differ that much. The distributions are again a bit more narrow compared to the initial setup, but are still quite similar. The relationships between the types generally match those found for the FHD, but there is no consistent large gap between the values of type B and the others for larger $n$.

The \textbf{SSIM} also presents a slight decrease in the mean values, but shows no noteworthy differences to the previous attacks.

The \textbf{generator} shows rather strong differences with respect to the \textbf{FHD}, which is consistently worse than for the regular datasets. The mean FHDs all increased by values between roughly $0.05$ and $0.1$, such that the relative differences on different challenge sizes and types as well as the gradual increase of the FHD proportional to $n$ is still quite similar, but shifted upwards significantly. This shift becomes even stronger when $n$ increases. The new best mean FHD increased from $0.093$ to $0.145$, both on the type B dataset of size $5 \times 5$, and the worst mean FHD from $0.278$ to $0.341$, previously for type C of size $15 \times 15$ and now type C of size $13 \times 13$. The earlier appearance of the worst mean FHD is likely only due to chance, since the corresponding dataset of size $15 \times 15$ shows a very similar FHD of $0.337$ and some variations due to the CRP sampling have already been observed before. This behavior can be found for the distributions as well, which in addition seem to follow the tendency to become a bit wider, especially on type B datasets, i.e. the predictions become more unstable. All previously observed relationships between the types in general still hold for the new dataset, i.e. the generator performs best on type B and worst on type C.

Both the \textbf{PC} and \textbf{SSIM} behave in a very similar way, showing a general drop compared to the previous setup. The same tendency of wider distributions as seen for the FHD can be found as well, and again, the already identified relationships between types are still valid.

It became apparent that there are significant differences on how the two approaches handle the new datasets. The quality of the predictions of the LR does not greatly differ with respect to the PC and SSIM, but there are noticeable differences on the FHD, which acts as the most important of the three metrics. These differences match those that can be found between the FHDs of the dataset evaluation. Here, a rather large drop in the FHD for smaller $n$, but a slight increase for larger $n$ could be observed, which directly correlates to the increases and decreases in performance of the LR. On the other hand, the generator shows a significant overall decrease in its predictive abilities. The correlation with the observations on the datasets can still be found for the DL model as well, since the deterioration becomes even stronger when $n$ increases.

It seems that the new simulation setup can actually be more difficult to model for larger $n$, but, at least in the case of the LR, makes it easier when $n$ is small. The significantly worse results of the generator indicate that either the architecture or the hyperparameter configuration does not quite match the new datasets. Since DL models have a lot more hyperparameters that can be adjusted and are usually highly sensitive to small changes, better results can be expected after fine-tuning the model. This indicates that in general, the linear attack can be better transferred onto new domains, whereas the DL attack is more sensitive to such changes and might require reconfigurations.

\subsection{Attacks on cropped responses}\label{sec:croppedAttack}
As already explained in chapter \ref{sec:implementation}, the speckle pattern of the simulated PUF only emerges in the center of the response, while the rest is filled with a large dark area. For the computation of the FHDs, a center square of $128 \times 128$ pixels is cut out of the response so as to not distort the value due to the constantly dark parts of the images. While this does catch the largest part of the relevant area in most cases, due to the varying sizes and the circular nature of the speckle pattern, it is inevitable that sometimes parts of the outer pattern are cut off. For this reason, the attacks up to this point have been applied with the intent to produce the full image, so the models have to predict the whole response and therefore the whole speckle pattern as well. However, this approach also forces the models to ``predict'' the large unused dark area. From the view of an attacker, this is unnecessary, since only the resulting bitstrings from the Gabor transformation have to match for the attack to be successful. Therefore, it may be worth to take a look at whether, and if so, how the performances differ if from the beginning, only the center square used in the FHD calculation is given to the models. Note that both the PC and SSIM will be expected to be significantly lower, since the dark area already increases the correlation and similarity between two responses, without taking the speckle pattern into account. Therefore, the basis on which these two metrics are computed is completely different for this new approach, so no comparisons to the previous results will be made. The FHD, however, directly correlates with the change in performance, since the basis of computation is still the same.

This new approach was only carried out for the generator, since there is in fact no difference for the LR models. As previously stated, the actual model that is referred to by \textit{LR} is a multivariate linear regression, which, as explained in chapter \ref{sec:backgroundML}, computes a standard multiple linear regression for each of the dependent variables, i.e. for each pixel of the response. Here, each of these standard regressions has its own set of regression coefficients. Removing a part of the images is therefore equivalent to removing the multiple linear regression that correspond to these pixels, which does not have any impact on the computations of the pixels that are left.

To run the new experiments, slight alterations to the previous architecture of the DL model from table \ref{tab:generatorLayers} had to be made to match the new images sizes. Two different models have been tested out on this account: For the first model, simply the first two layers of the regular model have been removed and the third layer was changed to act as the input layer. The number of filters of each layer did not change, such that the overall number of filters is lower, but consistent with the corresponding layers of the initial model. For the second model, the last two layers were removed, changing the third to last layer into the output layer. This way, the number of filters are identical for the input and the following layers, but are cut off into the final output a bit earlier to make up for the missing two layers. The architecture of both models can be found in table \ref{tab:generatorCropLayers}. The first model will also be referred to as \textit{cropped generator type 1} and the second as \textit{cropped generator type 2}.

With the \textbf{first model}, there is a clear decrease in the \textbf{FHD} for almost all datasets. The average mean decreased to $0.189$ (previously: $0.199$) and the ranges of the means to $[0.082,0.246]$ (previously: $[0.093,0.278]$). The distributions tend to shrink and simultaneously shift downwards across all datasets. Apparently, the DL approach does actually quite consistently perform better without having to predict the dark area, even if a considerably lower overall number of filters is used. The mean \textbf{PCs} and \textbf{SSIMs} did decrease as expected and the distributions largely grew in size at the lower boundary. This shows that there are still a lot of predictions that are not much worse than when predicting the full response, but also a significant number of predictions that are actually worse when only observing the ``relevant'' part. In general, only the absolute mean values and distributions of all metrics changed, but the relationships between the results for different challenges sizes and types still hold.

The \textbf{second model} shows even better results, outperforming the previous one on every dataset and decreasing the average mean \textbf{FHD} to $0.166$, with all means in the interval $[0.065,0.219]$. Except for this, there are no noticeable differences to the previous models. The distributions shrink and shift downwards even more than for the first model, but all relative relationships again still hold and apparently no other effects are taking place. Again, the mean \textbf{PCs} and \textbf{SSIMs} are significantly worse than for the initial generator, but both greatly improved compared to the previous one.

Apparently, the DL attack works significantly better when only the relevant segment of the response is used and the model does not have to additionally focus on parts that will not be used for the final response. Investigating how the model performs on the whole response may be interesting in an experimental setup to draw certain comparisons, but in a real-world scenario, the adversary will in general only strive to be able to predict the right binary response to the challenge. Since both new approaches greatly improved the performance of the DL attack, they should be preferred in all cases. Also, except for more computational resources that are necessary to run the attacks, there are no downsides in using the second model over the first one, since this one consistently performs better. All further analyses will therefore only relate to the performance of the cropped generator type 2.

Since the LR performs the same on the cropped responses, the new model changes the relationships between the DL and the LR attacks. Previously, the LR started to outperform the DL model starting with the $9 \times 9$ datasets and gradually built up its advantage with an increase in $n$. Now, this advantage only starts with the $11 \times 11$ datasets and is significantly smaller than for the initial model, even for larger $n$. Most of the time, this decrease in the FHD does not even exceed a value of $0.01$. Additionally, while the DL attacks show better results for the type B datasets, the LR initially still outperformed the DL model starting with the $13 \times 13$ size. This now shifted upwards to $15 \times 15$, however, the difference in the FHD between both models here is only $0.002$, so it is practically negligible. Also, the predictions of the DL models are now even more stable than before, which can be seen in the significantly smaller distributions. This effect holds on until the $15 \times 15$ datasets, where the size of the distributions start to align for both models.
Altogether, the new approach shows that there was still a lot of space left for improvements on the DL attack side and resulted in a modification of the initial generator that outperforms the previous one in all aspects. However, note again that this only accounts for the investigated challenge sizes and number of CRPs.

\begin{center}
\begin{table}[htbp]
{\small
\begin{center}
\begin{tabular}[center]{cccccrr}
\toprule
\multicolumn{5}{c}{}{}{} & type 1 & type 2 \\
Nr. &  Layer & kernel & stride & padding & parameters & parameters \\
\midrule
1 & T-Conv2D, BN, LeakyReLU & 4 & 2 & 1 & 102,912 & 411,648 \\
2 & T-Conv2D, BN, LeakyReLU & 4 & 2 & 1 & 524,544 & 8,389,632 \\
3 & T-Conv2D, BN, LeakyReLU & 4 & 2 & 1 & 131,200 & 2,097,664 \\
4 & T-Conv2D, BN, LeakyReLU & 8 & 4 & 2 & 131,136 & 2,097,408 \\
5 & T-Conv2D, Tanh & 8 & 4 & 2 & 2,048 & 8,192\\
\bottomrule
& & & & & 891,840 & 13,004,544 \\
\end{tabular}
\end{center}
}
\caption[Structure of the cropped generators.]{Structure of the generators for the attacks on the cropped responses. \label{tab:generatorCropLayers}}
\end{table}
\end{center}

\section{New machine learning attacks}
Both linear and non-linear attacks seem to have their own advantages and downsides and should be preferred under different circumstances. At the same time, both approaches have raised further questions. The LR seems to be rather stable and is not greatly influenced by $n$, but there is still room for optimization that could lead to an even lower general FHD. On the other hand, while the generator produces good results for a smaller $n$, as soon as $n$ increases, the prediction quality drops significantly. To further explore if even better modeling attacks on the simulated PUF are possible, additional experiments with new approaches were carried out. Note that for each of the following experiments, both Gabor transformations were applied, but just as for the regular attacks, there was no noteworthy difference except for a generally lower FHD using the second transformation. Therefore, all numerical results relate to only the first Gabor transformation. Furthermore, the focus will be shifted even more towards only the FHD, since both the PC and SSIM mostly supported the observations that could be made for this metric and did not introduce any new relevant aspects. On this account, if nothing is explicitly stated, the PC and SSIM match the results described for the FHD. Lastly, all models will from now on only be trained on the cropped responses introduced in chapter \ref{sec:croppedAttack}, since the approach is equivalent for LR-based models and showed to improve the results for the DL models.

Note that in the following, only the initial simulation setup defined in chapter \ref{sec:Concept} will be subject to the ML attacks. However, all of the following attacks have also been carried out for the earlier introduced larger simulation setup. Here, the same behavior as previously discovered still applies, i.e. there are not many differences on the linear attacks and all deep learning models generally perform worse, but all relevant relationship still hold. Therefore, for reasons of simplicity, it will be refrained from explicitly presenting this data. However, all results can be found in the appendix under \ref{sec:appendix_attacks}.

\subsection{Ridge regression}\label{sec:Ridge}
Ridge regression (RR) is a modification of the linear regression, where the coefficients are estimated by an augmented version of OLS, the so called \textit{ridge estimator}. This adds a penalty term of the squared value of the magnitude of the coefficients, which is also referred to as \textit{L2 regularization}. The loss functions turns into
$$\Vert y - X \beta \Vert^2 + \lambda \Vert \hat{\beta} \Vert^2$$
and, solving this for $\beta$, the estimator becomes
$$\hat{\beta}_{ridge} = (X ^{\mathsf{T}} X + \lambda I)^{-1}X ^{\mathsf{T}}y,$$
where $\lambda$ is a positive constant that controls the size of the penalty and $I$ is the $k \times k$ identity matrix for $k$ coefficients. Effectively, this penalty term forces the coefficients to shrink towards $0$ and is often used to compensate multicollinearity. \cite{hoerl1970ridge} Obviously, no set of variables in this setup can actually be multicollinear as there is no correlation between the states of individual bits in a challenge, but RR may still reduce the complexity of the model and improve the performance. Additionally, since there is a large number of independent variables that each have a rather small influence, it may actually be desired to have coefficients that are closer to zero, especially for the datasets with larger $n$.

As already seen in chapter \ref{sec:eval_lr}, the LR performs worse when the number of independent variables increases. The previously explained challenge transformation, which transfers a challenge into an equivalent challenge with larger $n$, was carried out and used to transform the training sets for the RR as well. The results show that the RR is not afflicted by this behavior in the same manner. While there is a tendency to an increase in FHD as well, the difference is in the order of around $0.0001$ in the most extreme case and therefore completely negligible. Since the only difference between these two regressions is the L2 regularization, the differences in the magnitudes of the coefficients are most likely the reason for this observation. An investigation on these magnitudes showed that while it was not uncommon for the coefficients of the LR to exceed sizes of $10^{13}$, the coefficients of the RR tend to move in a range of $[-1, 1]$. It is therefore likely that this observed behavior of the LR directly correlates to the sizes of the coefficients and can therefore be avoided by instead using the RR to shift these values towards $0$.

The RR was run on all datasets with a specific parameter $\lambda$ that was empirically evaluated individually for each dataset. The results were not too different from those of the LR, showing only minor deviations most of the time. The overall mean FHD fell from $0.195$ to $0.192$. However, the majority of the smaller deviations and all comparably larger differences showed an improvement to the LR, which becomes even stronger for larger $n$. This would match the observation of increasingly worse results for a larger number of independent variables using a LR model, which can be avoided via the regularization term of the RR. Interestingly, increasingly large differences on only the type B datasets starting with the size $9 \times 9$ could be found, increasing up to a difference of even $0.03$ for the size $15 \times 15$. No such influence could be inferred from the increase in the number of independent variables, so the RR seems to outperform the LR on these types of datasets, with an increasing impact for larger $n$. The regularization, therefore, seems to help especially for challenge compositions where bits are permanently deactivated, i.e. where some independent variables are never used.

Altogether, the RR did not significantly outperform the LR, but showed some minor improvements for the datasets with larger $n$. The regularization seems to increase the performance, which probably at least partly relates to the finding that a larger number of independent variables does not deteriorate the results, as seen for the LR. While there is some additional overhead due to the computation of the $\lambda$ parameter, the RR should be preferred over the LR to improve the prediction quality.

\subsection{Linear attacks with quadratic dependence}\label{sec:opr_attack}
Optical PUFs following the principle of the PUF of Pappu et. al. \cite{pappu} use a \textit{linear} scattering medium. The simulation used in this thesis is no exception; all processes that take place are completely linear. However, this only applies for the electrical field that emerges when the light passes through the PUF. Since cameras are only sensitive to intensities, the phase can be neglected in this case. However, for a given amplitude $E_i$, the corresponding intensity is defined as $I_i = |E_i|^2$, so the recorded speckle patterns only hold information about the absolute values of the electrical field and are therefore actually a quadratic function of the challenge. Since the responses of the simulation only measure the intensity as well, this also applies to the simulated data. Rührmair et. al. \cite{ruhrmair2013optical} have elaborated an analysis on how to turn this into a linear relationship: Assuming the challenges are of size $n$, the intensity at the cell $i$ is defined as
$$I_i = |E_i|^2 = |\sum_{j=1}^n T_{ij}b_j|^2 = \sum_{j=1}^n\sum_{k=1}^n T_{ij}T_{ik}^*b_jb_k,$$
where $T_{ij} \in \mathbb{C}$ and $b_{ij}$ is a binary variable that indicates whether block $j$ of the challenge mask is turned on or off. By defining a matrix $R_{i,j \cdot n + k} = T_{ij}T_{ik}^*$ and the vector $\beta_{j \cdot n + k} = b_jb_k$ of suitable dimensions, the intensity can be written as
$$I_i = \sum_{l=1}^{n^2} R_{il}\beta_l,$$
such that the responses are now linear in $\beta$. \cite{ruhrmair2013optical} In other words: Using this approach, the initial challenge, a bit vector of size $n$ where each bit refers to whether the corresponding block is turned on or off, is transformed into a new bit vector of size $n^2$, that consists of all products of all ordered pairs of the initial bit vector. This, however, is partly redundant, since there is no need to treat ordered pairs of challenge bits as distinct, as they refer to the same challenge configuration. By applying the same procedure, but instead only considering the products of all unordered pairs, the same result can be achieved with challenges of only size $\frac{n (n + 1)}{2}$.

First, this transformation of the challenges was applied to the datasets and used to train a new LR model, also referred to as \textbf{QLR}. As one would expect, the new attack shows a general improvement in the quality of the predictions, decreasing the mean FHD from $0.194$ to $0.171$, which is especially influenced by two extremes: The mean FHD of the predictions on all $5 \times 5$ datasets significantly improved by between $0.08$ and $0.1$, such that all of these are now below $0.1$. Except for the best DL model of chapter \ref{sec:croppedAttack} on the type B dataset, this by far outperforms all previously investigated models for this challenge size. The same holds for the type B dataset of size $7 \times 7$. The other extreme is the type B version of the $9 \times 9$ dataset, which, with a mean FHD of $0.295$, presents the highest value that was recorded across all attacks. All other datasets showed a small but noticeable decrease and shrinking distributions with a small downwards shift. The \textbf{PC} and \textbf{SSIM} again support these observations.

With these results, one can see that the new linear approach overall performed better, but the most significant improvement can be seen from the attacks on the datasets with the lowest $n$, i.e. where the least number of independent variables are used. As we have seen before in chapter \ref{sec:eval_lr}, the LR performs worse if the number of independent variables increases. For the new attack, this number is larger from the beginning and increases much faster, since a challenge of size $n$ is transformed into a new challenge of size $\frac{n (n + 1)}{2}$. It may be that the results are biased to some extent due to the deterioration that comes along with a larger $n$. The extreme outlier of the type B dataset of size $7 \times 7$ was further investigated, but no differences in neither the data nor the model that could explain the large deviation in performance could be found. It appears that some combination of the challenge size and the collected CRPs makes the approach fail to infer a fitting linear model when using LR.

Next, the same procedure was applied for a new RR model, referred to as \textbf{QRR}. Here, some further improvements with respect to the previous QLR could be achieved, decreasing the mean FHD further to $0.164$. The improvements are larger for the datasets with smaller $n$ and gradually decrease when $n$ increases. Note that while the development of these improvements is opposite to what has been observed for the RR on the regular challenges of chapter \ref{sec:Ridge}, they are not directly comparable due to the different number of independent variables for each challenge size. As an example: while the dataset of size $5 \times 5$ of the regular RR creates $25$ independent variables, the new approach increases this number to $325$ following the previously mentioned formula. In addition, the outlier that was previously found using the LR does not appear here any more and instead the results on this dataset are now similar to those of the others of the same size. This indicates that the QLR simply failed to infer a fitting model for this dataset, which can be counteracted by adding the regularization term of the QRR.

This new approach for linear models showed some further improvements in the quality of the predictions, especially for the datasets with smaller $n$. This also introduces a gradual increase in the mean FHD proportional to $n$, similar to how it has been observed for the DL models before. The approach now even outperforms the initial generator on almost all occasions, even for type B datasets and those with a small $n$, where the generator performed better than all linear models so far. However, the modified version of the generator of chapter \ref{sec:croppedAttack} still performs better on a few datasets: all datasets of size $7 \times 7$, all datasets of type B up to size $11 \times 11$ and the datasets of type A and C of size $9 \times 9$, however, the differences on the last two are rather small. Similar to how it has already been observed for the initial generator and LR in chapter \ref{sec:comparison_attacks}, the new linear approach therefore outperforms the DL approach for larger $n$, but now additionally for the smallest $n$ and a few other occasions as well. Needless to say, in terms of prediction performance, this new approach should be preferred over the regular LR and RR and, as has been observed for the regular models as well, preferably using the QRR instead of QLR.

\subsection{Advanced generator architecture}\label{sec:advancedGenerator}
As a last new ML attack, the architecture of the generator was altered. While a modified version has already been introduced in chapter \ref{sec:croppedAttack}, the underlying architecture did not greatly differ except for the removal of two layers to fit the new images sizes. On the contrary, the following DL model, also referred to as \textit{advanced generator}, was altered on several locations.

First, the input layer was changed to a fully connected layer, which receives the number of challenge bits $n$ as input and maps these onto $w \cdot d^2$ values, such that the resulting tensor can be reorganized into the shape $(w, d, d)$, which is handed over to a transposed convolutional layer. Here, $d$ refers to the height and width and $c$ to the number of channels of the abstract image representation that is used for the convolutions. These parameters were empirically chosen to be $d=32$ and $c=20$, i.e. each challenge is transformed into a tensor of shape $(20, 32, 32)$ before any convolutions are carried out. Afterwards, several transposed convolutional layers follow, alternating between kernels of size $3 \times 3$ with a stride of 1 and $4 \times 4$ with a stride of 2. A batch normalization is carried out after each convolution and LeakyReLU with a slope of $0.2$ was chosen as activation function. Further details can be found in table \ref{tab:advancedGeneratorLayers}. This new type of model was trained on all datasets for 300 epochs, using a batch size of $16$ and the ADAM optimizer with the values $\alpha = 0.01$, $\beta_1 = 0.8$ and $\beta_2 = 0.9$. Only the MSE was chosen as loss function.

The mean FHD fell to $0.036$ and all means now move in the interval $[0.023,0.066]$. Again, a gradual increase in the mean FHD proportional to $n$ can be observed. This time, however, the distributions actually become slightly wider proportional to $n$ instead of more narrow or no changes at all, as has been observed for all models so far. Therefore, the model performs better \textit{and} is more stable for smaller $n$. Most of the time, the best performance can be seen on the type B datasets, which becomes even more noticeable for larger $n$, i.e. the gradual deterioration of the mean FHD is not as strong for type B as for the others. However, no clear pattern regarding type A, C and D can be found and their results stick very close together consistently.

As can be seen easily, this new model greatly outperforms all previously investigated models on all the siumlated data. The mean FHD across all datasets decreased by $0.163$ and $0.159$ compared to the initial generator and LR as well as by $0.130$ and $0.128$ compared to the modified generator of chapter \ref{sec:croppedAttack} and QRR, which is an unrivaled improvement. During the development of the architecture, several slightly altered models were tested as well, and the corresponding results indicated that the fully connected layer has the largest influence on this major improvement. Neglecting computational resources and training time, there are no downsides using the new DL model instead of all previously investigated models. Therefore, for all the in this thesis investigated cases, this approach should always be preferred to maximize the prediction quality of the attacks.

\begin{center}
\begin{table}[htbp]
{\small
\begin{center}
\begin{tabular}[center]{cccccr}
\toprule
Nr. &  Layer & kernel & stride & padding & parameters \\
\midrule
1 & FCL, Reshape & --- & --- & --- & 532,480 \\
2 & T-Conv2D, BN, LeakyReLU & 3 & 1 & 1 & 46,592 \\
3 & T-Conv2D, BN, LeakyReLU & 4 & 2 & 1 & 524,544 \\
4 & T-Conv2D, BN, LeakyReLU & 3 & 1 & 1 & 147,712 \\
5 & T-Conv2D, BN, LeakyReLU & 4 & 2 & 1 & 131,200 \\
6 & T-Conv2D, BN, LeakyReLU & 3 & 1 & 1 & 576 \\
\bottomrule
& & & & & 1,383,104 \\
\end{tabular}
\end{center}
}
\caption[Structure of the advanced generator.]{Structure of the advanced generator for the DL attack. \label{tab:advancedGeneratorLayers}}
\end{table}
\end{center}

\section{Attacks on a real optical PUF}
So far, all attacks have only been carried out on the simulated data, which works under ideal conditions that are not met in the real world. This includes for example shot noise from the varying number of collected photons, dark noise from the current within the camera sensor or simple inaccuracies during the measuring process or across several measuring sessions. As previously explained in chapter \ref{sec:ml_eval}, due to these circumstances, Pappu et. al \cite{pappu} defined a threshold for the FHD of $0.41$, below which two responses are regarded as the same response. This value leads itself perfectly to be used as a reference for when a ML attack can be viewed as successful. However, this is not the case for the completely deterministic simulation, where no external influences are possible and two responses from the same challenge will always have a FHD of $0$. Since there is no trivial solution to this, to at least get a feeling on how the ML attacks would relate to real-world data, the in chapter \ref{sec:dataset_eval} mentioned optical PUF of the Italian research group was attacked with all applicable ML models that have been introduced so far. The only attack that is excluded from this is the initial generator, since it acts on images of $512 \times 512$ pixels, but the responses of this real optical PUF are only of size $200 \times 200$ pixels. For this PUF, following the principle of the cross-over point of the like and unlike distributions of Pappu et. al. \cite{pappu}, the threshold, for when two responses are regarded as identical due to the noise of the system, was found to be at around $0.31$ using the \textit{second Gabor transformation}. The available data amounts to 2000 CRPs, from which 1800 were used for training and 200 for testing. To fit the DL model architectures, only the center square of $128 \times 128$ pixels is used for both training and evaluation for all attacking approaches. Note that none of the hyperparameters or architectures of the models was altered, except for RR-based models, for which the penalty parameter $\lambda$ changes, as this parameter is directly inferred from the used training set. This means that better results could be expected when the models are fine-tuned for the new dataset, especially the DL models, which are more sensitive to the hyperparameters. For reasons of simplicity, only the FHD will be analysed in the following. The FHD here will be computed using the second Gabor transformation, since the threshold was defined using this transformation as well, and the previous evaluation of the dataset in chapter \ref{sec:dataset_eval} showed that this transformation creates a higher FHD on the dataset itself. All results can be seen in table \ref{tab:realPUFAttack}.

The evaluation shows that all of the \textbf{linear attacks} almost perform identical. The LR scores a mean FHD of $0.208$, RR of $0.207$, and $0.206$ for both models with the quadratic challenge approach. While these results follow the previously found trend with respect to their performance, the differences are so small that they can be seen as negligible. The largest FHDs that were recorded follow the same pattern, with the largest FHD across all datasets at $0.284$ for the LR, i.e. all predicted responses lie below the found threshold.

The \textbf{DL attacks} show the same pattern that has already been seen on the simulated data: The cropped generator of type 2 performs better with a mean FHD of $0.208$ than the one of type 1 with $0.231$, but the advanced generator performs significantly better than the other two with a value of $0.149$. Again, the largest recorded FHDs follow this pattern as well, such that the largest across all three attacks was found for the type 1 cropped generator with a value of $0.298$. The FHDs for all other predictions are lower, i.e. all predicted responses do not exceed the threshold.

It was found that all of the attacks can be regarded as successful on this setup, since the FHDs for all predicted responses lie below the threshold to identify identical responses. While there were close to no differences at all for all linear attacks, the differences on the DL attacks match these that were found for the simulated data, indicating that there are some differences in performance due to the architectures themselves, and not only the type of data they are presented with. While the cropped generator of type 1 on average performs worse, the one of type 2 is on par with the linear attacks. The advanced generator, however, again outperforms all the other attacks by a rather large amount.

\begin{center}
\begin{table}[htbp]
{\small
\begin{center}
\begin{tabular}[center]{l|c|c}
\toprule
Real optical PUF & Mean FHD & Max FHD \\
\midrule
Linear regression & 0.208 & 0.284 \\
Ridge regression & 0.207 & 0.281 \\
Quadratic Linear regression & 0.206 & 0.277 \\
Quadratic Ridge regression & 0.206 & 0.277\\
\midrule
Cropped generator 1  & 0.231 & 0.298 \\
Cropped generator 2  & 0.208 & 0.284 \\
Advanced generator & 0.149 & 0.204\\
\bottomrule
\end{tabular}
\end{center}
\caption[Mean FHDs of the ML attacks on a real optical PUF.]{The mean FHDs for the second Gabor transformation of all applicable introduced ML attacks on a real optical PUF. The threshold below which two responses are regarded as the same lies at $0.31$.}
\label{tab:realPUFAttack}}
\end{table}
\end{center}

\section{Attacks with a larger number of CRPs}
As a last additional investigation, it was decided to examine how much space there is left for improvements if a significantly larger number of simulated CRPs is used to train the models. Note that this experiment is supposed to find out how well the simulated PUF can be modelled, and not to investigate the different types of datasets, which has already been the subject of the majority of the previous chapters. Therefore, only challenge compositions without any restrictions, i.e. type A datasets, will be taken into account. Additionally, due to the vast number of computational resources that would be necessary to do this for all of the type A datasets, it was decided to only use the first Gabor transformation and to select only two representatives. As to not use the two extremes regarding the challenge size, but to still compare datasets with smaller and larger challenge sizes, the $7 \times 7$ and $13 \times 13$ datasets of type A were chosen for this experiment. For each of these, 10000 CRPs were generated and split into training sets of 9000 as well as test sets of 1000 CRPs. All models that have been introduced so far were trained using these two datasets, keeping all hyperparameters constant, such that only the number of CRPs is different between the attacks. The only exceptions are again the RR-based models, for which the penalty parameter $\lambda$ was changed. In the following, all analyses will only focus on the mean FHDs of the results. The percentages for the improvements are calculated using the relative difference, i.e. an improvement of 100\% would be equivalent to a new FHD of 0. All results can be found in table \ref{tab:smallLargeAttackComparison}.

For the \textbf{linear attacks}, both the standard LR and RR improved only very slightly on the $7 \times 7$ dataset by $2\%$. On the other hand, both variants that use the quadratic challenge transformation of chapter \ref{sec:opr_attack} significantly improved by $59\%$ and $55\%$, such that both now score a mean FHD of $0.075$ on the test data. However, for the $13 \times 13$ dataset, both LR and RR improved noticeably more, showing rates of $13\%$ and $12\%$, while QLR and QRR here only improved by $20\%$ and $29\%$.

Especially noticeable here is the very low improvement of only $2\%$ for the standard LR and RR on the $7 \times 7$ dataset, which may even be deemed negligible, especially given that the amount of data was increased by a factor of 10. It appears that these models are already close to a threshold at which they will not improve any more, even if more data is provided. Interestingly, this does not seem to be the case for the $13 \times 13$ dataset, where both models improved a lot more and even scored better final results than for the $7 \times 7$ dataset. For the training with only 1000 CRPs, this was actually the other way around. It would be possible that the LR and RR have a threshold for improvements on both datasets, which just happens to be lower for the $13 \times 13$ variant. In this case, they would have already been closer to this value for the $7 \times 7$ dataset with only 1000 CRPs than for the corresponding set of size $13 \times 13$, which is why the models seem to improve differently when more training data is provided. However, with the current data at hand, it is not possible to give a clear answer to this assumption. To do so, new experiments with more variations in the number of training samples would be necessary.

The data from the QLR and QRR do not give much space for an assumption of such a threshold. However, it is quite noticeable that the gap between the mean FHDs for both datasets greatly changed when increasing the number of CRPs, such that the predictions for the $7 \times 7$ dataset are significantly better than for the $13 \times 13$ dataset, while the previous difference was rather low. Even though there are large differences in the total values on both datasets, for both attacks on both datasets, the amount of improvement is quite significant.

The results for the \textbf{DL attacks} are rather straightforward: each model has significantly improved in performance. For all models, the amount of improvement is higher in both absolute and relative values for the $13 \times 13$ dataset, where the initial results were worse for 1000 CRPs training sets than for the $7 \times 7$ dataset. However, the relationship between both datasets did not change, i.e. the models still perform better on the $7 \times 7$ variant. The only exception here is the advanced generator, but this exception is likely only due to the already immensely low FHD and some differences in the CRPs due to the random sampling process, and therefore does not show any fundamental differences between the behaviors on these two datasets. The largest improvements can be found for the second model of the cropped generator of chapter \ref{sec:croppedAttack} and the advanced generator of chapter \ref{sec:advancedGenerator}, such that the former showed the highest decrease in the mean FHD on the $7 \times 7$ and the latter on the $13 \times 13$ dataset. However, since the advanced generator already scored a mean FHD as low as $0.026$ on the $7 \times 7$ dataset, there was not too much space left for improvements in the first place, which largely influences why the cropped generator improved more in this case.

Essentially, the results for the DL attacks did not provide any surprises and are in fact exactly what could be expected. All models have significantly improved with more training data and those models that have previously already performed better, perform even better now. The relative order of prediction performance did not change either, i.e. the initial generator still produces the results with the highest, while the advanced generator produces those with the lowest mean FHD.

Altogether, there are rather large differences between the linear and DL attacks with respect to an increase in the number of training CRPs. The standard LR and RR showed close to no improvement at all and therefore produce the worst results of all attacks for the new training sets. However, the QLR and QRR did show some improvements, but there are large differences in the amounts between both datasets. On the other hand, the DL attacks all significantly improved and previously better models even increased the gap in performance to the other models.

\begin{center}
\begin{table}[htbp]
{\small
\begin{center}
\begin{tabular}[center]{l|cc|c|c}
\toprule
$7 \times 7$ dataset (type A) & FHD (1k CRPs) & FHD (10k CRPs) & Diff. & Improvement  \\
\midrule
Linear regression & 0.184 & 0.181 & 0.003 & 2\% \\
Ridge regression & 0.184 & 0.181 & 0.003 & 2\% \\
Quadratic Linear regression & 0.181 & 0.075 & 0.106 & 59\% \\
Quadratic Ridge regression & 0.168 & 0.075 & 0.093 & 55\% \\
\midrule
Generator & 0.171 & 0.127 & 0.044 & 26\% \\
Cropped generator 1 & 0.157 & 0.119 & 0.038 & 24\% \\
Cropped generator 2 & 0.140 & 0.062 & 0.078 & 56\% \\
Advanced generator & 0.026 & 0.017 & 0.009 & 35\% \\
\toprule
$13 \times 13$ dataset (type A) & \multicolumn{4}{l}{} \\
\midrule
Linear regression & 0.196 & 0.171 & 0.025 & 13\% \\
Ridge regression & 0.194 & 0.171 & 0.023 & 12\% \\
Quadratic Linear regression & 0.191 & 0.153 & 0.038 & 20\% \\
Quadratic Ridge regression & 0.191 & 0.136 & 0.055 & 29\% \\
\midrule
Generator & 0.246 & 0.177 & 0.069 & 28\% \\
Cropped generator 1 & 0.233 & 0.150 & 0.083 & 36\% \\
Cropped generator 2 & 0.214 & 0.090 & 0.124 & 58\% \\
Advanced generator & 0.046 & 0.015 & 0.031 & 67\% \\
\bottomrule
\end{tabular}
\end{center}
\caption[Comparison of mean FHDs of attacks on simulated data with more CRPs.]{A comparison of the mean FHDs of all attacks on the $7 \times 7$ and $13 \times 13$ datasets of type A with 1000 and 10000 CRPs. An improvement of 100\% is equivalent to decreasing the FHD to 0.}
\label{tab:smallLargeAttackComparison}}
\end{table}
\end{center}

    \chapter{Conclusion}
In this thesis, several linear and non-linear modeling attacks on integrated optical PUFs were introduced and investigated. On this account, a simulated version of an integrated optical PUF based on one of the models defined by Rührmair et. al. \cite{ruhrmair2013optical} was implemented and used for the generation of multiple datasets that vary in their challenge sizes $n$. During the setup of the simulation, it was found that for certain challenges, the corresponding responses of the simulation showed quite large correlations. This was especially noticeable for challenges with a large number of activated blocks, since the emerging speckle pattern turned out to be almost identical with the responses for closely related challenge, i.e. these, which had most of the blocks activated as well. This led to the introduction of certain restrictions on the composition of the challenges, such that for each $n$, multiple datasets with varying restrictions were generated, also referred to as the \textit{type} of the dataset. These restrictions include both the deactivation of fixed bits within the challenge as well as upper boundaries for the allowed number of simultaneously activated bits.

First, the datasets were investigated using the fractional Hamming distance (FHD) and Shannon entropy. To get a feeling for how they would compare to real-world setups, these values were also compared to those of real integrated optical PUFs. Here, it was found that the datasets show slightly worse values for the FHD than what would be ideal, but are still on par with one of the investigated real optical PUFs. Next, a linear regression model (LR) and a convolution-based deep learning model (generator) were used to attack the simulated PUF using all available datasets, to see whether, and if so how they behave differently on varying $n$ and types. It was found that the LR performs in a quite stable manner across all datasets, whereas the generator performs significantly better for the datasets with lower $n$, but at the same time significantly worse for those with larger $n$. The types of the datasets did not seem to play a major role in the performance, except for unexpected side effects which even facilitated the attacking process of the generator on some datasets.

In the following, an additional simulation setup that increases the size and the number of scatterers of the PUF was introduced and used to generate the same kinds of datasets as before. An analysis on these datasets showed that the FHD falls for smaller $n$, but slightly increases for larger $n$ compared to the initial setup. The setup can therefore not be seen as a simple improvement to the initial one, but as a completely separate PUF with its own advantages and downsides. This new simulation was attacked as well, presenting only minor differences on the LR, but significantly worse values on the generator. However, since the generator was not specifically optimized on the new data, this was to some degree expected, and further adjustments with respect to the new setup would likely lead to better performances. Additionally, some further minor differences which correlate to the results that have been found during the dataset analysis could be observed as well for both models.

The investigations on the effects of different input data led to the idea to examine how the ML attacks behave on the so-called \textit{cropped responses}. This term simply refers to a cut version of the image of the speckle pattern, where the response was cut in a way that the whole image is filled with the pattern. Since the initial response only contains a circular speckle pattern in the center of the image, only this cropped version was used for the FHD calculation in the first place and thus could also be used for the training of the models. This experiment was only carried out for the generator, as LR acts on every pixel distinctly, so removing some of the pixels will not change the results of the others. Since the generator only works on the initial image sizes, two alterations to the initial model were suggested, which originated from the removal of different layers of the initial generator. However, they fundamentally only differed in the number of filters that are used in each layer of the network. The model with fewer filters was referred to as \textit{cropped generator type 1} and the one with more as \textit{cropped generator type 2}. It was found that while both models performed better than the initial generator, the latter showed a much more significant improvement, even outperforming the LR on some datasets, on which the initial generator performed worse.

Subsequently, the focus was shifted towards completely new ML attacks. In the course of this, the \textit{ridge regression} (RR) was introduced, which adds a penalty term to the LR to force the coefficients to stay close to a value of $0$. However, the RR attacks only showed some very minor improvements compared to the LR, which may even be deemed negligible. As a further attempt to enhance the linear models, an algorithm that transforms the challenges into a quadratic representation, such that they more closely resemble the actual physical circumstances, was presented. This method has already been explained in detail by Rührmair et. al. \cite{ruhrmair2013optical}. An additional model for both the LR and RR was trained using this new approach, which showed quite noticeable improvements compared to their standard counterparts. On the non-linear attacking side, a new DL model was implemented, for which parts of the convolutional layers were changed and a fully connected layer was added. This new generator presented values quite close to the optimal value and significantly outperformed all of the previous models on all of the datasets.

After presenting several linear and non-linear ML attacks on the \textit{simulated} data, it was investigated how these ML attacks will perform if they are instead used on data from a real PUF. On this account, the data from the real integrated optical PUF, which has already been investigated before, was used to train all of the presented models. Due to several factors that influence the measurement accuracy of this PUF, a threshold for the FHD was defined until which two responses are regarded as identical. This threshold was used as the criterion to classify the ML attacks as either successful or unsuccessful. It could be observed that all of the investigated models did not exceed this value and can therefore be regarded as successful with respect to this criterion. Furthermore, almost no differences on the linear attacks were found, while there were rather large differences on the DL attack side. Here, the cropped generator of type 1 performed worse than the linear approaches, whereas the one of type 2 scored on a par with these. The newest DL model, which already outperformed all the others on the simulated data, did so on the real data as well.

Lastly, the subject switched one more time back to the simulated data, to see how the model's performances change if a significantly larger number of CRPs is used during the training process. Due to the large number of computational resources required, only two datasets with 10 times the previous number of CRPs were generated. Here, both standard linear approaches showed only minor improvements, whereas the models that use the quadratic challenge transformation improved by a fair amount more. On the other hand, quite significant increases in performance could be seen for all DL models, especially for the type 2 cropped generator. However, the newest generator still performs the best by far, such that the responses could be predicted up to an average FHD of $0.015$.

Overall, the investigations of this thesis presented numerous quite interesting differences between linear and non-linear attacks on integrated optical PUFs. However, at the same time, many questions are still left unanswered and even further questions have arisen. One aspect that requires further analyses is the simulation of the optical PUF. The setup that was presented in this thesis only investigated a very small subset of the possible output and thus does not give any answers regarding the quality of the \textit{whole} challenge space. Furthermore, more thorough research should be done on certain edge cases of the simulation, in particular the cases where almost all blocks of the challenge mask are either turned on or off. This has already been identified to strongly influence the complexity of the outputs, but did not make many differences due to the low number of CRPs used per simulated dataset. Also, the location of the activated blocks or the influence of single blocks under varying circumstances have not been investigated. Lastly, the simulation used a rather straightforward setup and there is still a lot of potential being left for increasing the complexity by adding further components or by even replacing it with a completely different implementation.

Not only the simulation, but also the ML attacks left some questions unanswered. To the best of our current knowledge, no similar comparisons between both linear and non-linear attacks on optical PUFs have been published so far, so the results in this thesis may serve as the cornerstone for further investigations on this topic. While it was possible to explain some of the differences under specific circumstances, for other cases, it is still unclear why exactly these differences could be found and why one of the DL approaches could consistently perform better than the linear ones on \textit{all} investigated datasets of both simulated and real optical PUFs. Furthermore, it was seen that there are quite large differences between some of the models when the number of available CRPs is increased, which raises the questions how many improvements can be expected for which model, and when a point of saturation is reached. Also, the influence of noise on the performance of the ML model has only briefly been seen on the attacks on the real optical PUF, but no further investigations have been conducted. Since real optical PUFs are usually subject to some kind of noise, the results of the attacks on the simulated data cannot easily be transferred to a real-world scenario, which also applies the other way around. There are still multiple questions regarding the impact of noise and especially whether, and if so how the linear and non-linear approaches handle this influence differently. Last but not least, from a physical point of view, it may also be interesting to investigate how the DL models correspond to given physical circumstances. In this thesis, only convolutional networks were investigated, which handled the PUF itself as a complete black box and only took the inputs and the outputs into consideration. However, there is still a whole bandwidth of possible other architectures that could be used, some of which may even try to actually model the physical processes that take place within the PUF. Broadly speaking, the structure of the internal processes of an optical PUF leads itself quite well to be reconstructed using a feedforward neural network. Thus, it might be worth to take the internal structure of the PUF into consideration as well and not to only take the view of an outside adversary, i.e. to see if not only the input-output behavior can be modelled, but also the PUF itself.

Altogether, even though the number of publications in the recent years has been rather low, there is still a lot of space for further research on the topic of optical PUFs. This includes both the design of the PUFs themselves as well as adversary attacks on the system. With the quickly advancing field of ML, new techniques are developed at an incredibly fast rate, which allows for further investigations on already established attacks as well as on completely new approaches. We think that the interplay between designing secure optical PUFs and breaking them with new attacks can lead to further fascinating findings, which may be used to develop further security systems and further attacking processes, and, on this account hope, that this thesis will be able to contribute to the progress in these fields.

%
%
    \appendix
%
%
\chapter{Appendix}

\section{Plots of the analysis of the generated datasets}\label{sec:appendix_datasets}
\begin{figure}
\centering
  \includegraphics[width=1\linewidth]{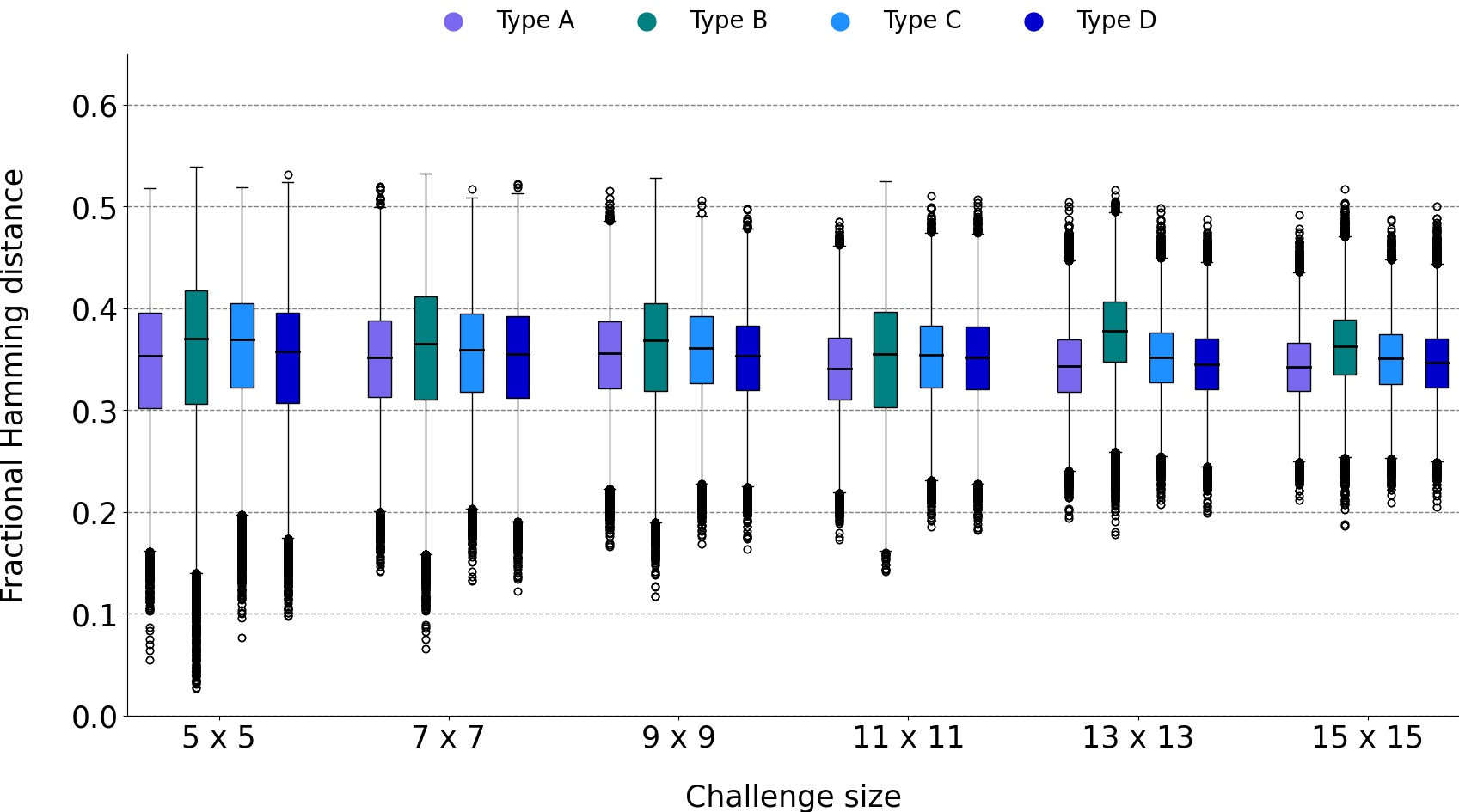}
\caption[FHDs of the initial generated datasets using the second Gabor transformation.]{The FHDs of all the initially generated datasets using the second Gabor transformation.}
\label{fig:x1_fhd_g2}
\end{figure}

\begin{figure}
\centering
  \includegraphics[width=1\linewidth]{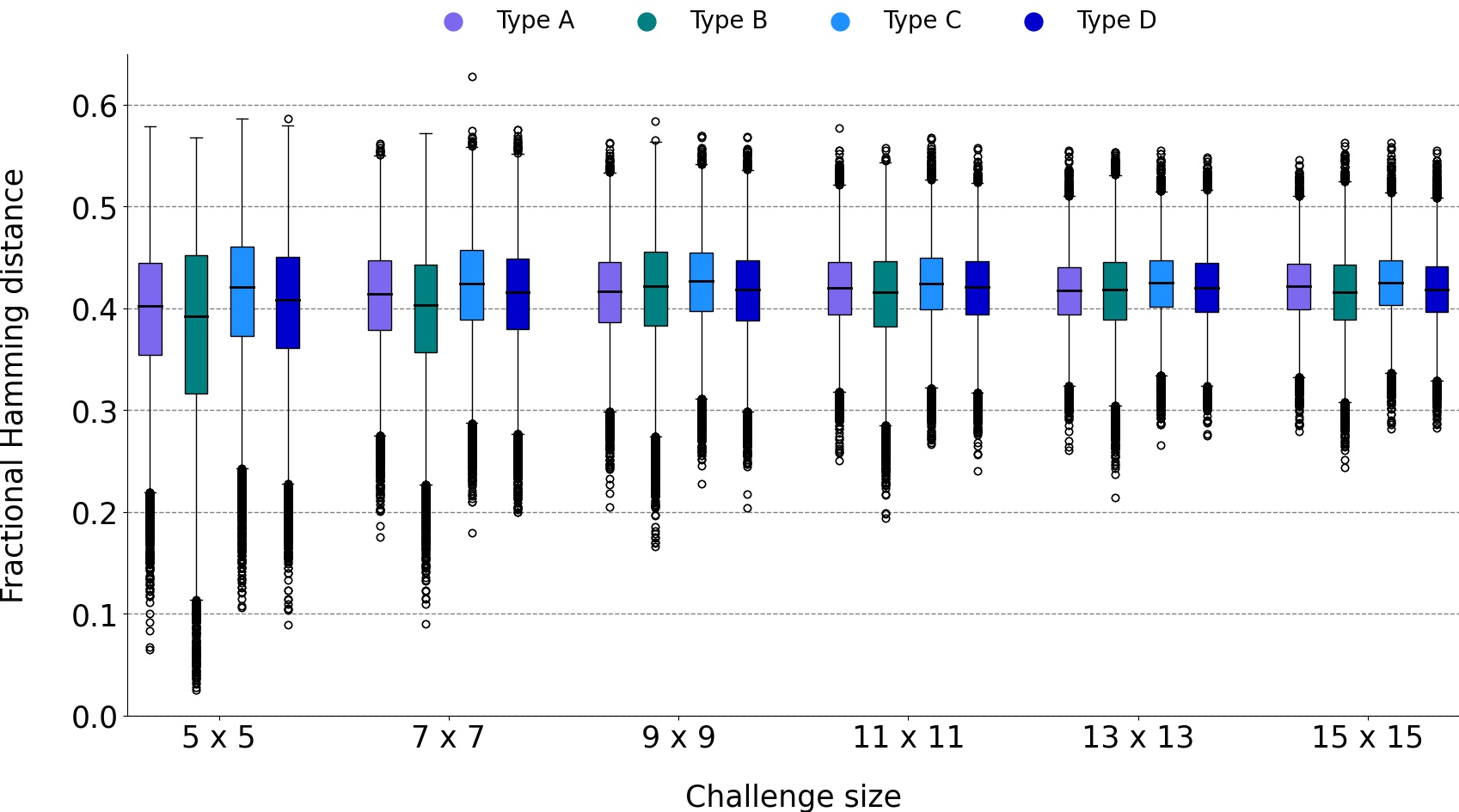}
  \caption[FHDs of the generated datasets for the larger simulation using the first Gabor transformation.]{The FHDs of all the generated datasets for the larger simulation using the first Gabor transformation.}
\label{fig:x2_fhd_g1}
\end{figure}

\begin{figure}
\centering
  \includegraphics[width=1\linewidth]{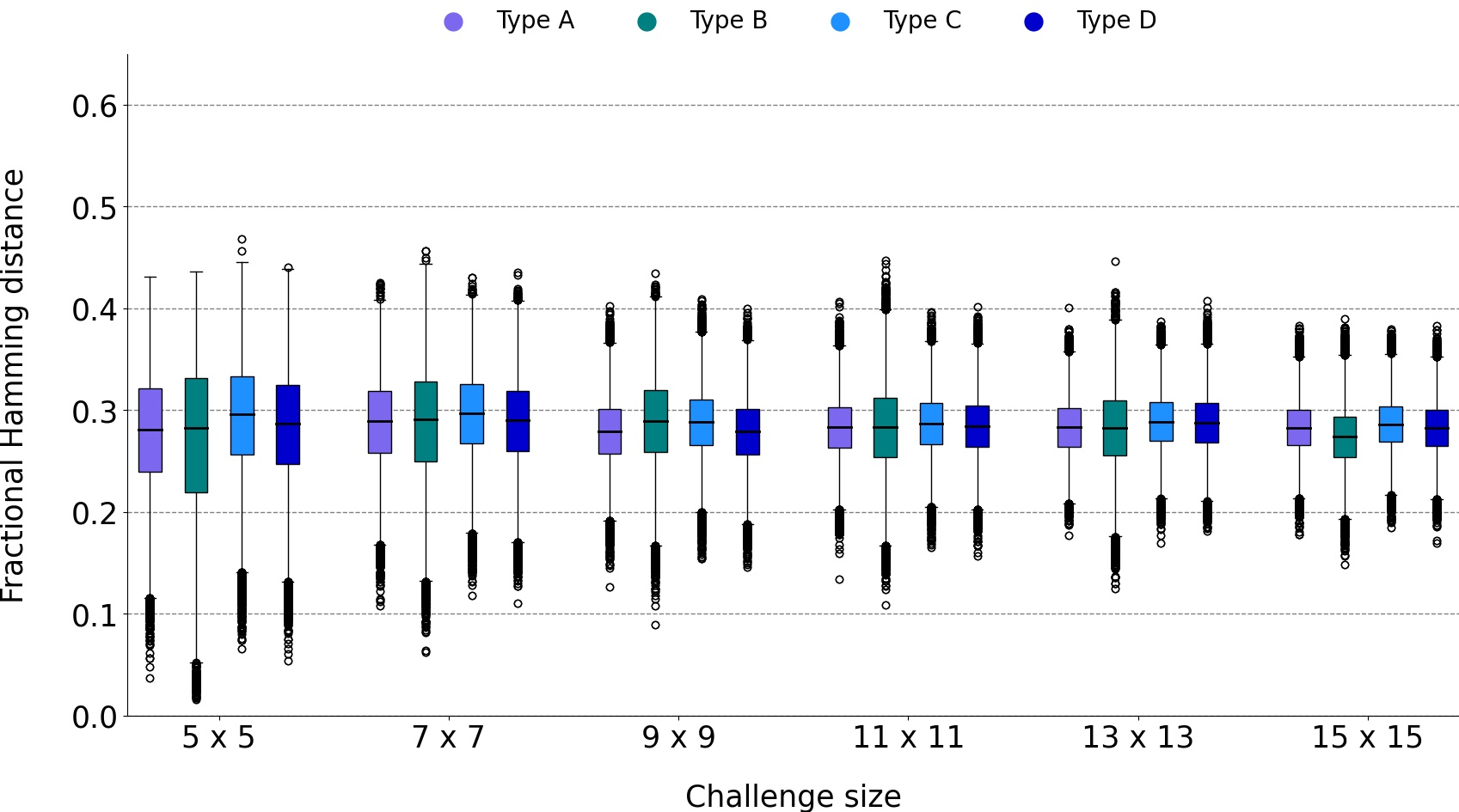}
  \caption[FHDs of the generated datasets for the larger simulation using the second Gabor transformation.]{The FHDs of all the generated datasets for the larger simulation using the second Gabor transformation.}
\label{fig:x2_fhd_g2}
\end{figure}

\begin{figure}
\centering
  \includegraphics[width=1\linewidth]{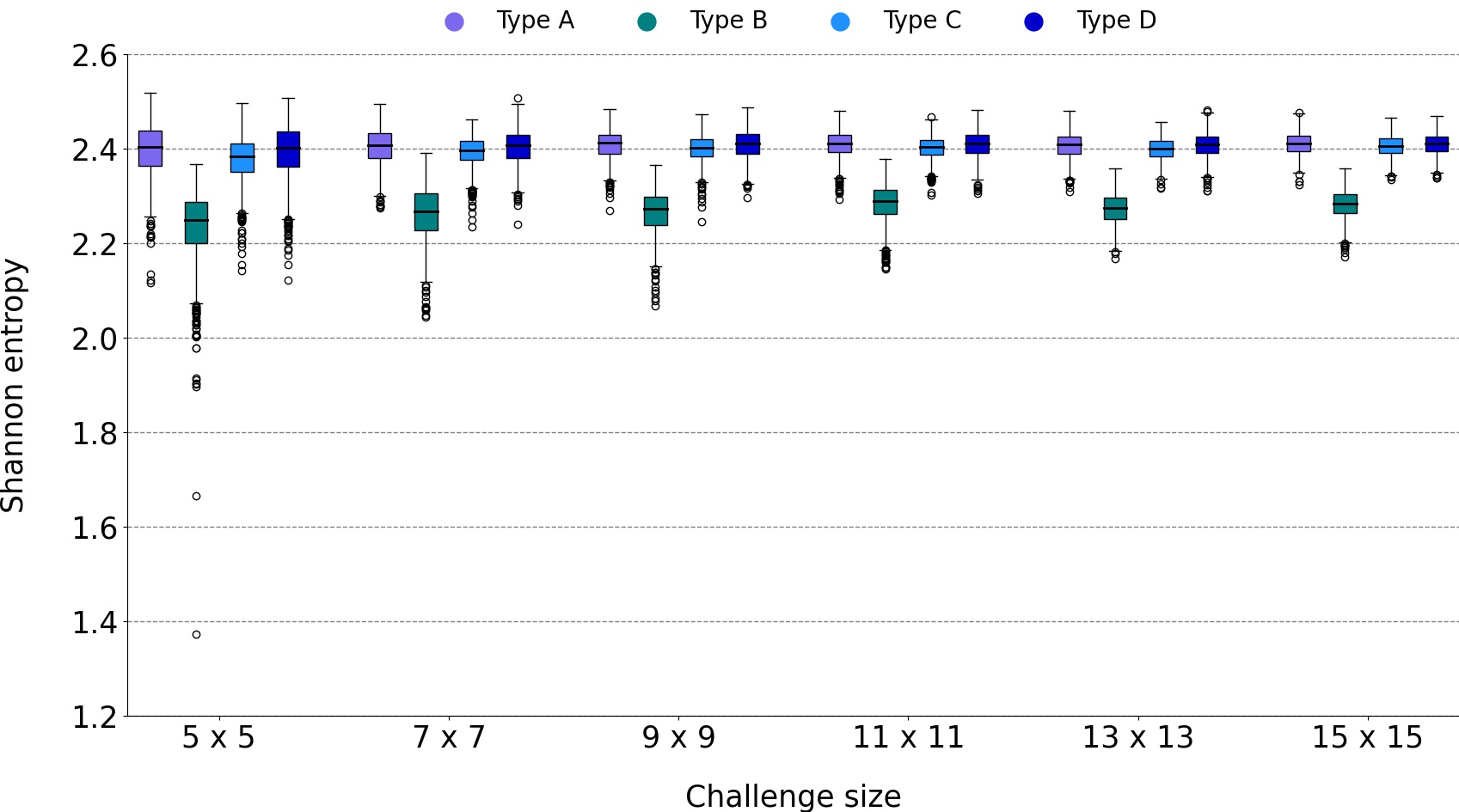}
  \caption[Shannon entropies of the generated datasets for the larger simulation.]{The Shannon entropies of all the generated datasets for the larger simulation.}
\label{fig:x2_entropy}
\end{figure}

Here, the plots of the metrics for the dataset evaluation that have not been previously shown are displayed. The FHDs using the second Gabor transformation for the in chapter \ref{sec:x1_results} generated datasets can be found in figure \ref{fig:x1_fhd_g2}. Furthermore, for the in chaper \ref{sec:datasetx2} introduced larger simulation, the FHDs using the first Gabor transformation can be found in figure \ref{fig:x2_fhd_g1} and these using the second Gabor transformation in figure \ref{fig:x2_fhd_g2}. Lastly, the Shannon entropies for this simulation are portrayed in figure \ref{fig:x2_entropy}.

\section{Plots of the performance of the machine learning attacks}\label{sec:appendix_attacks}
\begin{figure}
\centering
  \includegraphics[width=1\linewidth]{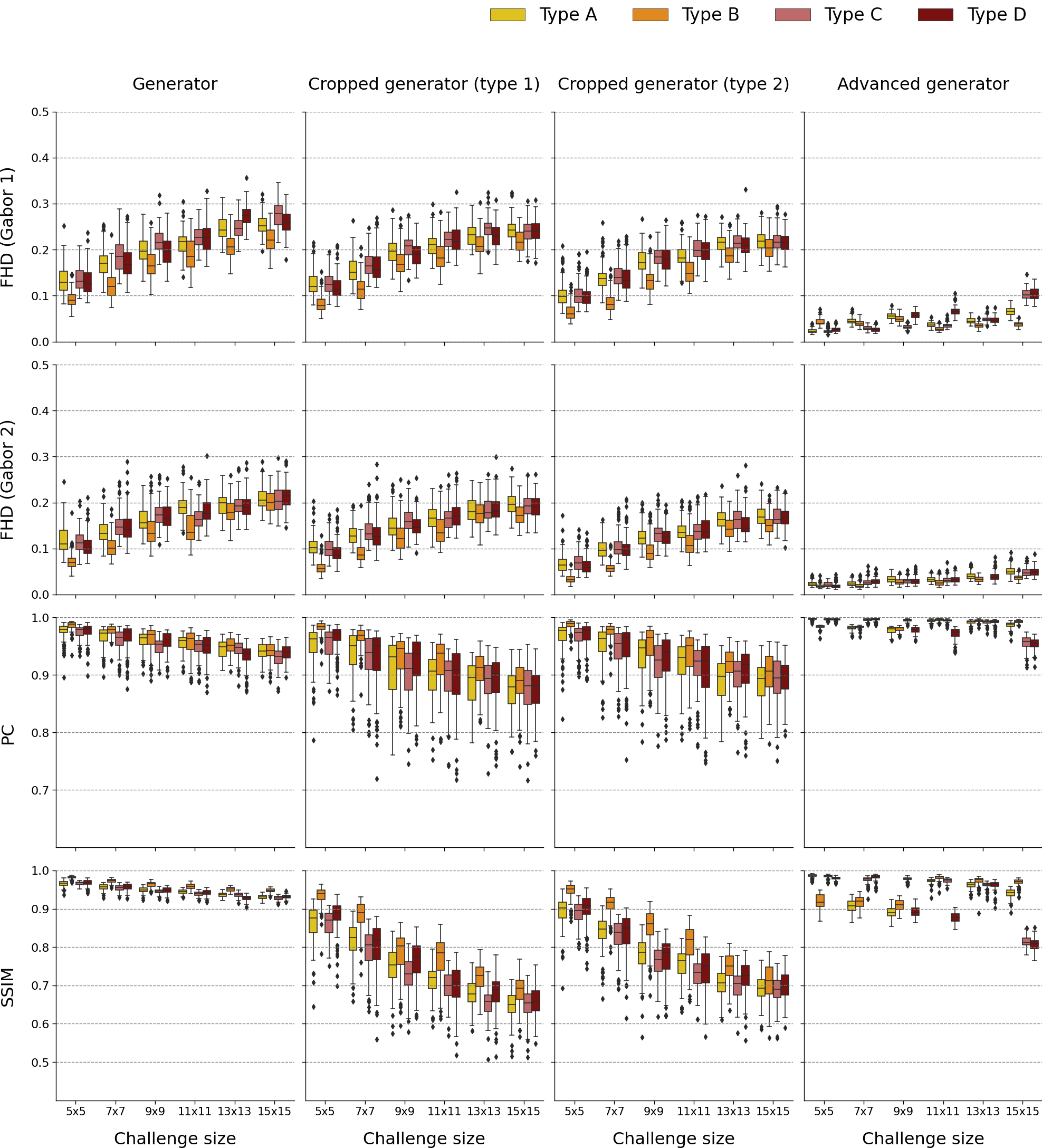}
\caption[Metrics of the DL attacks on the standard simulation.]{The data for all metric for all DL attacks on the standard simulation of the PUF.}
\label{fig:dl_x1}
\end{figure}

\begin{figure}
\centering
  \includegraphics[width=1\linewidth]{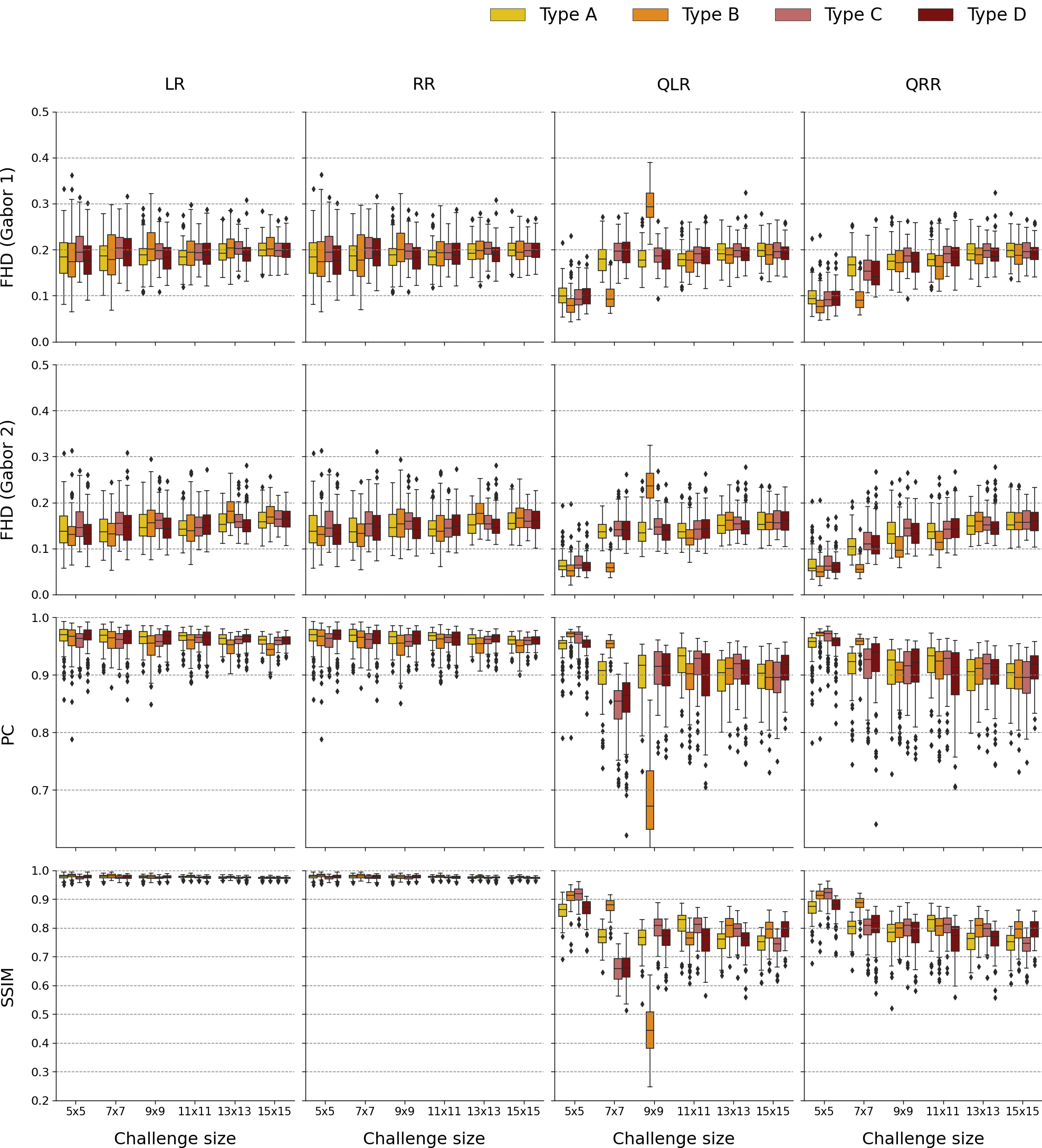}
\caption[Metrics of the linear attacks on the standard simulation.]{The data for all metric for all linear attacks on the standard simulation of the PUF.}
\label{fig:linear_x1}
\end{figure}

\begin{figure}
\centering
  \includegraphics[width=1\linewidth]{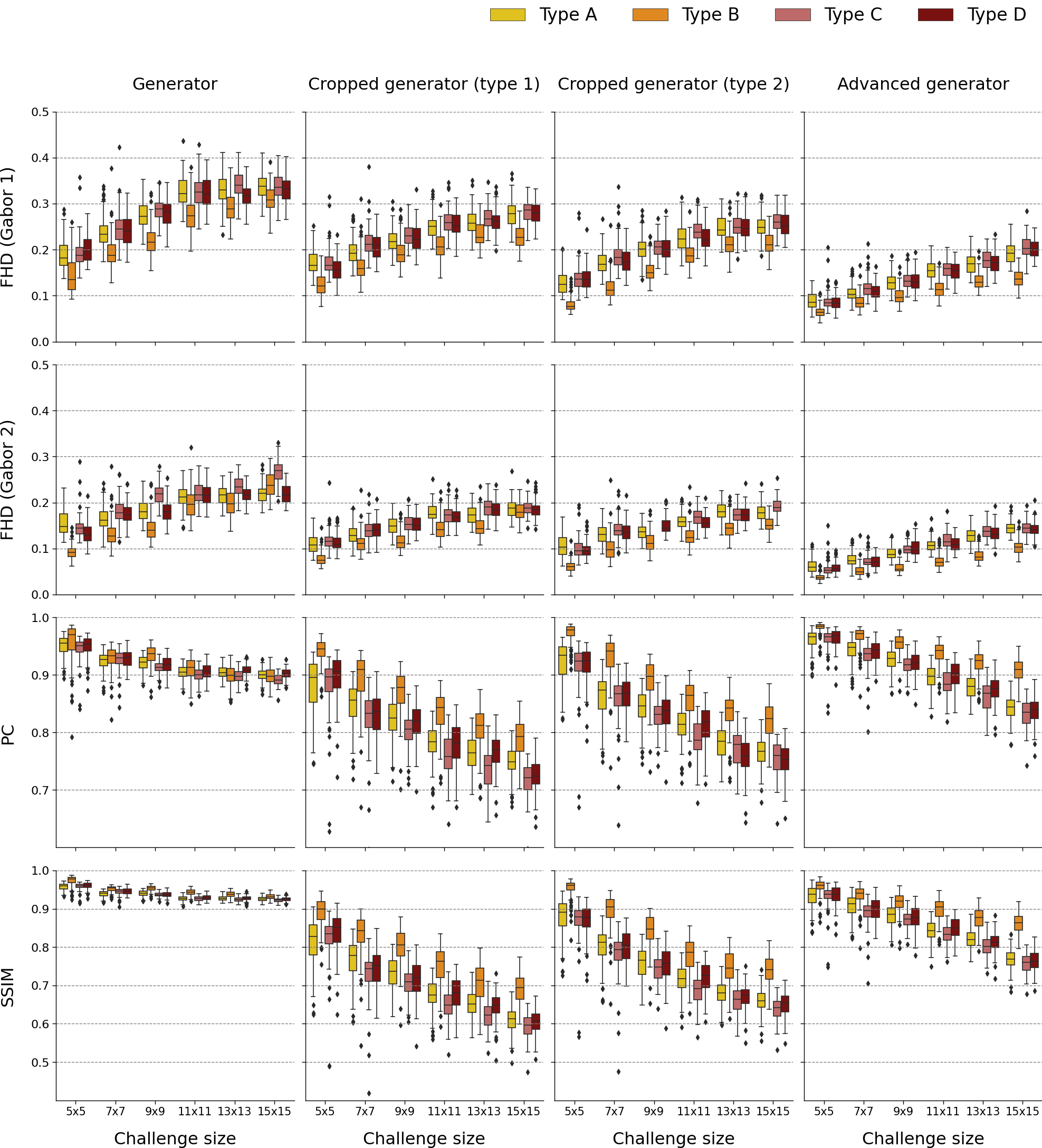}
\caption[Metrics of the DL attacks on the larger simulation.]{The data for all metric for all DL attacks on the larger simulation of the PUF.}
\label{fig:dl_x2}
\end{figure}

\begin{figure}
\centering
  \includegraphics[width=1\linewidth]{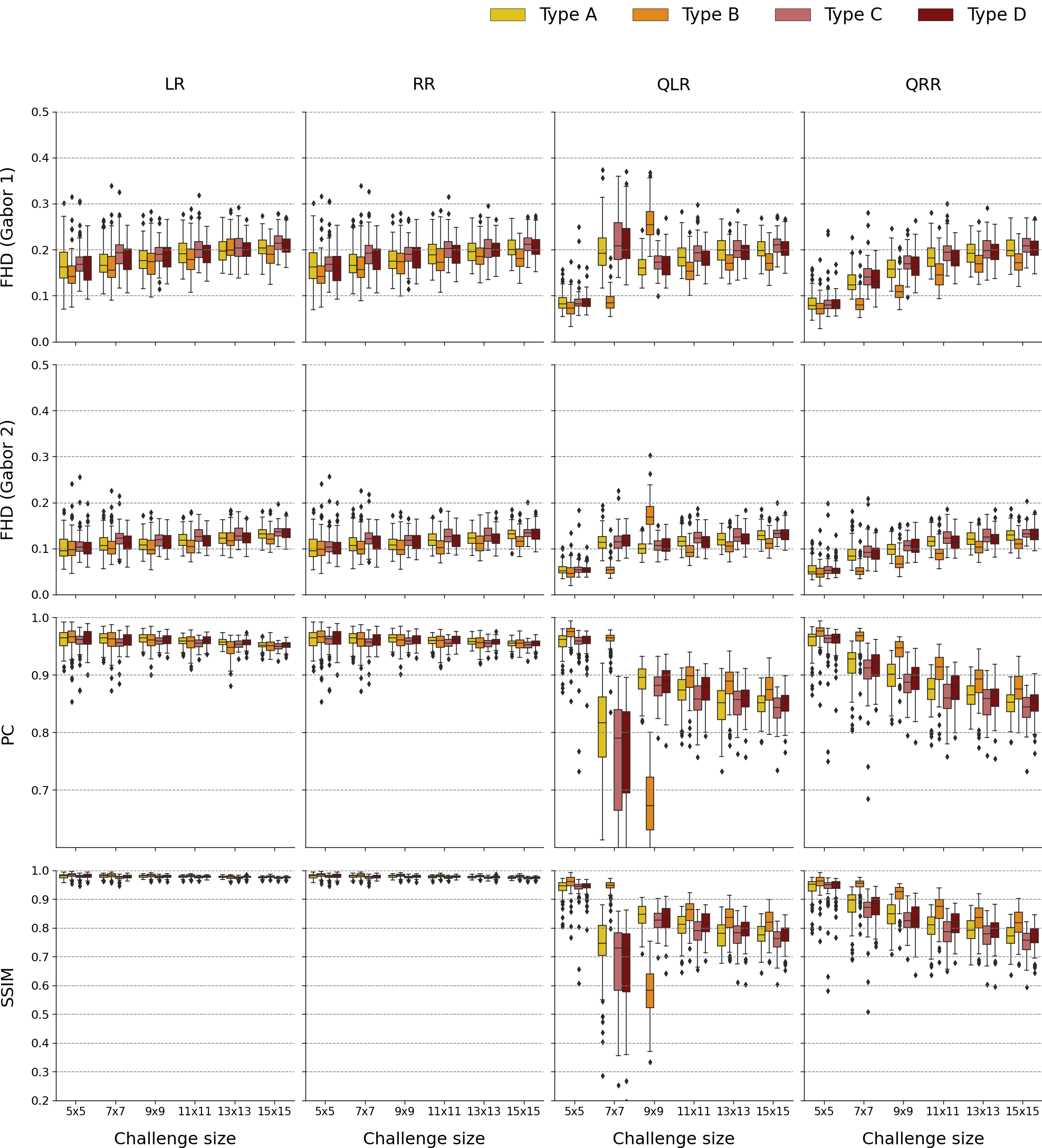}
\caption[Metrics of the linear attacks on the larger simulation.]{The data for all metric for all linear attacks on the larger simulation of the PUF.}
\label{fig:linear_x2}
\end{figure}

This part contains several plots which enable the comparison of the machine learning attacks that have been introduced. For the initial simulation, the metrics for the DL attacks can be found in figure \ref{fig:dl_x1} and these for the linear attacks in figure \ref{fig:linear_x1}. Respectively, the same accounts for the larger simulation in the figures \ref{fig:dl_x2} and \ref{fig:linear_x2}.

%
%
    \backmatter
    \listoffigures                                
    \listoftables                                 
%
%
    \begin{spacing}{0.9}                          
       \bibliographystyle{ieeetr}               
       \bibliography{bibliography}                

\begin{thebibliography}{100}

\bibitem{eisenbarth2008power}
T.~Eisenbarth, T.~Kasper, A.~Moradi, C.~Paar, M.~Salmasizadeh, and M.~T.~M.
  Shalmani, ``On the power of power analysis in the real world: A complete
  break of the keeloq code hopping scheme,'' in {\em Annual International
  Cryptology Conference}, pp.~203--220, Springer, 2008.

\bibitem{kasper2010all}
T.~Kasper, M.~Silbermann, and C.~Paar, ``All you can eat or breaking a
  real-world contactless payment system,'' in {\em International Conference on
  Financial Cryptography and Data Security}, pp.~343--350, Springer, 2010.

\bibitem{boneh1998attack}
D.~Boneh, G.~Durfee, and Y.~Frankel, ``An attack on rsa given a small fraction
  of the private key bits,'' in {\em International Conference on the Theory and
  Application of Cryptology and Information Security}, pp.~25--34, Springer,
  1998.

\bibitem{dainty2013laser}
J.~C. Dainty, {\em Laser speckle and related phenomena}, vol.~9.
\newblock Springer science \& business Media, 2013.

\bibitem{pappu}
R.~Pappu, B.~Recht, J.~Taylor, and N.~Gershenfeld, ``Physical one-way
  functions,'' {\em Science}, vol.~297, no.~5589, pp.~2026--2030, 2002.

\bibitem{ruhrmair2013optical}
U.~R{\"u}hrmair, C.~Hilgers, S.~Urban, A.~Weiersh{\"a}user, E.~Dinter,
  B.~Forster, and C.~Jirauschek, ``Optical pufs reloaded,'' {\em Eprint. Iacr.
  Org}, 2013.

\bibitem{atakhodjaev2018machine}
I.~Atakhodjaev, {\em Machine Learning Attacks on Optical Physical Unclonable
  Functions}.
\newblock PhD thesis, Johns Hopkins University, 2018.

\bibitem{petkovic2007security}
P.~Tuyls and B.~Skoric, ``Strong authentication with physical unclonable
  functions,'' in {\em Security, Privacy and Trust in Modern Data Management}
  (M.~Petkovic and W.~Jonker, eds.), ch.~10, pp.~132--148, Springer, 2007.

\bibitem{tuyls2007secure}
P.~Tuyls, G.-J. Schrijen, F.~Willems, T.~Ignatenko, and B.~Skoric, ``Secure key
  storage with pufs,'' {\em Security with Noisy Data-On Private Biometrics,
  Secure Key Storage and Anti-Counterfeiting}, pp.~269--292, 2007.

\bibitem{tuyls2007security}
P.~Tuyls, B.~{\v{S}}koric, and T.~Kevenaar, {\em Security with noisy data: on
  private biometrics, secure key storage and anti-counterfeiting}.
\newblock Springer Science \& Business Media, 2007.

\bibitem{skoric2007experimental}
B.~Skoric, G.-J. Schrijen, W.~Ophey, R.~Wolters, N.~Verhaegh, and J.~van
  Geloven, ``Experimental hardware for coating pufs and optical pufs,'' in {\em
  Security with Noisy Data}, pp.~255--268, Springer, 2007.

\bibitem{kursawe2009reconfigurable}
K.~Kursawe, A.-R. Sadeghi, D.~Schellekens, B.~Skoric, and P.~Tuyls,
  ``Reconfigurable physical unclonable functions-enabling technology for
  tamper-resistant storage,'' in {\em 2009 IEEE International Workshop on
  Hardware-Oriented Security and Trust}, pp.~22--29, IEEE, 2009.

\bibitem{horstmeyer2015physically}
R.~Horstmeyer, S.~Assawaworrarit, U.~Ruhrmair, and C.~Yang, ``Physically secure
  and fully reconfigurable data storage using optical scattering,'' in {\em
  2015 IEEE International Symposium on Hardware Oriented Security and Trust
  (HOST)}, pp.~157--162, IEEE, 2015.

\bibitem{jacinto2021utilizing}
H.~S. Jacinto, A.~M. Smith, and N.~I. Rafla, ``Utilizing a fully optical and
  reconfigurable puf as a quantum authentication mechanism,'' {\em OSA
  Continuum}, vol.~4, no.~2, pp.~739--747, 2021.

\bibitem{lu2018cmos}
X.~Lu, L.~Hong, and K.~Sengupta, ``Cmos optical pufs using noise-immune
  process-sensitive photonic crystals incorporating passive variations for
  robustness,'' {\em IEEE Journal of Solid-State Circuits}, vol.~53, no.~9,
  pp.~2709--2721, 2018.

\bibitem{mesaritakis2018physical}
C.~Mesaritakis, M.~Akriotou, A.~Kapsalis, E.~Grivas, C.~Chaintoutis, T.~Nikas,
  and D.~Syvridis, ``Physical unclonable function based on a multi-mode optical
  waveguide,'' {\em Scientific reports}, vol.~8, no.~1, pp.~1--12, 2018.

\bibitem{dolev2016optical}
S.~Dolev, {\L}.~Krzywiecki, N.~Panwar, and M.~Segal, ``Optical puf for
  non-forwardable vehicle authentication,'' {\em Computer Communications},
  vol.~93, pp.~52--67, 2016.

\bibitem{dachowicz2018optical}
A.~Dachowicz, M.~Atallah, and J.~H. Panchal, ``Optical puf design for
  anti-counterfeiting in manufacturing of metallic goods,'' in {\em ASME 2018
  International Design Engineering Technical Conferences and Computers and
  Information in Engineering Conference}, American Society of Mechanical
  Engineers Digital Collection, 2018.

\bibitem{maiti2009improving}
A.~Maiti and P.~Schaumont, ``Improving the quality of a physical unclonable
  function using configurable ring oscillators,'' in {\em 2009 International
  Conference on Field Programmable Logic and Applications}, pp.~703--707, IEEE,
  2009.

\bibitem{yin2009temperature}
C.-E. Yin and G.~Qu, ``Temperature-aware cooperative ring oscillator puf,'' in
  {\em 2009 IEEE International Workshop on Hardware-Oriented Security and
  Trust}, pp.~36--42, IEEE, 2009.

\bibitem{arbiterPUF}
J.~W. {Lee}, {Daihyun Lim}, B.~{Gassend}, G.~E. {Suh}, M.~{van Dijk}, and
  S.~{Devadas}, ``A technique to build a secret key in integrated circuits for
  identification and authentication applications,'' in {\em 2004 Symposium on
  VLSI Circuits. Digest of Technical Papers (IEEE Cat. No.04CH37525)},
  pp.~176--179, 2004.

\bibitem{gassend2002silicon}
B.~Gassend, D.~Clarke, M.~Van~Dijk, and S.~Devadas, ``Silicon physical random
  functions,'' in {\em Proceedings of the 9th ACM Conference on Computer and
  Communications Security}, pp.~148--160, 2002.

\bibitem{xu2015reliable}
X.~Xu, A.~Rahmati, D.~E. Holcomb, K.~Fu, and W.~Burleson, ``Reliable physical
  unclonable functions using data retention voltage of sram cells,'' {\em IEEE
  Transactions on Computer-Aided Design of Integrated Circuits and Systems},
  vol.~34, no.~6, pp.~903--914, 2015.

\bibitem{holcomb2008power}
D.~E. Holcomb, W.~P. Burleson, and K.~Fu, ``Power-up sram state as an
  identifying fingerprint and source of true random numbers,'' {\em IEEE
  Transactions on Computers}, vol.~58, no.~9, pp.~1198--1210, 2008.

\bibitem{chen2011bistable}
Q.~Chen, G.~Csaba, P.~Lugli, U.~Schlichtmann, and U.~R{\"u}hrmair, ``The
  bistable ring puf: A new architecture for strong physical unclonable
  functions,'' in {\em 2011 IEEE International Symposium on Hardware-Oriented
  Security and Trust}, pp.~134--141, IEEE, 2011.

\bibitem{chen2009analog}
Q.~Chen, G.~Csaba, X.~Ju, S.~B. Natarajan, P.~Lugli, M.~Stutzmann,
  U.~Schlichtmann, and U.~R{\"u}hrmair, ``Analog circuits for physical
  cryptography,'' in {\em Proceedings of the 2009 12th International symposium
  on integrated circuits}, pp.~121--124, IEEE, 2009.

\bibitem{ruehrmair2012method}
U.~Ruehrmair, M.~Stutzmann, J.~Finley, C.~Jirauschek, G.~Csaba, P.~Lugli,
  E.~Biebl, R.~Dietmueller, K.~Mueller, and H.~Langhuth, ``Method for security
  purposes,'' July~5 2012.
\newblock US Patent App. 13/250,534.

\bibitem{ruhrmair2010towards}
U.~R{\"u}hrmair, Q.~Chen, M.~Stutzmann, P.~Lugli, U.~Schlichtmann, and
  G.~Csaba, ``Towards electrical, integrated implementations of simpl
  systems,'' in {\em IFIP International Workshop on Information Security Theory
  and Practices}, pp.~277--292, Springer, 2010.

\bibitem{jaeger2010random}
C.~Jaeger, M.~Algasinger, U.~R{\"u}hrmair, G.~Csaba, and M.~Stutzmann, ``Random
  pn-junctions for physical cryptography,'' {\em Applied Physics Letters},
  vol.~96, no.~17, p.~172103, 2010.

\bibitem{katzenbeisser2011recyclable}
S.~Katzenbeisser, {\"U}.~Kocaba{\c{s}}, V.~Van Der~Leest, A.-R. Sadeghi, G.-J.
  Schrijen, and C.~Wachsmann, ``Recyclable pufs: Logically reconfigurable
  pufs,'' {\em Journal of Cryptographic Engineering}, vol.~1, no.~3,
  pp.~177--186, 2011.

\bibitem{delvaux2019machine}
J.~Delvaux, ``Machine-learning attacks on polypufs, ob-pufs, rpufs, lhs-pufs,
  and puf--fsms,'' {\em IEEE Transactions on Information Forensics and
  Security}, vol.~14, no.~8, pp.~2043--2058, 2019.

\bibitem{nedospasov2013invasive}
D.~Nedospasov, J.-P. Seifert, C.~Helfmeier, and C.~Boit, ``Invasive puf
  analysis,'' in {\em 2013 Workshop on Fault Diagnosis and Tolerance in
  Cryptography}, pp.~30--38, IEEE, 2013.

\bibitem{koeberl2013memristor}
P.~Koeberl, {\"U}.~Kocaba{\c{s}}, and A.-R. Sadeghi, ``Memristor pufs: a new
  generation of memory-based physically unclonable functions,'' in {\em 2013
  Design, Automation \& Test in Europe Conference \& Exhibition (DATE)},
  pp.~428--431, IEEE, 2013.

\bibitem{chatterjee2018rf}
B.~Chatterjee, D.~Das, S.~Maity, and S.~Sen, ``Rf-puf: Enhancing iot security
  through authentication of wireless nodes using in-situ machine learning,''
  {\em IEEE Internet of Things Journal}, vol.~6, no.~1, pp.~388--398, 2018.

\bibitem{mazady2015memristor}
A.~Mazady, M.~T. Rahman, D.~Forte, and M.~Anwar, ``Memristor puf—a security
  primitive: Theory and experiment,'' {\em IEEE Journal on Emerging and
  Selected Topics in Circuits and Systems}, vol.~5, no.~2, pp.~222--229, 2015.

\bibitem{zhang2014survey}
J.-L. Zhang, G.~Qu, Y.-Q. Lv, and Q.~Zhou, ``A survey on silicon pufs and
  recent advances in ring oscillator pufs,'' {\em Journal of computer science
  and technology}, vol.~29, no.~4, pp.~664--678, 2014.

\bibitem{shi2019approximation}
J.~Shi, Y.~Lu, and J.~Zhang, ``Approximation attacks on strong pufs,'' {\em
  IEEE transactions on computer-aided design of integrated circuits and
  systems}, vol.~39, no.~10, pp.~2138--2151, 2019.

\bibitem{maes2008intrinsic}
R.~Maes, P.~Tuyls, and I.~Verbauwhede, ``Intrinsic pufs from flip-flops on
  reconfigurable devices,'' in {\em 3rd Benelux workshop on information and
  system security (WISSec 2008)}, vol.~17, p.~2008, 2008.

\bibitem{yu2016lockdown}
M.-D. Yu, M.~Hiller, J.~Delvaux, R.~Sowell, S.~Devadas, and I.~Verbauwhede, ``A
  lockdown technique to prevent machine learning on pufs for lightweight
  authentication,'' {\em IEEE Transactions on Multi-Scale Computing Systems},
  vol.~2, no.~3, pp.~146--159, 2016.

\bibitem{delvaux2013side}
J.~Delvaux and I.~Verbauwhede, ``Side channel modeling attacks on 65nm arbiter
  pufs exploiting cmos device noise,'' in {\em 2013 IEEE International
  Symposium on Hardware-Oriented Security and Trust (HOST)}, pp.~137--142,
  IEEE, 2013.

\bibitem{grubel2017silicon}
B.~C. Grubel, B.~T. Bosworth, M.~R. Kossey, H.~Sun, A.~B. Cooper, M.~A. Foster,
  and A.~C. Foster, ``Silicon photonic physical unclonable function,'' {\em
  Optics Express}, vol.~25, no.~11, pp.~12710--12721, 2017.

\bibitem{horstmeyer2013physical}
R.~Horstmeyer, B.~Judkewitz, I.~M. Vellekoop, S.~Assawaworrarit, and C.~Yang,
  ``Physical key-protected one-time pad,'' {\em Scientific reports}, vol.~3,
  no.~1, pp.~1--6, 2013.

\bibitem{buchanan2005fingerprinting}
J.~D. Buchanan, R.~P. Cowburn, A.-V. Jausovec, D.~Petit, P.~Seem, G.~Xiong,
  D.~Atkinson, K.~Fenton, D.~A. Allwood, and M.~T. Bryan,
  ``‘fingerprinting’documents and packaging,'' {\em Nature}, vol.~436,
  no.~7050, pp.~475--475, 2005.

\bibitem{dejean2007rf}
G.~DeJean and D.~Kirovski, ``Rf-dna: Radio-frequency certificates of
  authenticity,'' in {\em International Workshop on Cryptographic Hardware and
  Embedded Systems}, pp.~346--363, Springer, 2007.

\bibitem{lakafosis2010rf}
V.~Lakafosis, A.~Traille, H.~Lee, E.~Gebara, M.~M. Tentzeris, G.~R. DeJean, and
  D.~Kirovski, ``Rf fingerprinting physical objects for anticounterfeiting
  applications,'' {\em IEEE Transactions on Microwave Theory and Techniques},
  vol.~59, no.~2, pp.~504--514, 2010.

\bibitem{zalivaka2018reliable}
S.~S. Zalivaka, A.~A. Ivaniuk, and C.-H. Chang, ``Reliable and modeling attack
  resistant authentication of arbiter puf in fpga implementation with trinary
  quadruple response,'' {\em IEEE Transactions on Information Forensics and
  Security}, vol.~14, no.~4, pp.~1109--1123, 2018.

\bibitem{liu2017acro}
C.~Q. Liu, Y.~Cao, and C.~H. Chang, ``Acro-puf: A low-power, reliable and
  aging-resilient current starved inverter-based ring oscillator physical
  unclonable function,'' {\em IEEE Transactions on Circuits and Systems I:
  Regular Papers}, vol.~64, no.~12, pp.~3138--3149, 2017.

\bibitem{john2021halide}
R.~A. John, N.~Shah, S.~K. Vishwanath, S.~E. Ng, B.~Febriansyah,
  M.~Jagadeeswararao, C.-H. Chang, A.~Basu, and N.~Mathews, ``Halide perovskite
  memristors as flexible and reconfigurable physical unclonable functions,''
  {\em Nature Communications}, vol.~12, no.~1, pp.~1--11, 2021.

\bibitem{rosenfeld2010sensor}
K.~Rosenfeld, E.~Gavas, and R.~Karri, ``Sensor physical unclonable functions,''
  in {\em 2010 IEEE international symposium on hardware-oriented security and
  trust (HOST)}, pp.~112--117, IEEE, 2010.

\bibitem{rose2013hardware}
G.~S. Rose, J.~Rajendran, N.~McDonald, R.~Karri, M.~Potkonjak, and B.~Wysocki,
  ``Hardware security strategies exploiting nanoelectronic circuits,'' in {\em
  2013 18th Asia and South Pacific Design Automation Conference (ASP-DAC)},
  pp.~368--372, IEEE, 2013.

\bibitem{tang2016securing}
J.~Tang, R.~Karri, and J.~Rajendran, ``Securing pressure measurements using
  sensorpufs,'' in {\em 2016 IEEE International Symposium on Circuits and
  Systems (ISCAS)}, pp.~1330--1333, IEEE, 2016.

\bibitem{majzoobi2012slender}
M.~Majzoobi, M.~Rostami, F.~Koushanfar, D.~S. Wallach, and S.~Devadas,
  ``Slender puf protocol: A lightweight, robust, and secure authentication by
  substring matching,'' in {\em 2012 IEEE Symposium on Security and Privacy
  Workshops}, pp.~33--44, IEEE, 2012.

\bibitem{rostami2014robust}
M.~Rostami, M.~Majzoobi, F.~Koushanfar, D.~S. Wallach, and S.~Devadas, ``Robust
  and reverse-engineering resilient puf authentication and key-exchange by
  substring matching,'' {\em IEEE Transactions on Emerging Topics in
  Computing}, vol.~2, no.~1, pp.~37--49, 2014.

\bibitem{maes2009soft}
R.~Maes, P.~Tuyls, and I.~Verbauwhede, ``A soft decision helper data algorithm
  for sram pufs,'' in {\em 2009 IEEE international symposium on information
  theory}, pp.~2101--2105, IEEE, 2009.

\bibitem{lugli2013physical}
P.~Lugli, A.~Mahmoud, G.~Csaba, M.~Algasinger, M.~Stutzmann, and
  U.~R{\"u}hrmair, ``Physical unclonable functions based on crossbar arrays for
  cryptographic applications,'' {\em International journal of circuit theory
  and applications}, vol.~41, no.~6, pp.~619--633, 2013.

\bibitem{ruhrmair2010applications}
U.~R{\"u}hrmair, C.~Jaeger, M.~Bator, M.~Stutzmann, P.~Lugli, and G.~Csaba,
  ``Applications of high-capacity crossbar memories in cryptography,'' {\em
  IEEE Transactions on Nanotechnology}, vol.~10, no.~3, pp.~489--498, 2010.

\bibitem{ruhrmair2012simpl}
U.~R{\"u}hrmair, ``Simpl systems as a keyless cryptographic and security
  primitive,'' in {\em Cryptography and Security: From Theory to Applications},
  pp.~329--354, Springer, 2012.

\bibitem{sauer2017sensitized}
M.~Sauer, P.~Raiola, L.~Feiten, B.~Becker, U.~R{\"u}hrmair, and I.~Polian,
  ``Sensitized path puf: A lightweight embedded physical unclonable function,''
  in {\em Design, Automation \& Test in Europe Conference \& Exhibition (DATE),
  2017}, pp.~680--685, IEEE, 2017.

\bibitem{kappelhoff2022strong}
F.~Kappelhoff, R.~Rasche, D.~Mukhopadhyay, and U.~R{\"u}hrmair, ``Strong puf
  security metrics: Response sensitivity to small challenge perturbations,'' in
  {\em 2022 23rd International Symposium on Quality Electronic Design (ISQED)},
  pp.~1--10, IEEE, 2022.

\bibitem{langhuth2011strong}
H.~Langhuth, S.~Fr{\'e}d{\'e}rick, M.~Kaniber, J.~J. Finley, and
  U.~R{\"u}hrmair, ``Strong photoluminescence enhancement from colloidal
  quantum dot near silver nano-island films,'' {\em Journal of fluorescence},
  vol.~21, no.~2, pp.~539--543, 2011.

\bibitem{csaba2010application}
G.~Csaba, X.~Ju, Z.~Ma, Q.~Chen, W.~Porod, J.~Schmidhuber, U.~Schlichtmann,
  P.~Lugli, and U.~R{\"u}hrmair, ``Application of mismatched cellular nonlinear
  networks for physical cryptography,'' in {\em 2010 12th International
  Workshop on Cellular Nanoscale Networks and their Applications (CNNA 2010)},
  pp.~1--6, IEEE, 2010.

\bibitem{ruhrmair2015virtual}
U.~R{\"u}hrmair, J.~Martinez-Hurtado, X.~Xu, C.~Kraeh, C.~Hilgers,
  D.~Kononchuk, J.~J. Finley, and W.~P. Burleson, ``Virtual proofs of reality
  and their physical implementation,'' in {\em 2015 IEEE Symposium on Security
  and Privacy}, pp.~70--85, IEEE, 2015.

\bibitem{ruhrmair2011simpl}
U.~R{\"u}hrmair, ``Simpl systems, or: can we design cryptographic hardware
  without secret key information?,'' in {\em International Conference on
  Current Trends in Theory and Practice of Computer Science}, pp.~26--45,
  Springer, 2011.

\bibitem{ruhrmair2009simpl}
U.~R{\"u}hrmair, ``Simpl systems: On a public key variant of physical
  unclonable functions,'' {\em Cryptology ePrint Archive}, 2009.

\bibitem{jin2020erasable}
C.~Jin, W.~Burleson, M.~van Dijk, and U.~R{\"u}hrmair, ``Erasable pufs: formal
  treatment and generic design,'' in {\em Proceedings of the 4th ACM Workshop
  on Attacks and Solutions in Hardware Security}, pp.~21--33, 2020.

\bibitem{jin2015playpuf}
C.~Jin, X.~Xu, W.~Burleson, U.~R{\"u}hrmair, and M.~van Dijk, ``Playpuf:
  programmable logically erasable pufs for forward and backward secure key
  management,'' {\em Cryptology ePrint Archive}, 2015.

\bibitem{orosa2022spyhammer}
L.~Orosa, U.~R{\"u}hrmair, A.~G. Yaglikci, H.~Luo, A.~Olgun, P.~Jattke,
  M.~Patel, J.~Kim, K.~Razavi, and O.~Mutlu, ``Spyhammer: Using rowhammer to
  remotely spy on temperature,'' {\em arXiv preprint arXiv:2210.04084}, 2022.

\bibitem{eliezer2022exploiting}
Y.~Eliezer, U.~Ruhrmair, N.~Wisiol, S.~Bittner, and H.~Cao, ``Exploiting
  structural nonlinearity of a reconfigurable multiple-scattering system,''
  {\em arXiv preprint arXiv:2208.08906}, 2022.

\bibitem{jin2022programmable}
C.~Jin, W.~Burleson, M.~van Dijk, and U.~R{\"u}hrmair, ``Programmable
  access-controlled and generic erasable puf design and its applications,''
  {\em Journal of Cryptographic Engineering}, pp.~1--20, 2022.

\bibitem{chen2011circuit}
Q.~Chen, G.~Csaba, P.~Lugli, U.~Schlichtmann, M.~Stutzmann, and U.~Ruehrmair,
  ``Circuit-based approaches to simpl systems,'' {\em Journal of Circuits,
  Systems, and Computers}, vol.~20, no.~01, pp.~107--123, 2011.

\bibitem{gao2018efficient}
Y.~Gao, C.~Jin, J.~Kim, H.~Nili, X.~Xu, W.~Burleson, O.~Kavehei, M.~van Dijk,
  D.~C. Ranasinghe, and U.~R{\"u}hrmair, ``Efficient erasable pufs from
  programmable logic and memristors,'' {\em Cryptology ePrint Archive}, 2018.

\bibitem{csaba2009chip}
G.~Csaba, X.~Ju, Q.~Chen, W.~Porod, J.~Schmidhuber, U.~Schlichtmann, P.~Lugli,
  and U.~R{\"u}hrmair, ``On-chip electric waves: An analog circuit approach to
  physical uncloneable functions,'' {\em Cryptology ePrint Archive}, 2009.

\bibitem{pavanello2021recent}
F.~Pavanello, I.~O’Connor, U.~R{\"u}hrmair, A.~C. Foster, and D.~Syvridis,
  ``Recent advances in photonic physical unclonable functions,'' in {\em 2021
  IEEE European Test Symposium (ETS)}, pp.~1--10, IEEE, 2021.

\bibitem{ruhrmair2014pufs}
U.~R{\"u}hrmair and D.~E. Holcomb, ``Pufs at a glance,'' in {\em 2014 Design,
  Automation \& Test in Europe Conference \& Exhibition (DATE)}, pp.~1--6,
  IEEE, 2014.

\bibitem{ruhrmair2019towards}
U.~R{\"u}hrmair, ``Towards secret-free security,'' {\em Cryptology ePrint
  Archive}, 2019.

\bibitem{ruhrmair2020sok}
U.~Ruhrmair, ``Sok: Towards secret-free security,'' in {\em Proceedings of the
  4th ACM Workshop on Attacks and Solutions in Hardware Security}, pp.~5--19,
  2020.

\bibitem{ruhrmair2022secret}
U.~R{\"u}hrmair, ``Secret-free security: A survey and tutorial,'' {\em Journal
  of Cryptographic Engineering}, pp.~1--26, 2022.

\bibitem{ruhrmair2012security}
U.~R{\"u}hrmair, S.~Devadas, and F.~Koushanfar, ``Security based on physical
  unclonability and disorder,'' in {\em Introduction to Hardware Security and
  Trust}, pp.~65--102, Springer, 2012.

\bibitem{ruhrmair2009foundations}
U.~R{\"u}hrmair, J.~S{\"o}lter, and F.~Sehnke, ``On the foundations of physical
  unclonable functions,'' {\em Cryptology ePrint Archive}, 2009.

\bibitem{searle2016linear}
S.~R. Searle and M.~H. Gruber, {\em Linear models}.
\newblock John Wiley \& Sons, 2016.

\bibitem{mccullagh2019generalized}
P.~McCullagh, ``Generalized linear models,'' 2019.

\bibitem{rencher2008linear}
A.~C. Rencher and G.~B. Schaalje, {\em Linear models in statistics}.
\newblock John Wiley \& Sons, 2008.

\bibitem{roehe2000estimation}
R.~Roehe, E.~Kalm, {\em et~al.}, ``Estimation of genetic and environmental risk
  factors associated with pre-weaning mortality in piglets using generalized
  linear mixed models,'' {\em ANIMAL SCIENCE-GLASGOW-}, vol.~70, no.~2,
  pp.~227--240, 2000.

\bibitem{pirinen2013efficient}
M.~Pirinen, P.~Donnelly, C.~C. Spencer, {\em et~al.}, ``Efficient computation
  with a linear mixed model on large-scale data sets with applications to
  genetic studies,'' {\em The Annals of Applied Statistics}, vol.~7, no.~1,
  pp.~369--390, 2013.

\bibitem{chursin2017linear}
A.~Chursin, P.~Drogovoz, T.~Sadovskaya, and V.~Shiboldenkov, ``A linear model
  of economic and technological shocks in science-intensive industries,'' {\em
  Jour. App. Econ. Scien.}, vol.~6, no.~52, 2017.

\bibitem{henshaw1966application}
R.~C. Henshaw~Jr, ``Application of the general linear model to seasonal
  adjustment of economic time series,'' {\em Econometrica: Journal of the
  Econometric Society}, pp.~381--395, 1966.

\bibitem{brehmer1994psychology}
B.~Brehmer, ``The psychology of linear judgement models,'' {\em Acta
  Psychologica}, vol.~87, no.~2-3, pp.~137--154, 1994.

\bibitem{dawes1974linear}
R.~M. Dawes and B.~Corrigan, ``Linear models in decision making.,'' {\em
  Psychological bulletin}, vol.~81, no.~2, p.~95, 1974.

\bibitem{serban2015hierarchical}
I.~V. Serban, A.~Sordoni, Y.~Bengio, A.~Courville, and J.~Pineau,
  ``Hierarchical neural network generative models for movie dialogues,'' {\em
  arXiv preprint arXiv:1507.04808}, vol.~7, no.~8, pp.~434--441, 2015.

\bibitem{serban2016building}
I.~Serban, A.~Sordoni, Y.~Bengio, A.~Courville, and J.~Pineau, ``Building
  end-to-end dialogue systems using generative hierarchical neural network
  models,'' in {\em Proceedings of the AAAI Conference on Artificial
  Intelligence}, vol.~30, 2016.

\bibitem{mosser2017reconstruction}
L.~Mosser, O.~Dubrule, and M.~J. Blunt, ``Reconstruction of three-dimensional
  porous media using generative adversarial neural networks,'' {\em Physical
  Review E}, vol.~96, no.~4, p.~043309, 2017.

\bibitem{rivenson2018phase}
Y.~Rivenson, Y.~Zhang, H.~G{\"u}nayd{\i}n, D.~Teng, and A.~Ozcan, ``Phase
  recovery and holographic image reconstruction using deep learning in neural
  networks,'' {\em Light: Science \& Applications}, vol.~7, no.~2,
  pp.~17141--17141, 2018.

\bibitem{wang2019generative}
G.~Wang, A.~Ledwoch, R.~M. Hasani, R.~Grosu, and A.~Brintrup, ``A generative
  neural network model for the quality prediction of work in progress
  products,'' {\em Applied Soft Computing}, vol.~85, p.~105683, 2019.

\bibitem{khodayar2018interval}
M.~Khodayar, J.~Wang, and M.~Manthouri, ``Interval deep generative neural
  network for wind speed forecasting,'' {\em IEEE Transactions on Smart Grid},
  vol.~10, no.~4, pp.~3974--3989, 2018.

\bibitem{gregor2015draw}
K.~Gregor, I.~Danihelka, A.~Graves, D.~Rezende, and D.~Wierstra, ``Draw: A
  recurrent neural network for image generation,'' in {\em International
  Conference on Machine Learning}, pp.~1462--1471, PMLR, 2015.

\bibitem{yan2016attribute2image}
X.~Yan, J.~Yang, K.~Sohn, and H.~Lee, ``Attribute2image: Conditional image
  generation from visual attributes,'' in {\em European Conference on Computer
  Vision}, pp.~776--791, Springer, 2016.

\bibitem{taigman2016unsupervised}
Y.~Taigman, A.~Polyak, and L.~Wolf, ``Unsupervised cross-domain image
  generation,'' {\em arXiv preprint arXiv:1611.02200}, 2016.

\bibitem{oord2016conditional}
A.~v.~d. Oord, N.~Kalchbrenner, O.~Vinyals, L.~Espeholt, A.~Graves, and
  K.~Kavukcuoglu, ``Conditional image generation with pixelcnn decoders,'' {\em
  arXiv preprint arXiv:1606.05328}, 2016.

\bibitem{nguyen2015efficient}
P.~H. Nguyen, D.~P. Sahoo, R.~S. Chakraborty, and D.~Mukhopadhyay, ``Efficient
  attacks on robust ring oscillator puf with enhanced challenge-response set,''
  in {\em 2015 Design, Automation \& Test in Europe Conference \& Exhibition
  (DATE)}, pp.~641--646, IEEE, 2015.

\bibitem{modellingRuehrmair}
U.~R\"{u}hrmair, F.~Sehnke, J.~S\"{o}lter, G.~Dror, S.~Devadas, and
  J.~Schmidhuber, ``Modeling attacks on physical unclonable functions,'' in
  {\em Proceedings of the 17th ACM Conference on Computer and Communications
  Security}, CCS '10, (New York, NY, USA), p.~237–249, Association for
  Computing Machinery, 2010.

\bibitem{modellingRuehrmair2}
U.~{Rührmair}, J.~{Sölter}, F.~{Sehnke}, X.~{Xu}, A.~{Mahmoud},
  V.~{Stoyanova}, G.~{Dror}, J.~{Schmidhuber}, W.~{Burleson}, and S.~{Devadas},
  ``Puf modeling attacks on simulated and silicon data,'' {\em IEEE
  Transactions on Information Forensics and Security}, vol.~8, no.~11,
  pp.~1876--1891, 2013.

\bibitem{ruhrmair2014puf}
U.~R{\"u}hrmair and J.~S{\"o}lter, ``Puf modeling attacks: An introduction and
  overview,'' in {\em 2014 Design, Automation \& Test in Europe Conference \&
  Exhibition (DATE)}, pp.~1--6, IEEE, 2014.

\bibitem{sehnke2010policy}
F.~Sehnke, C.~Osendorfer, J.~S{\"o}lter, J.~Schmidhuber, and U.~R{\"u}hrmair,
  ``Policy gradients for cryptanalysis,'' in {\em International Conference on
  Artificial Neural Networks}, pp.~168--177, Springer, 2010.

\bibitem{sideChannelRuehrmair}
U.~R{\"u}hrmair, X.~Xu, J.~S{\"o}lter, A.~Mahmoud, M.~Majzoobi, F.~Koushanfar,
  and W.~Burleson, ``Efficient power and timing side channels for physical
  unclonable functions,'' in {\em Cryptographic Hardware and Embedded Systems
  -- CHES 2014} (L.~Batina and M.~Robshaw, eds.), (Berlin, Heidelberg),
  pp.~476--492, Springer Berlin Heidelberg, 2014.

\bibitem{sideChannelTajik}
S.~{Tajik}, H.~{Lohrke}, F.~{Ganji}, J.~{Seifert}, and C.~{Boit}, ``Laser fault
  attack on physically unclonable functions,'' in {\em 2015 Workshop on Fault
  Diagnosis and Tolerance in Cryptography (FDTC)}, pp.~85--96, 2015.

\bibitem{ruhrmair2013power}
U.~R{\"u}hrmair, X.~Xu, J.~S{\"o}lter, A.~Mahmoud, F.~Koushanfar, and
  W.~Burleson, ``Power and timing side channels for pufs and their efficient
  exploitation,'' {\em Cryptology ePrint Archive}, 2013.

\bibitem{ruhrmair2014special}
U.~R{\"u}hrmair, U.~Schlichtmann, and W.~Burleson, ``Special session: How
  secure are pufs really? on the reach and limits of recent puf attacks,'' in
  {\em 2014 Design, Automation \& Test in Europe Conference \& Exhibition
  (DATE)}, pp.~1--4, IEEE, 2014.

\bibitem{helfmeier2013cloning}
C.~Helfmeier, C.~Boit, D.~Nedospasov, and J.-P. Seifert, ``Cloning physically
  unclonable functions,'' in {\em 2013 IEEE International Symposium on
  Hardware-Oriented Security and Trust (HOST)}, pp.~1--6, IEEE, 2013.

\bibitem{ruhrmair2013pufs}
U.~R{\"u}hrmair and M.~van Dijk, ``Pufs in security protocols: Attack models
  and security evaluations,'' in {\em 2013 IEEE symposium on security and
  privacy}, pp.~286--300, IEEE, 2013.

\bibitem{ruhrmair2010strong}
U.~R{\"u}hrmair, H.~Busch, and S.~Katzenbeisser, ``Strong pufs: models,
  constructions, and security proofs,'' in {\em Towards hardware-intrinsic
  security}, pp.~79--96, Springer, 2010.

\bibitem{ruhrmair2010oblivious}
U.~R{\"u}hrmair, ``Oblivious transfer based on physical unclonable functions,''
  in {\em International Conference on Trust and Trustworthy Computing},
  pp.~430--440, Springer, 2010.

\bibitem{ruhrmair2013practical}
U.~R{\"u}hrmair and M.~van Dijk, ``On the practical use of physical unclonable
  functions in oblivious transfer and bit commitment protocols,'' {\em Journal
  of Cryptographic Engineering}, vol.~3, no.~1, pp.~17--28, 2013.

\bibitem{ruhrmair2012practical}
U.~R{\"u}hrmair and M.~v. Dijk, ``Practical security analysis of puf-based
  two-player protocols,'' in {\em International Workshop on Cryptographic
  Hardware and Embedded Systems}, pp.~251--267, Springer, 2012.

\bibitem{ruhrmair2011physical}
U.~R{\"u}hrmair, ``Physical turing machines and the formalization of physical
  cryptography,'' {\em Cryptology ePrint Archive}, 2011.

\bibitem{van2014protocol}
M.~van Dijk and U.~R{\"u}hrmair, ``Protocol attacks on advanced puf protocols
  and countermeasures,'' in {\em 2014 Design, Automation \& Test in Europe
  Conference \& Exhibition (DATE)}, pp.~1--6, IEEE, 2014.

\bibitem{ruhrmair2016security}
U.~R{\"u}hrmair, ``On the security of puf protocols under bad pufs and
  pufs-inside-pufs attacks,'' {\em Cryptology ePrint Archive}, 2016.

\bibitem{dachman2014feasibility}
D.~Dachman-Soled, N.~Fleischhacker, J.~Katz, A.~Lysyanskaya, and
  D.~Schr{\"o}der, ``Feasibility and infeasibility of secure computation with
  malicious pufs,'' in {\em Annual Cryptology Conference}, pp.~405--420,
  Springer, 2014.

\bibitem{ostrovsky2013universally}
R.~Ostrovsky, A.~Scafuro, I.~Visconti, and A.~Wadia, ``Universally composable
  secure computation with (malicious) physically uncloneable functions,'' in
  {\em Annual International Conference on the Theory and Applications of
  Cryptographic Techniques}, pp.~702--718, Springer, 2013.

\bibitem{gabor_daugman}
J.~G. Daugman, ``Uncertainty relation for resolution in space, spatial
  frequency, and orientation optimized by two-dimensional visual cortical
  filters,'' {\em JOSA A}, vol.~2, no.~7, pp.~1160--1169, 1985.

\bibitem{shalev2014understanding}
S.~Shalev-Shwartz and S.~Ben-David, {\em Understanding machine learning: From
  theory to algorithms}.
\newblock Cambridge university press, 2014.

\bibitem{sathya2013comparison}
R.~Sathya and A.~Abraham, ``Comparison of supervised and unsupervised learning
  algorithms for pattern classification,'' {\em International Journal of
  Advanced Research in Artificial Intelligence}, vol.~2, no.~2, pp.~34--38,
  2013.

\bibitem{cunningham2008supervised}
P.~Cunningham, M.~Cord, and S.~J. Delany, ``Supervised learning,'' in {\em
  Machine learning techniques for multimedia}, pp.~21--49, Springer, 2008.

\bibitem{gross2012linear}
J.~Gro{\ss}, {\em Linear regression}, vol.~175.
\newblock Springer Science \& Business Media, 2012.

\bibitem{yuan2007dimension}
M.~Yuan, A.~Ekici, Z.~Lu, and R.~Monteiro, ``Dimension reduction and
  coefficient estimation in multivariate linear regression,'' {\em Journal of
  the Royal Statistical Society: Series B (Statistical Methodology)}, vol.~69,
  no.~3, pp.~329--346, 2007.

\bibitem{priddy2005artificial}
K.~L. Priddy and P.~E. Keller, {\em Artificial neural networks: an
  introduction}, vol.~68.
\newblock SPIE press, 2005.

\bibitem{idoko2020deep}
J.~B. IDOKO, {\em DEEP LEARNING-BASED SIGN LANGUAGE TRANSLATION SYSTEM}.
\newblock PhD thesis, NEAR EAST UNIVERSITY, 2020.

\bibitem{albawi2017understanding}
S.~Albawi, T.~A. Mohammed, and S.~Al-Zawi, ``Understanding of a convolutional
  neural network,'' in {\em 2017 International Conference on Engineering and
  Technology (ICET)}, pp.~1--6, Ieee, 2017.

\bibitem{dumoulin2016guide}
V.~Dumoulin and F.~Visin, ``A guide to convolution arithmetic for deep
  learning,'' {\em arXiv preprint arXiv:1603.07285}, 2016.

\bibitem{Goodfellow-et-al-2016}
I.~Goodfellow, Y.~Bengio, and A.~Courville, {\em Deep Learning}.
\newblock MIT Press, 2016.
\newblock \url{http://www.deeplearningbook.org}.

\bibitem{shannon2001mathematical}
C.~E. Shannon, ``A mathematical theory of communication,'' {\em ACM SIGMOBILE
  mobile computing and communications review}, vol.~5, no.~1, pp.~3--55, 2001.

\bibitem{lee1988thirteen}
J.~Lee~Rodgers and W.~A. Nicewander, ``Thirteen ways to look at the correlation
  coefficient,'' {\em The American Statistician}, vol.~42, no.~1, pp.~59--66,
  1988.

\bibitem{wang2004image}
Z.~Wang, A.~C. Bovik, H.~R. Sheikh, and E.~P. Simoncelli, ``Image quality
  assessment: from error visibility to structural similarity,'' {\em IEEE
  transactions on image processing}, vol.~13, no.~4, pp.~600--612, 2004.

\bibitem{goodfellow2014generative}
I.~J. Goodfellow, J.~Pouget-Abadie, M.~Mirza, B.~Xu, D.~Warde-Farley, S.~Ozair,
  A.~Courville, and Y.~Bengio, ``Generative adversarial networks,'' {\em arXiv
  preprint arXiv:1406.2661}, 2014.

\bibitem{diffractio}
L.~M.~S. Brea, ``Diffractio.''
  \url{https://diffractio.readthedocs.io/en/latest/} v. 10.7.0.0.11.
\newblock accessed: 2020-03-09.

\bibitem{poon2017engineering}
T.-C. Poon and T.~Kim, {\em Engineering optics with Matlab{\textregistered}}.
\newblock World Scientific Publishing Company, 2017.

\bibitem{rukhin2001statistical}
A.~Rukhin, J.~Soto, J.~Nechvatal, M.~Smid, and E.~Barker, ``A statistical test
  suite for random and pseudorandom number generators for cryptographic
  applications,'' tech. rep., Booz-allen and hamilton inc mclean va, 2001.

\bibitem{marsaglia2008marsaglia}
G.~Marsaglia, ``The marsaglia random number cdrom including the diehard battery
  of tests of randomness,'' {\em http://www.stat.fsu.edu/pub/diehard/}, 2008.

\bibitem{10.1145/1268776.1268777}
P.~L'Ecuyer and R.~Simard, ``Testu01: A c library for empirical testing of
  random number generators,'' {\em ACM Trans. Math. Softw.}, vol.~33, Aug.
  2007.

\bibitem{hoerl1970ridge}
A.~E. Hoerl and R.~W. Kennard, ``Ridge regression: Biased estimation for
  nonorthogonal problems,'' {\em Technometrics}, vol.~12, no.~1, pp.~55--67,
  1970.

\end{thebibliography}
    \end{spacing}
\end{document}